\documentclass[pdftex, 11pt]{article}
\usepackage[utf8]{inputenc}
\usepackage{longtable}
\usepackage[pdftex]{graphicx}
\usepackage[pdftex,                %%% hyper-references for pdflatex
bookmarks=true,%                   %%% generate bookmarks ...
bookmarksnumbered=true,%           %%% ... with numbers
hypertexnames=false,%              %%% needed for correct links to figures !!!
breaklinks=true,%                  %%% break links if exceeding a single line
linkbordercolor={0 0 1}]{hyperref} %%% blue frames around links
\usepackage{tabularx}    
\usepackage{import}      
\usepackage{amsmath, amssymb} 
\usepackage{pdfpages}
\usepackage{xcolor,colortbl}
\usepackage{multirow}
\usepackage[24hr,mmddyyyy]{datetime}
\usepackage{hyphenat}			% supression hyphenation in a word or phrase
\usepackage{lineno}
\usepackage{booktabs}			%\midrule
\usepackage{textcmds}	%% \mdash used by BibTex
\usepackage{xspace}
\usepackage{longtable}	%% for acronyms appendix
\usepackage{bm}		%% bold math
\usepackage{afterpage}
\usepackage[most]{tcolorbox}

\newcommand{\gsim}{\raisebox{-0.13cm}{~\shortstack{$>$ \\[-0.07cm]
      $\sim$}}~}

\newcommand\arcsec{\hbox{$^{\prime\prime}$}}

\usepackage[margin=0pt,font=small,labelfont=bf,labelsep=colon]{caption}

\usepackage[citestyle=authoryear,doi=false,natbib=true,backend=bibtex]{biblatex}

\usepackage{aas_macros}

\def\bfseries{\fontseries \bfdefault \selectfont \boldmath}

\usepackage[margin=0.25in]{caption}

    \usepackage[outermarks]{titlesec}
    \titleformat{\section}
      {\normalfont\Large\bfseries}{\thesection}{1em}{}
    \titleformat{\subsection}
      {\normalfont\large\bfseries}{\thesubsection}{1em}{}
    \titleformat{\subsubsection}
      {\normalfont\normalsize\bfseries}{\thesubsubsection}{1em}{}
    \titleformat{\paragraph}
      {\normalfont\normalsize\bfseries}{\theparagraph}{1em}{}
    \titleformat{\subparagraph}
      {\normalfont\normalsize\bfseries}{\thesubparagraph}{1em}{}

    \titlespacing*{\section}      {0pt}{3.5ex plus 1ex minus .2ex} {2.3ex plus .2ex}
    \titlespacing*{\subsection}   {0pt}{3.25ex plus 1ex minus .2ex}{1.5ex plus .2ex}
    \titlespacing*{\subsubsection}{0pt}{3.25ex plus 1ex minus .2ex}{1.5ex plus .2ex}
    \titlespacing*{\paragraph}    {0pt}{3.25ex plus 1ex minus .2ex}{1.5ex plus .2ex}
    \titlespacing*{\subparagraph} {0pt}{3.25ex plus 1ex minus .2ex}{1.5ex plus .2ex}

\newenvironment{subsec}[1][]{\par\noindent\begin{center}\begin{minipage}{16cm}}{\end{minipage}\end{center}}

\newcommand{\nai}{\mbox{Na\,{\scshape i}}}
\newcommand{\ki}{\mbox{K\,{\scshape i}}}
\newcommand{\hi}{\mbox{H\,{\scshape i}}}
\newcommand{\hii}{\mbox{H\,{\scshape ii}}}
\newcommand{\caii}{\mbox{Ca\,{\scshape ii}}}
\newcommand{\oii}{\mbox{O\,{\scshape ii}}}
\newcommand{\oiii}{\mbox{O\,{\scshape iii}}}
\newcommand{\cii}{\mbox{C\,{\scshape ii}}}
\newcommand{\ciii}{\mbox{C\,{\scshape iii}}}
\newcommand{\civ}{\mbox{C\,{\scshape iv}}}
\newcommand{\ariv}{\mbox{Ar\,{\scshape iv}}}
\newcommand{\neii}{\mbox{Ne\,{\scshape ii}}}
\newcommand{\neiii}{\mbox{Ne\,{\scshape iii}}}
\newcommand{\nii}{\mbox{N\,{\scshape ii}}}
\newcommand{\niii}{\mbox{N\,{\scshape iii}}}
\newcommand{\niv}{\mbox{N\,{\scshape iv}}}
\newcommand{\nv}{\mbox{N\,{\scshape v}}}
\newcommand{\sii}{\mbox{S\,{\scshape ii}}}
\newcommand{\siii}{\mbox{Si\,{\scshape ii}}}
\newcommand{\siiii}{\mbox{Si\,{\scshape iii}}}
\newcommand{\siiv}{\mbox{Si\,{\scshape iv}}}
\newcommand{\hei}{\mbox{He\,{\scshape i}}}
\newcommand{\heii}{\mbox{He\,{\scshape ii}}}
\newcommand{\feii}{\mbox{Fe\,{\scshape ii}}}
\newcommand{\mgii}{\mbox{Mg\,{\scshape ii}}}
\newcommand{\ovi}{\mbox{O\,{\scshape vi}}}

\definecolor{Gray}{gray}{0.85}
\definecolor{LightCyan}{rgb}{0.88,1,1}
\newcolumntype{a}{>{\columncolor{LightCyan}}c}
\newcolumntype{b}{>{\columncolor{white}}c}

\usepackage{fancybox}

\begin{document}
\begin{titlepage}

\begin{tcolorbox}[minipage,colback=white,arc=2mm,outer arc=2mm,colframe=blue,left=30mm,right=30mm]

\begin{flushleft}
\hspace{-3cm}\includegraphics[width=5cm]{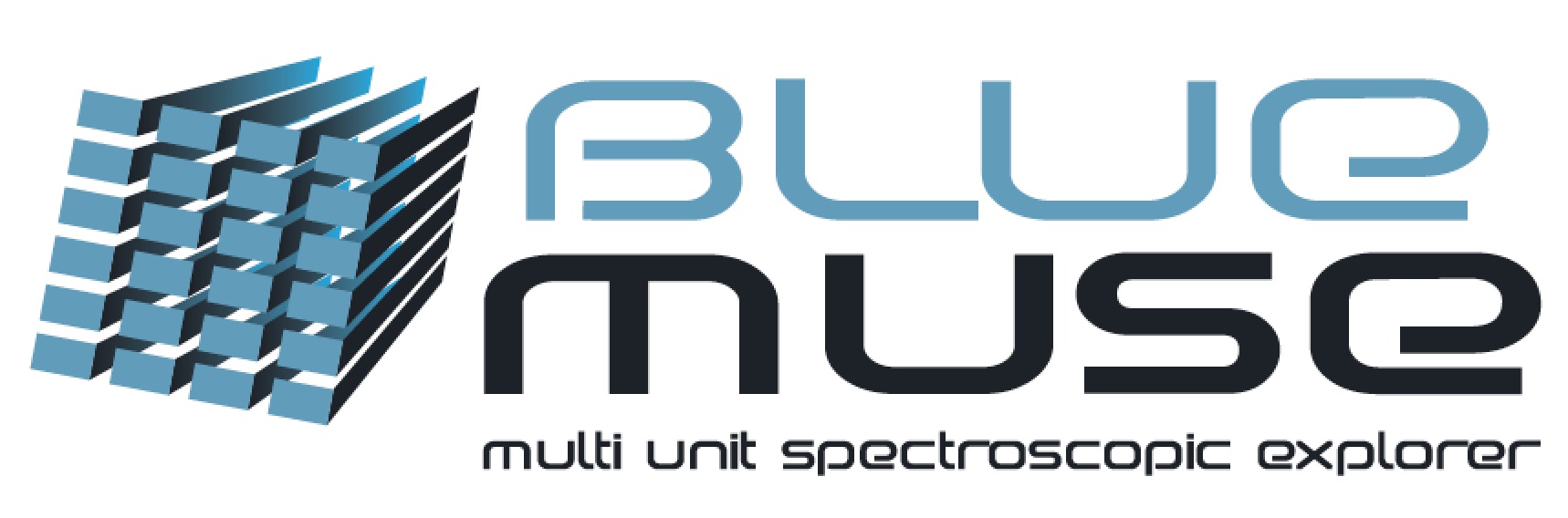}
\end{flushleft}

\centering
\begin{center}
\begin{Huge}
\textbf{BlueMUSE}
\end{Huge}
\vspace{1cm}

\begin{LARGE}
Project Overview and
\medskip\par
Science Cases
\end{LARGE}
\vspace{1cm}

\begin{Large}
June 4th, 2019
\vspace{1cm}

\end{Large}
\end{center}
\end{tcolorbox}

\begin{center}
Johan Richard$^{1}$, Roland Bacon$^1$, Jérémy Blaizot$^1$, Samuel Boissier$^2$, Alessandro Boselli$^2$, Nicolas Bouché$^1$, Jarle Brinchmann$^{3,4}$, Norberto Castro$^{5}$, Laure Ciesla$^2$, Paul Crowther$^{6}$, Emanuele Daddi$^7$, Stefan Dreizler$^8$, Pierre-Alain Duc$^9$, David Elbaz$^7$, Benoit \'Epinat$^2$, Chris Evans$^{10}$, Matteo Fossati$^{11}$, Michele Fumagalli$^{11}$, Miriam Garcia$^{12}$, Thibault Garel$^{1,13}$, Matthew Hayes$^{14}$, Angela Adamo$^{14}$, Artemio Herrero$^{15,16}$, Emmanuel Hugot$^2$, Andrew Humphrey$^{3}$, Pascale Jablonka$^{17}$, Sebastian Kamann$^{18}$, Lex Kaper$^{19}$, Andreas Kelz$^{5}$, Jean-Paul Kneib$^{17}$, Alex de Koter$^{19,20}$, Davor Krajnovi\'c$^{5}$, Rolf-Peter Kudritzki$^{21}$, Norbert Langer$^{22}$, Carmela Lardo$^{17}$, Floriane Leclercq$^{13}$, Danny Lennon$^{15}$, Guillaume Mahler$^{23}$, Fabrice Martins$^{24}$, Richard Massey$^{11}$, Peter Mitchell$^{4}$, Ana Monreal-Ibero$^{15,16}$,  Paco Najarro$^{12}$, Cyrielle Opitom$^{25}$, Polychronis Papaderos$^{3,26}$, Céline Péroux$^{28,2}$, Yves Revaz$^{17}$, Martin M. Roth$^{5}$, Philippe Rousselot$^{29}$, Andreas Sander$^{30}$, Charlotte Simmonds Wagemann$^{13}$,  Ian Smail$^{11}$, Anthony Mark Swinbank$^{11}$, Frank Tramper$^{31}$, Tanya Urrutia$^{5}$, Anne Verhamme$^{13}$, Jorick Vink$^{30}$, Jeremy Walsh$^{28}$, Peter Weilbacher$^5$, Martin Wendt$^{32}$, Lutz Wisotzki$^{5}$, Bin Yang$^{25}$.

\end{center}

\begin{center}
\textbf{\Large{Abstract}}    
\end{center}
We present the concept of BlueMUSE, a blue-optimised, medium spectral resolution, panoramic integral field spectrograph based on the MUSE concept and proposed for the Very Large Telescope. With an optimised transmission down to 350 nm, a larger FoV (1.4 $\times$ 1.4 arcmin$^2$) and a higher spectral resolution compared to MUSE, BlueMUSE will open up a new range of galactic and extragalactic science cases allowed by its specific capabilities, beyond those possible with MUSE. For example a survey of massive stars in our galaxy and the Local Group will increase the known population of massive stars by a factor $>100$, to answer key questions about their evolution. Deep field observations with BlueMUSE will also significantly increase samples of Lyman-$\alpha$ emitters, spanning the era of Cosmic Noon. This will revolutionise the study of the distant Universe: allowing the intergalactic medium to be detected unambiguously in emission, enabling the study of 
the exchange of baryons between galaxies and their surroundings.

%BlueMUSE builds upon the heritage of MUSE, but includes both obvious and novel improvements (e.g., better temperature control and alignment processes). To achieve the larger field of view, the spectrograph optical design exploits recent successful development in curved detectors to provide an excellent image quality and throughput while keeping the same number (24) and format (4k 15μm pixel) CCDs and overall volume as MUSE. Curved detectors development will have a major significance for future wide-field instrumentations, especially for ELT. 

By 2030, at a time when the focus of most of the new large facilities (ELT, JWST) will be on the infra-red, BlueMUSE will be a unique facility, outperforming any ELT instrument in the Blue/UV. It will have a strong synergy with ELT, JWST as well as ALMA, SKA, \textit{Euclid} and \textit{Athena}.

\newpage

\thispagestyle{empty}
\noindent\par
{\footnotesize \noindent$^1$Univ Lyon, Univ Lyon1, Ens de Lyon, CNRS, Centre de Recherche Astrophysique de Lyon UMR5574, F-69230, Saint-Genis-Laval, France,}\\
{\footnotesize $^2$Aix Marseille Univ, CNRS, CNES, LAM, Marseille, France,}\\
{\footnotesize $^3$Instituto de Astrofísica e Ciências do Espaço, Universidade do Porto, CAUP, Rua das Estrelas, PT4150-762 Porto, Portugal, }\\
{\footnotesize $^4$Leiden Observatory, Leiden University, P.O. Box 9513, 2300
RA, Leiden, The Netherlands,}\\
{\footnotesize $^5$Leibniz-Institut f\"ur Astrophysik Potsdam (AIP), An der Sternwarte 16, D-14482 Potsdam, Germany,}\\
{\footnotesize $^6$Department of Physics and Astronomy, University of Sheffield, Sheffield, S3 7RH, United Kingdom,}\\
{\footnotesize $^7$CEA, IRFU, DAp, AIM, Universit\'e Paris-Saclay, Universit\'e de Paris,  CNRS, F-91191 Gif-sur-Yvette, France,}\\ 
{\footnotesize $^8$Institut f\"ur Astrophysik, Georg-August-Universit\"at G\"ottingen, Friedrich-Hund-Platz 1, 37077 G\"ottingen, Germany,
}\\
{\footnotesize $^9$Universit\'e de Strasbourg, CNRS, Observatoire astronomique de Strasbourg, UMR 7550, F-67000 Strasbourg, France, }\\
{\footnotesize $^{10}$UK Astronomy Technology Centre, Royal Observatory Edinburgh, Blackford Hill, Edinburgh, EH9 3HJ, United Kingdom, }\\
{\footnotesize $^{11}$Centre for Extragalactic Astronomy, Department of Physics, Durham University, South Road, Durham DH1 3LE, United Kingdom, }\\
{\footnotesize $^{12}$Centro de Astrobiolog\'{\i}a, CSIC-INTA. Crtra. de Torrej\'on a Ajalvir km 4. 28850 Torrej\'on de Ardoz (Madrid), Spain,}\\ 
{\footnotesize $^{13}$Observatoire de Gen\`eve, Universit\'e de Gen\`eve, 51 Ch. des Maillettes, 1290, Versoix, Switzerland, }\\
{\footnotesize $^{14}$Stockholm University, Department of Astronomy and Oskar Klein Centre for Cosmoparticle Physics, AlbaNova University Centre, SE-10691, Stockholm, Sweden.,}\\  
{\footnotesize $^{15}$Instituto de Astrofisica de Canarias, C/ Via Lactea s/n, E-38205 La Laguna, Spain,}\\
{\footnotesize $^{16}$Universidad de La Laguna, Avda. Astrofisico Francisco Sanchez, 2, E-38206 La Laguna, Spain,}\\
{\footnotesize $^{17}$Laboratoire d’Astrophysique, Ecole Polytechnique F\'ed\'erale de Lausanne (EPFL), Observatoire de Sauverny, CH-1290 Versoix, Switzerland,}\\
{\footnotesize $^{18}$Astrophysics Research Institute, Liverpool John Moores University, 146 Brownlow Hill, Liverpool L3 5RF, United Kingdom,}\\
{\footnotesize $^{19}$Anton Pannekoek Institute for Astronomy, University of Amsterdam, Science Park 904, 1098 XH Amsterdam, The Netherlands,}\\ 
{\footnotesize $^{20}$Leuven, Institute of Astrophysics, Universiteit Leuven, Celestijnenlaan 200 D, 3001 Leuven, Belgium,}\\
{\footnotesize $^{21}$University Observatory Munich
Scheinerstrasse 1, 81679 Munich, Germany,}\\
{\footnotesize $^{22}$Bonn, Argelander-Institut für Astronomie, Universität Bonn, Auf dem Hügel 71, D-53121 Bonn, Germany,}\\
{\footnotesize $^{23}$Department of Astronomy, University of Michigan, 1085 S. University Ave., Ann Arbor, MI 48109, USA,}\\
{\footnotesize $^{24}$LUPM, Universit\'e de Montpellier, CNRS, Place Eug\`ene Bataillon, F-34095 Montpellier, France,}\\
{\footnotesize $^{25}$European Southern Observatory, Alonso de Cordova 3107, Vitacura, Santiago Chile,}\\
{\footnotesize $^{26}$Instituto de Astrofísica e Ciências do Espaço, Universidade de Lisboa, OAL, Tapada da Ajuda, PT1349-018 Lisbon, Portugal,}\\
{\footnotesize $^{27}$Departamento de Física, Faculdade de Ciências, Universidade de Lisboa, Edifício C8, Campo Grande, PT1749-016 Lisbon, Portugal,}\\
{\footnotesize $^{28}$European Southern Observatory, Karl-Schwarzschild-Str. 2, 85748 Garching near Munich, Germany, }\\
{\footnotesize $^{29}$Institut UTINAM UMR 6213, CNRS, Univ. Bourgogne Franche-Comté, OSU THETA, BP 1615, 25010 Besan\c on Cedex,
France,}\\
{\footnotesize $^{30}$Armagh Observatory, College Hill, BT61 9DG, Armagh, Northern Ireland,}\\
{\footnotesize $^{31}$Institute for Astronomy, Astrophysics, Space Applications \& Remote Sensing, National Observatory of Athens, Vas. Pavlou and I.Metaxa, Penteli 15236, Greece,}\\
{\footnotesize $^{32}$Institut f\"ur Physik und Astronomie, Universit\"at Potsdam, Karl-Liebknecht-Str. 24/25, 14476 Golm, Germany
}

\end{titlepage}

\newpage
\setcounter{secnumdepth}{3}%%  numbered sections down to \suparagraph level - cjb 
\setcounter{tocdepth}{3}	%% TEMPORARY -- down to \subparagraph in TOC - cjb 

\pagenumbering{roman}		%% roman for toc
\tableofcontents
\cleardoublepage

\pagestyle{headings}		%% section name in header - cjb 
\pagenumbering{arabic}		%% arabic for body - cjb 

\hypertarget{overview}{%
\section{Overview}\label{overview}}

BlueMUSE is a proposed optical seeing-limited,
blue-optimised, medium spectral resolution, panora\-mic integral field
spectrograph for the ESO Very Large Telescope (VLT). The project is an evolution of the technology used on the
very successful MUSE instrument, with a similar architecture and many
similar systems (and so is low risk), but with a new and distinct
science case enabled by its unique blue spectral coverage.

BlueMUSE will cover in one setting the 350-600 nm spectral range at
R$\sim$4000, expanding the MUSE spectral range (480-930 nm)
towards the blue and near-UV at twice the spectral resolution. With a
field of view of 2 arcmin$^{2}$, BlueMUSE will also double
the sky area presently offered by MUSE. As with MUSE, BlueMUSE will
achieve exquisite end-to-end throughput including telescope and atmosphere 
(e.g., 35\% at 450 nm, 17\% at 350 nm), 100\% sky coverage, high stability and high efficiency.

BlueMUSE will offer new and unique science opportunities in many fields
of astrophysics, beyond those possible with MUSE. For example, a survey
of massive stars in the Milky Way and the Local Group will increase by
$>100\times$ the known population of massive stars and provide a complete census for stars in young, star forming clusters, to answer key
questions about their evolution, test the hypothesis of massive
Population III stars, search for spectroscopic binaries as progenitors
for gravitational waves-emitting black-hole binaries, and map the chemical abundance in
galaxies in relation to their environment. Other examples of Galactic
and planetary science include the study of ionised nebulae and their
light element abundances, the investigation of multi-populations in globular clusters, the study of the morphology of comets,
including the origin of chemical elements and the properties of their
nuclei.

In the field of nearby galaxies, BlueMUSE will probe the physical
conditions in extreme starburst galaxies, quantifying the interplay
between the populations of massive stars (supernovae, stellar winds and
ionizing radiation) and their surroundings. It will also measure the
opacity to Lyman continuum and Lyman-$\alpha$ radiation. Other key science
goals concern the study of low surface brightness galaxies, the role of
environment in local clusters on galaxy evolution. 

The study of the distant Universe will also be revolutionised with
BlueMUSE: allowing the intergalactic medium to be detected unambiguously
in emission, enabling the study of the exchange of baryons between
galaxies and their surroundings. The evolution of the circum-galactic
medium properties at the critical peak in cosmic star formation
will be probed. Deep field observations with BlueMUSE will
significantly increase samples of Lyman-$\alpha$ emitters (and spanning the era
of Cosmic Noon), allowing statistical samples of Lyman continuum
emitters to be constructed, to yield critical constraints on the Lyman
continuum leakage processes. These studies will be further boosted by
exploiting gravitational lens clusters, probing the faint end of the
luminosity function and measuring the Lyman-$\alpha$ haloes at the sub-kpc
scale. At the same time, the high surface densities of sources achieved by BlueMUSE
will also enable these observations to optimally constrain the dark
matter distribution in the lensing clusters. BlueMUSE will also study
the emergence of the first galaxy clusters by giving crucial insights
into both cold accretion onto the most massive early structures, and
galaxy evolution models.

The science cases highlighted here are unique to BlueMUSE and make the
best use of its exceptional performance in the blue. They, however, only
scratch the surface of the diversity of science programs which can
benefit from it. Like MUSE, BlueMUSE will have a broad impact, touching
many different fields of astrophysics by opening a new area in parameter space. In addition BlueMUSE has a great
potential for serendipitous discoveries and will greatly enhance the
legacy of European observational astronomy.

BlueMUSE builds upon the heritage of MUSE, but includes both obvious and
novel improvements (e.g., better temperature control and alignment
processes). To achieve the larger field of view (2 arcmin$^{2}$), BlueMUSE will sample the seeing disk at
0.3\arcsec\ which matches the Paranal natural seeing at 350-600 nm. The
spectrograph optical design exploits recent successful developments in
curved detectors to provide an excellent image quality and throughput
while keeping the same number (24) and format (4K $\times$ 4K, 15$\mu$m pixel) CCDs and
overall volume and weight as MUSE. Such a development will have a major
significance for future wide-field instrumentation, especially for ELT.

The curved detector is the only significant risk in the project. The fall-back solution would be to step back to the original 0.2\arcsec\ sampling and 1 arcmin$^{2}$ field of view - like available in MUSE. The resulting
science impact will be limited as all the other specific BlueMUSE
characteristics (e.g., spectral coverage and resolution, throughput) are
untouched.

BlueMUSE will be unique. The nearest instrument in terms of performance
is the Keck IFU KCWI instrument, but BlueMUSE with 40
$\times$ larger field of view, better throughput and stability and an overall increased
efficiency, will be two orders of magnitude more efficient. 

In seven years from now, at a time when the focus of most of the new
large facilities (ELT, JWST) will be on the infra-red, BlueMUSE
will be a unique facility. It will clearly outperform
any ELT instrument in the blue/UV. Its synergy with ELT and JWST is
strong, but also with ALMA, SKA, \textit{Euclid} and \textit{Athena}.

\newpage

\hypertarget{expected-performance}{%
\section{Performance}\label{expected-performance}}

\begin{figure}
    \centering
    \includegraphics[width=13cm]{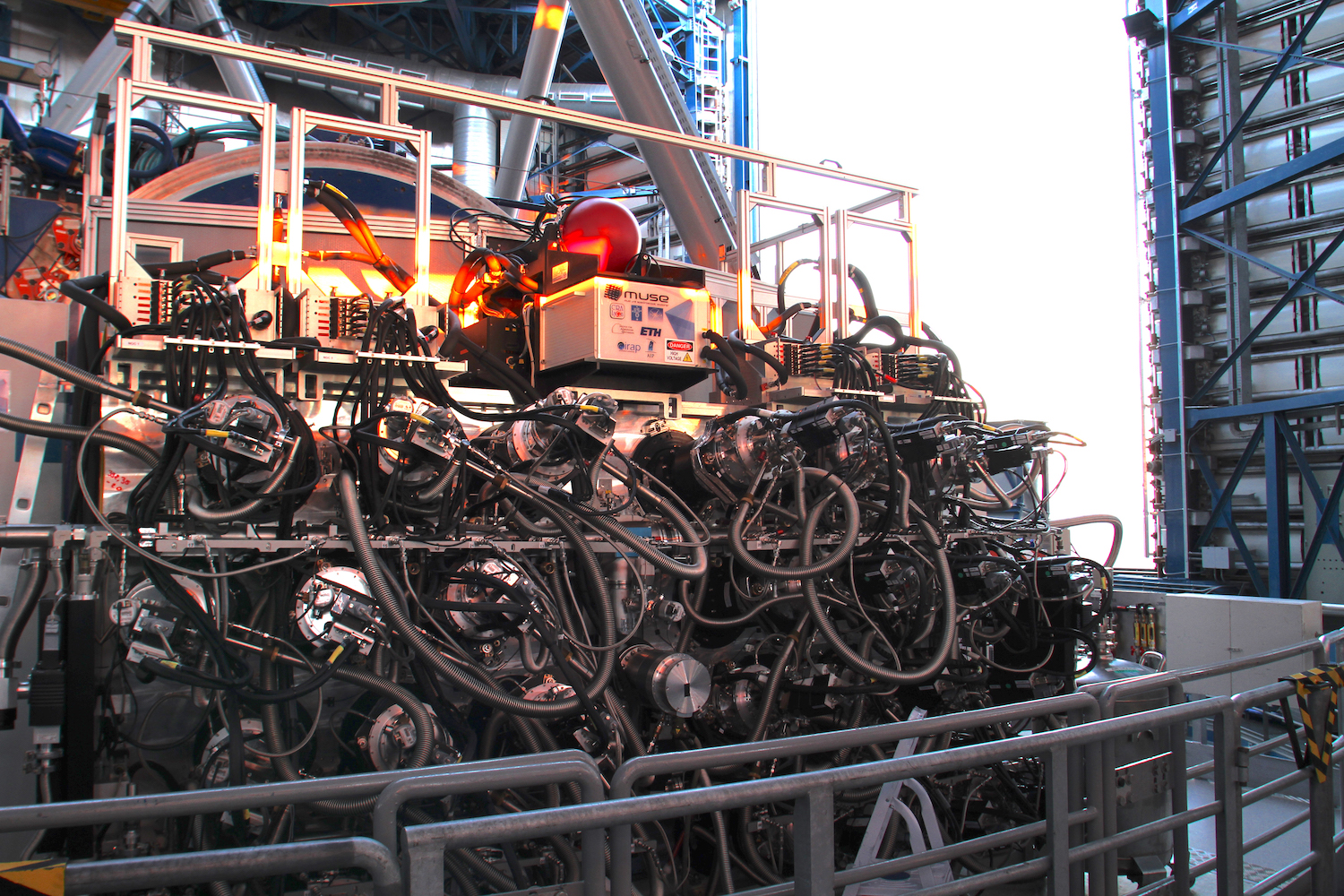}
    \caption{\label{fig:generalview}Overview of the MUSE instrument on the Nasmyth platform,
providing a good indication on the expected appearance of the BlueMUSE
instrument once at the telescope.
}
\end{figure}

BlueMUSE is largely based on the design and system architecture of MUSE
\citep{2010SPIE.7735E..08B}, with the same modular structure and two stages of
FoV slicing / splitting. It has a single mode of operation and a fixed
spectral and spatial format, which simplifies the overall fore-optics.
The instrument envelope will fit within the allocated space on the
Nasmyth platform as is the case for MUSE (Fig.\,\ref{fig:generalview}).

We summarise in Table \ref{characteristics} the main characteristics 
of BlueMUSE and detail below some of the main assumptions.

\begin{table}[ht]
    \centering
    \begin{tabular}{|a|b|}
         \hline
         \multirow{2}{*}{Wavelength range} & 
         \multirow{2}{*}{350 - 600 nm}
         \\[11pt]\hline
         Spectral resolution & 
         \begin{tabular}{@{}c@{}}
         $R>3000$, average $\sim3600$ \\ over the full wavelength range 
         \end{tabular}\\ \hline
         \multirow{2}{*}{Spectral sampling} & \multirow{2}{*}{0.58 \AA\ per spectral bin} \\[11pt] \hline
         \multirow{2}{*}{Field-of-view} & \multirow{2}{*}{1.4 arcmin $\times$ 1.4 arcmin}\\[11pt] \hline
         \multirow{2}{*}{Spatial sampling} & \multirow{2}{*}{0.3\arcsec$\times$0.3\arcsec\ per spaxel} \\[11pt] \hline
         \begin{tabular}{@{}c@{}}
         Throughput \\ (incl. telescope and atmosphere)
         \end{tabular}
         & \begin{tabular}{@{}c@{}}
         $>15\%$ and average $>25$\% \\ over the wavelength range
         \end{tabular}\\ \hline
         Image quality & \begin{tabular}{@{}c@{}}
         Max. 20\% degradation \\ under best seeing conditions (0.6\arcsec)
         \end{tabular}\\ \hline
    \end{tabular}
    \caption{  \label{characteristics}Main characteristics of BlueMUSE}
\end{table}

\begin{figure}
    \centering
    \includegraphics[width=16cm]{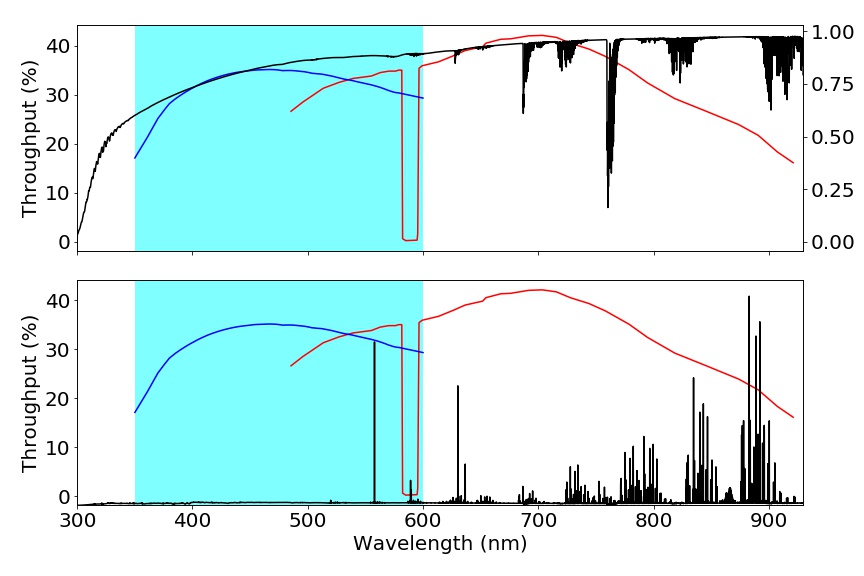}
    \caption{\label{throughput}Comparison between the end-to-end (including telescope and
atmosphere) BlueMUSE (blue curve) and MUSE (red curve) sensitivities,
plotted together with the atmospheric transmission (top panel) and sky emission
(bottom panel). The blue wavelength cut-off of BlueMUSE matches with an
expected atmospheric transmission $\sim$65\%. Apart from the two bright lines at 557.7 and 589.0 nm, the BlueMUSE wavelength range remains unaffected from strong night sky emission lines, which are indeed a limitation in the red.
}
\end{figure}

\textbf{Throughput}: As the type and number of optical systems in BlueMUSE is very similar to
MUSE, we have used the end-to-end MUSE transmission curve as a starting
point. On the basis of QE curves for commercially available CCDs, we expect excellent detector performance in the blue, and the overall shape of the VPH grating to be similar to MUSE around its peak
wavelength. The main difference at the blue end of the wavelength range is due to
the atmospheric transmission, which drops significantly, down to 65\% at
350 nm, as well as the detector quantum efficiency which slightly decreases below 400 nm. We use the Paranal extinction curve to account for the additional
atmospheric absorption at these wavelengths. In addition, we expect the glass 
transmission to be slightly lower at $\lambda<$400 nm. As a guideline we use the transmission curve measured for the Potsdam MRS
spectrograph \citep{2016SPIE.9912E..22M} which is based on the MUSE
spectrograph design but covers similar wavelengths as BlueMUSE.
The overall end-to-end BlueMUSE sensitivity we expect under these
assumptions is presented in Fig.\,\ref{throughput}, in comparison to the MUSE
sensitivity and the atmosphere transmission and emissions at these
wavelengths.

\newpage

\begin{figure}
    \centering
    \includegraphics[width=16cm]{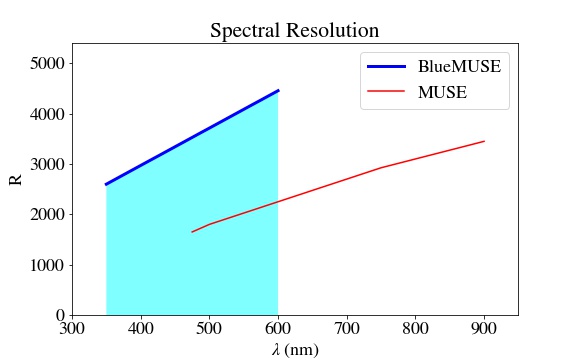}
    \caption{\label{fig:specres}Comparison of the BlueMUSE (in blue) and MUSE (in red) spectral resolution as a function of wavelength.
}
\end{figure}

\textbf{Spectral Resolution }: As for MUSE, the BlueMUSE image quality in the wavelength
direction will be dominated by the slice width on sky and the image
quality of the spectrograph system. The
predicted evolution of the spectral resolution as a function of
wavelength is presented Fig.\,\ref{fig:specres}. The average resolution is R=3600 and
the BlueMUSE spectral resolution is always two times larger than the
MUSE spectral resolution in their overlapping wavelength range.

\newpage

\hypertarget{our-galaxy-and-the-local-group}{%
\section{The Milky Way and the Local Group}\label{our-galaxy-and-the-local-group}}

  \hypertarget{key-science-case-massive-stars}{%
  \subsection{\texorpdfstring{\textbf{Key science case:} massive
  stars}{Key science case: massive stars}}\label{key-science-case-massive-stars}}

\begin{tcolorbox}[colback=blue!5!white,colframe=blue!75!black,title=Science Goals]
%\begin{quote}
%Homogeneous spectroscopic analyses of statistically %meaningful samples
%of individual massive stars and young clusters of %massive stars in the
%Milky Way, in the Local Group, and in nearby %galaxies.
%\end{quote}

\begin{itemize}
\item
  \begin{quote}
  Perform studies of the formation and \emph{{evolution of massive
  stars}} to provide empirical anchors to current theories, including
  the investigation of rotation, mass loss, overshooting, and the chemical composition; also finding emission line stars and peculiar
  objects such as WR, LBV, Be, B{[}e{]} stars etc.
  \end{quote}
\item
  \begin{quote}
  Test the hypothesis of very \emph{{massive Pop.\,III stars}} by
  probing the metallicity dependence of the upper IMF, or: \emph{Where
  are the very massive stars of the Local Group metal-poor dwarf
  galaxies}?
  \end{quote}
\item
  \begin{quote}
  Discover and categorize spectroscopic binaries as progenitors for
  \emph{{gravitational wave BH binaries}}
  \end{quote}
\item
  \begin{quote}
  \emph{{Map chemical abundances}} in galaxies as an alternative to \hii\ 
  regions and study massive stars \emph{{simultaneously with their
  environment}} (\hii\ regions, molecular clouds, pre-main sequence stars,
  ISM)
  \end{quote}
\end{itemize}
\end{tcolorbox}

\begin{figure}[ht]
    \centering
    \includegraphics[width=13.5cm]{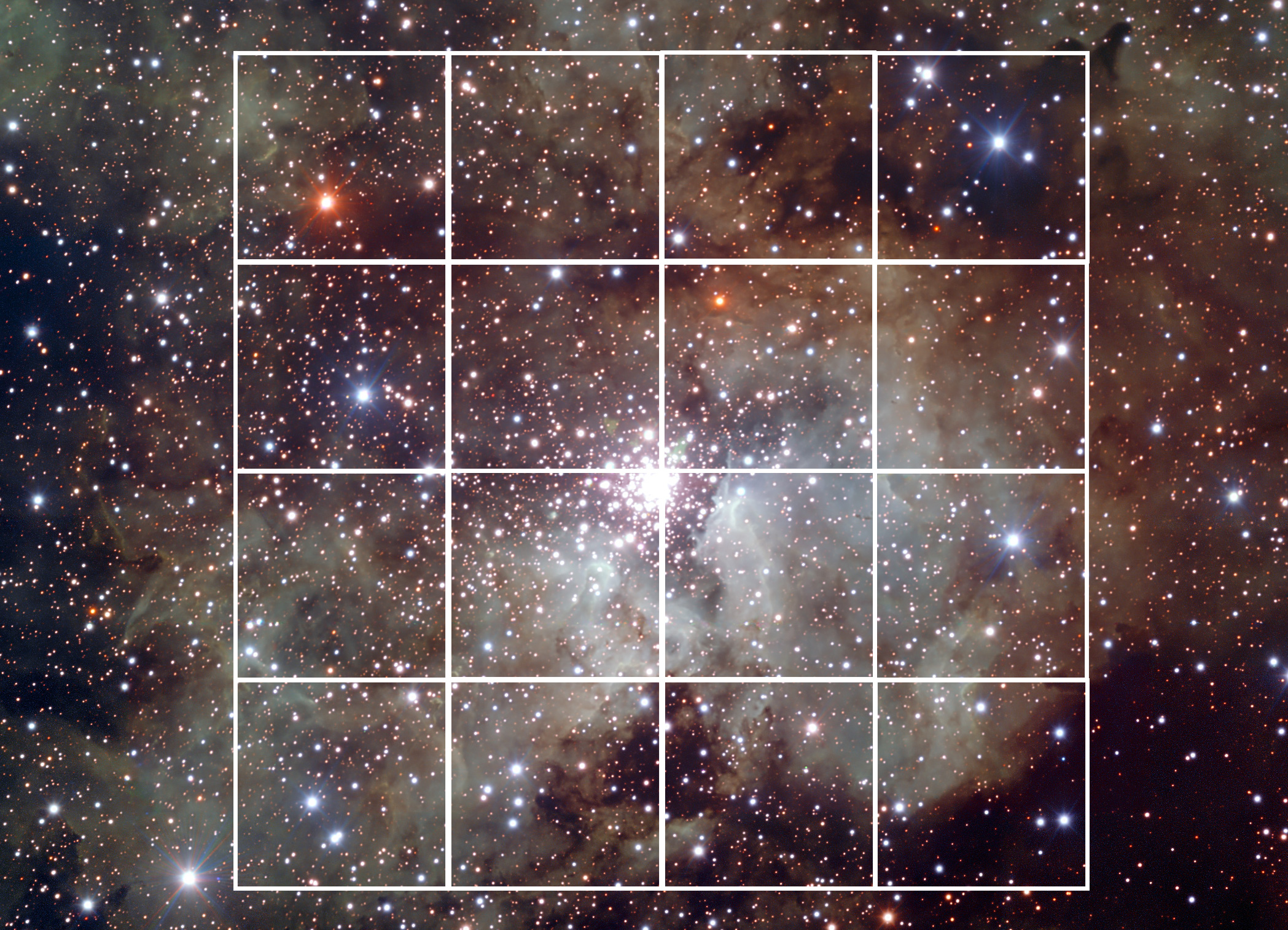}
    \caption{\label{NGC3606_IFU}NGC3603 covered by 4$\times$4 BlueMUSE pointings.
}
\end{figure}

\pagebreak

\textbf{Astrophysical context and relevance:}

Hot, massive stars are the most energetic ones and therefore dominate the
spectral energy distribution of galaxies at all redshifts. They are also contributing significantly to feedback processes \citep{2009ApJ...695..292C}: they possess strong stellar winds, and they are progenitors to
core-collapse supernovae. They are the ionizing sources for \hii\ regions and
thus relevant for estimating star formation rates, the diffuse ionized
gas (DIG), and Lyman-$\alpha$ radiation. They probe the abundances of
contemporary stellar populations as an alternative to strong line
nebular abundance determinations \citep{2009ApJ...700..309B}. Hot stars are
the constituents of super star clusters, and they are relevant for the
understanding of Population III stars, thus the study of the
re-ionization of the early Universe. More recently, the detection of
gravitational waves from stellar mass black hole (BH) mergers has
stimulated interest in understanding the origin of BH binaries, and
therefore massive star binaries as their progenitors (e.g., \citealt{2016A&A...588A..50M}).

The theory of massive stars poses challenges concerning stellar
evolution, the role of rotation, metallicity, stellar winds, overshooting, and
binarity \citep{2000A&A...361..101M,2012ARA&A..50..107L,2014A&A...570L..13C,2018A&A...615A.119V,2019A&A...622A..50H}.
However, hot massive stars are \emph{rare}. There are selection effects
within the Galaxy such as extinction and uncertain distances \citep{2009A&A...499..455C}. Massive stars need to be analyzed in different environments to
study the effects of metallicity in comparison with numerical models \citep{2009ApJ...700..309B,2018MNRAS.474L..66G}. Photometry is \emph{unable} to constrain stellar
parameters, which is why in the past a tedious two-step procedure of
finding hot stars with photometry and follow-up spectroscopy was
required \citep{2016AJ....152...62M}. This procedure is expensive and
incomplete. Also, spectroscopy of massive stars in star-forming regions
is notoriously difficult because of nebular contamination and crowding.
BlueMUSE will allow to overcome these limitations and offer an
unprecedented multiplex advantage as demonstrated already (e.g., \citealt{2018MNRAS.473.5591K,2018A&A...614A.147C,2018A&A...618A...3R}), however with
wavelength coverage for important diagnostic lines in the blue.
\medskip\par
\textbf{Why is BlueMUSE needed?}

\begin{figure}[ht]
    \centering
    \includegraphics[width=17cm]{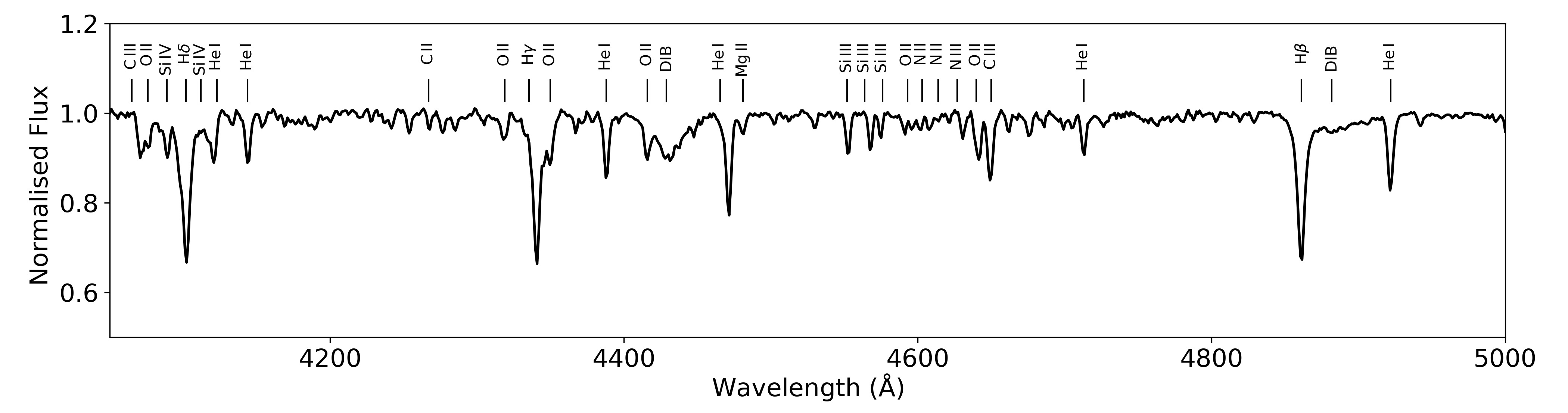}
    \caption{\label{diagnostic_lines} Early B-type spectrum with diagnostic lines in the blue relevant for hot, massive stars (from \citealt{2019AJ....157...53G}).
    }
\end{figure}

The canonical stellar transitions used for stellar atmosphere analysis,
chemical composition and spectral classification occur between 3500-5000 \AA\ \citep{1990PASP..102..379W,2018A&A...616A.135M}. The BlueMUSE wavelength range gives
access to the following spectral features constraining stellar
parameters, chemical composition and evolutionary stage of the stars (see also Fig.\,\ref{diagnostic_lines}):
\begin{itemize}
\item{Balmer lines as the principal surface gravity criteria in massive stars.}
\item{Balmer jump at 3646 \AA\ as effective temperature (T\textsubscript{eff}) criterion.}
\item{Wind and classification diagnostics from \heii~$\lambda$4686 and nearby CNO
lines.}
\item{\siiv~$\lambda4089,4116$ \AA, \siiii~triplet~$\lambda4552$ \AA,  \siii~$\lambda4128,4130$ \AA, as well as \hei\ (e.g., 4471 or 4387 \AA) and \heii\ (4200 and 4541 \AA) lines for T\textsubscript{eff} and/or helium abundance.}
\item{Crucial wavelength range for WR emission (``blue'' and ``red'' bumps), and \ovi~$\lambda3811$ \AA\ \citep{2007ARA&A..45..177C}.}
\end{itemize}

In the hottest stars, range $>45000$ K, where \hei\ lines
are weak or absent, optical lines of \niii, \niv\ and \nv\ are used as
temperature criteria in this blue optical range \citep{2012A&A...543A..95R}.
The 4000-5000 \AA\ wavelength range encloses many transitions to measure T$_{\rm eff}$ and log($g$): \cii, \ciii, \nii, \niii, \niv, \nv, \oii, \oiii, \siii, \siiii, \siiv\ and \mgii. These transitions also provide
the chemical composition of the stars and surrounding interstellar medium (ISM)  \citep{2015A&A...575A..34M}.
At lower temperatures (T$_{\rm eff}>$10000 K),
\feii\ lines are available and can be added to the chemical
composition and stellar parameter analysis.
The expected spectral resolution of BlueMUSE will be suitable to
estimate T\textsubscript{eff}, log(g) as well as abundances, as
demonstrated by earlier studies using FORS and LRIS-B (e.g., \citealt{2012ApJ...747...15K,2016ApJ...829...70K})
%, however with even lower spectral resolution.
\medskip\par
\textbf{Breakthrough Science:} 

BlueMUSE will enable unique science for two major reasons: (1)
\emph{{Multiplex+Sensitivity}}: The current state-of-the-art has been
set by the VLT-FLAMES Massive Star Survey \citep{2005Msngr.122...36E,2005A&A...437..467E} that
yielded a total of 803 spectra from an effort of more than 100~hrs VLT
time. BlueMUSE will be up to two orders of magnitude faster, depending
on the size of a cluster (albeit lower, however still acceptable,
spectral resolution), which has already been demonstrated with MUSE in
globular clusters, providing up to 1000 stellar spectra per pointing
\citep{2016A&A...588A.148H}. (2) \emph{{Crowding}}: 
Analogous to PSF-fitting CCD photometry \citep{1987PASP...99..191S}, the IFU concept is vital to deblend heavily crowded fields and yield cross-talk free spectra of stars with overlapping images  \citep[see][]{2013A&A...549A..71K}. Again,
the state-of-the-art can be appreciated from existing results obtained
with the VLT-FLAMES Tarantula Survey \citep{2011Msngr.145...33E,2011A&A...530A.108E} that has
provided multi-epoch fibre spectra of different spectral resolution for
more than 800 stars, however severely hampered by nebular contamination,
and completely unable to address crowded regions. In contrast, from four MUSE pointings on the R136 region with a
total exposure time of 2680 sec., \citet{2018A&A...614A.147C} were able to extract
2255 spectra, out of which 588 show a S/N$>$50. In the
foreseeable future, no other instrument will have such capability.
\medskip\par
\textbf{Examples and exposure time estimates:}

 Galactic clusters (e.g., NGC 3578, NGC 3603, or Westerlund 2) will
require mosaic observations with 1$\times$2\ldots\ 4$\times$4 pointings, each of
which will yield typically 500 spatially deblended spectra of individual
stars (Fig.\,\ref{NGC3606_IFU}). With an exposure time of 0.5~hr per pointing, the total exposure
time will range between 1~hr and 8~hrs. Clusters in the LMC, e.g. R136, will
require an exposure time of 1~hr per pointing, resulting in a total of 4~hrs
per 2$\times$2 mosaic. To cover an intermediate distance Local Group dwarf
galaxy (e.g., Sextans-A or NGC 3109) with much deeper exposures (4~hrs),
the total effort will be as high as 50-100~hrs. The expected number of
high quality spectra for objects more distant than the Galactic clusters
is also expected of order 500 per pointing \citep{2018A&A...618A...3R}.
\medskip\par

\textbf{Synergy with other facilities}:

$\bullet$ ``Classical'' MUSE: full coverage 350-930 nm, resolved stellar
population studies

$\bullet$ ALMA: comparison between locations and properties of molecular gas and massive stars

$\bullet$ Target selection for ELT-MOS: follow-up spectroscopy at higher
spectral resolution

$\bullet$ Target selection for JWST: imaging and spectroscopy of crowded and
obscured regions

$\bullet$ ERIS, MICADO, HARMONI: follow-up on red supergiants / nuclei of
clusters.

 \subsection{Globular clusters }\label{globular-clusters}

\begin{tcolorbox}[colback=blue!5!white,colframe=blue!75!black,title=Science Goals]
\begin{itemize}
\item
  Separate multiple populations based on dynamics as well as spectroscopic properties.
\item
  Extend the spectroscopic investigation of multiple populations down to the main sequence.
\item
  Understand the nature of multi-populations in Globular Clusters.
\item
  Use Globular Clusters as background sources for small scale ISM investigations.
\end{itemize}
\end{tcolorbox}

\begin{figure}[ht]
    \centering
    %\vspace{-80pt}
    \includegraphics[width=0.9\textwidth]{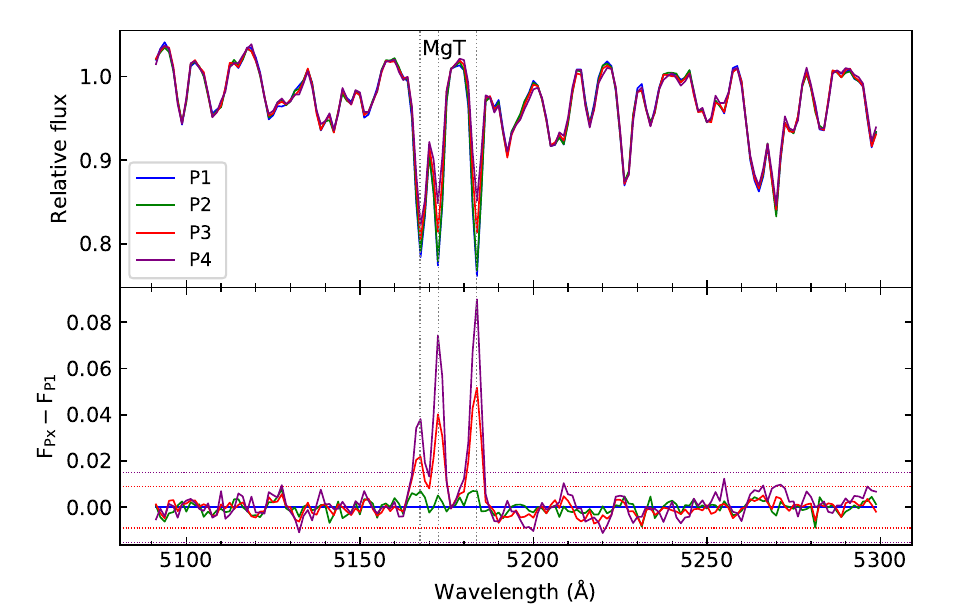}
    \caption{Abundance differences of multiple populations as seen by MUSE in NGC\,2808 (Latour et al. in preparation). Top: co-added MUSE spectra of Red Giant Branch stars belonging to different populations; bottom: spectral differences with respect to population P1.}
    \label{fig:multipop}
\end{figure}

MUSE allowed, for the first time, a detailed spectroscopic investigation for tens of thousands of individual stars in Galactic globular cluster and even in massive star clusters in the Magellanic Clouds. For most clusters, this corresponded to an increase of the spectroscopic samples by two orders of magnitude. Thanks to the integral-field nature of MUSE, it also became possible to advance to the heavily crowded cluster centres. This enabled detailed studies of the kinematics of the clusters \citep[e.g.][]{2018MNRAS.473.5591K,2018MNRAS.480.1689K}, binary searches \citep{2018MNRAS.475L..15G}, and the measurement of stellar parameters \citep{2016A&A...588A.148H}.

It is now well established that globular clusters have at least two distinct populations with differences in light elements like sodium or oxygen, and possibly helium \citep[see][for a review]{2018ARA&A..56...83B}. Some clusters show even more complex population patterns, including metallicity differences. The origin of these differences is still unknown and further studies will be required to understand how the clusters formed. So far high-resolution spectroscopy has been used to infer the abundance differences, hence the studies were restricted to rather small stellar samples in the outskirts of the clusters. With MUSE, abundance differences between the populations can be measured by co-adding the spectra of stars for which the population has been determined previously from precise photometry. The extremely high signal-to-noise ratios of the combined spectra compensate for the relatively low spectral resolution (Latour et al., in preparation, see Fig.\,\ref{fig:multipop}).

%Our view on Globular Clusters (GC) has seen a paradigm change over the last decade due to high-precision photometry, mainly with the Hubble Space Telescope \citep[e.g.][]{2015AJ....149...91P}, and detailed follow-up spectroscopy \citep[e.g.][]{2009A&A...505..117C}. It is now well established that GCs have at least two distinct populations with differences in light elements like sodium or oxygen, and possibly helium \citep[see][for a review]{2018ARA&A..56...83B}. Some clusters show an even more complex population pattern, including metallicity differences. High-resolution spectroscopy was used to infer the abundance differences, hence the studies were restricted to rather small stellar samples in the outskirts of the clusters. Owing to its ability to resolve stellar blends \citep{2013A&A...549A..71K}, MUSE opened up the possibility to increase the spectroscopic samples by two orders of magnitude and to advance to the heavily crowded cluster centres. This enabled detailed studies of the kinematics of the clusters and their populations \citep[e.g.][]{2018MNRAS.473.5591K}, binary searches \citep{2018MNRAS.475L..15G}, and the measurement of stellar parameters \citep{2016A&A...588A.148H}. Further, abundance differences between the populations can be measured by co-adding spectra of stars belonging to the same populations. The extremely high signal-to-noise ratios of the combined spectra compensate for the relatively low spectral resolution (Latour et al., in preparation, see Fig.\,\ref{fig:multipop}).

\begin{subsec}[ht]
\begin{tcolorbox}[colback=green!5!white,colframe=green!75!black,title=What is the need for an IFU in the blue?]

The main characteristics of BlueMUSE is to provide spectroscopy in the wavelength range 350-600 nm.

\begin{itemize}
\item
  \begin{quote}
  This wavelength range gives access to specific spectral lines in the local Universe, such as the K,H absorption lines and 4000 \AA\ break in the stellar continuum, important diagnostic absorption lines of hot massive stars, and the combination of [\oii]~$\lambda3727$ and [\oiii]~$\lambda5007$, as well as the temperature-sensitive auroral line [\oiii]~$\lambda4363$ (\S\ref{key-science-case-ism-and-hii-regions-extreme-starbursts}) for ionized nebulae.
  \end{quote}
\item
  \begin{quote}
 In the distant Universe, blue wavelengths allow to observe Lyman-$\alpha$ already at $z=2$, near the peak of the cosmic SFR, with an advantageous reduction in surface brightness dimming (\S\ref{key-science-case-gas-flows-around-and-between-galaxies})
  \end{quote}
\item
  \begin{quote}
By the time BlueMUSE goes on sky, there will be strong synergies with many other facilities that are optimized for red/NIR wavelengths, such as JWST and the ELT, with no sensitivity in the blue. (\S\ref{synergies}).
  \end{quote}
\end{itemize}
\end{tcolorbox}

\end{subsec}

BlueMUSE will be a significant improvement and therefore the next major step towards a more complete understanding of the formation and evolution of massive star clusters. Low-mass stellar spectra have significantly more spectral lines in the visual-near-UV spectral range covered by BlueMUSE compared to the red-near-IR range covered by MUSE, including strategic lines for spectral analysis like the higher Balmer lines, \caii\ H \& K, CN molecular lines, and many more. A spectral coverage down to 350\,nm will enable the separation of populations with low resolution spectroscopy \citep[e.g.][]{2017MNRAS.465L..39H}. Hence BlueMUSE will allow to extend investigations of multiple populations from red giants to turn-off and main sequence stars, thereby strongly increasing the number of accessible stars. The higher spectral resolution will not only improve the spectral analysis, but also the radial velocity (RV) precision, in particular considering the higher information content of the blue spectral range. The higher RV precision in combination with the improved sensitivity for spectral differences would allow for a chemo-dynamical separation and investigation of multiple populations in globular clusters. Furthermore, detailed studies of stellar rotation would become feasible. Stellar rotation has recently been confirmed to play a crucial role in shaping the colour magnitude diagrams of young and intermediate age clusters \citep[e.g.][]{2018MNRAS.480.1689K}.

Furthermore, BlueMUSE will allow detailed studies of the blue stars in GCs -- such as horizontal branch and potentially blue hook stars, extremely low mass white dwarfs,  or interacting binaries. As these objects probe the binary evolution in dense stellar populations their study is important for understanding the overall dynamical evolution of the cluster. 

The integral field of MUSE allows to detect numerous diffuse interstellar bands (DIBs) and neutral species such as \nai\ $\lambda$5890,5896 \AA\ (NaD) and \ki\ in the tens of thousands of stellar spectra per cluster. For the first time these ISM features were mapped in absorption toward a GC and 
revealed associated structures on unprecedented small scales of a few arcseconds \citep{2017A&A...607A.133W}.
%
%Johan: I commented the following sentence as we do not want to highlight specific teams' work in the document.
%The team at Potsdam University currently extends the analysis to the full sample of globular clusters. 
%
With BlueMUSE we would be capable of directly mapping the more prominent \caii\ H \& K lines in the Milky Way, as well as other atomic and molecular species, in particular the broad and strong DIB at 4430 \AA\ \citep{2017A&A...606A..76C}.

 \subsection{Ultra-Faint Dwarf Galaxies }\label{ultra-faint-dwarf-galaxies}

\begin{tcolorbox}[colback=blue!5!white,colframe=blue!75!black,title=Science Goals]
\begin{itemize}
\item
  Measure the dynamics of UFDs to determine their nature, assess their dynamical state and study their dark matter content 
\item
  Study the chemical abundances of UFDs to characterize their chemical evolution and constrain their star formation history.
\end{itemize}
\end{tcolorbox}

There are several lines of arguments that point to the existence of dark
matter on a range of scales from the Cosmic Microwave Background to tiny
dwarf galaxies, but the nature of this dark matter is still not
established as searches in Earth-based detectors have not yet found
clear evidence of dark matter particles.

Thus we need to use astronomical systems to constrain the nature of dark
matter, and the most natural place to do this is to study the most
dark-matter dominated systems we know: the Ultra-Faint Dwarf galaxies (UFDs). In
these very faint, M$_{_V}>-8$, systems, baryons
might only make up 1/500\textsuperscript{th} of the total mass -- the
rest is dark matter of some sort or the other (e.g., \citealt{2012AJ....144....4M}).

Not only are these UFDs a great laboratory for studying dark matter,
they are also stellar systems in which the impact of supernovae and massive
stars feedback are expected to be largest, thus making them crucial laboratories
for the study of baryonic physics in galaxy formation. Since UFDs have such weak gravitational potentials they only experience short periods of star formation. This means that their stellar chemical abundances contain the imprint of very few, in some cases possibly only one, supernova(e). Their small mass also make them susceptible to gravitational tidal forces and may cause their dynamics to be more complex than typically assumed \citep{2018MNRAS.480.2609L,2018ApJ...863...25M}.

UFD candidates are now identified with great efficiency in wide-field imaging surveys such as SDSS, DES or Pan-STARRS \citep{2013NewAR..57..100B,2015ApJ...813..109D}.  The Large Synoptic Survey Telescope (LSST) is expected to greatly increase the  number of photometrically identified UFD candidates. However to confirm  their nature and to study their dark matter content and the abundance
patterns of their constituent stars, spectroscopic follow-up is mandatory. Traditionally this has been done with multi-object spectrographs (MOS, \citealt{2017ApJ...838....8L}), but such follow-up requires a) pre-selection of member stars, and b) significant distances between the targeted stars to avoid slit/fibre collisions.

Because of their distances but also because their red giant branch (RGB) is hardly populated,  substantial stellar samples in UFDs are only accessible by reaching faint magnitudes ($r>21$ and typically $r\sim 24$ or even fainter) at which point pre-selection is very inefficient due to confusion with compact galaxies. Furthermore, the most important regions to study for dark
matter constraints are the very central regions where differences
between dark matter models are the largest but where fibre collisions
are most severe. Likewise, to study the dynamical properties of these dwarfs it is  essential to densely sample stars over several effective radii down to fairly faint magnitudes for which fibre or slit spectrographs are inefficient.

A wide-field integral field spectrograph offers the perfect alternative
to a MOS since one can get away from  pre-selection and the density of
spectra is orders of magnitudes higher than for a MOS. Moreover, the PSF-fitting crowed field technique described above (\S\ref{key-science-case-massive-stars}, \ref{globular-clusters}) allows to deblend overlapping stellar images which would not be possible with any other spectroscopic technique.

\textbf{Requirements:} The typical velocity dispersion of Ultra-Faint
Dwarfs is of the order of a few km\,s$^{-1}$. To resolve this
dispersion with the small samples of stars expected in UFDs, we need a
spectral resolution ideally $R>4000$ with a well sampled LSF.
Since many spectral features also lie at wavelengths $<5000$ \AA,
a good sampling of this range is also essential.

\begin{figure}
    \centering
    \includegraphics[width=17cm]{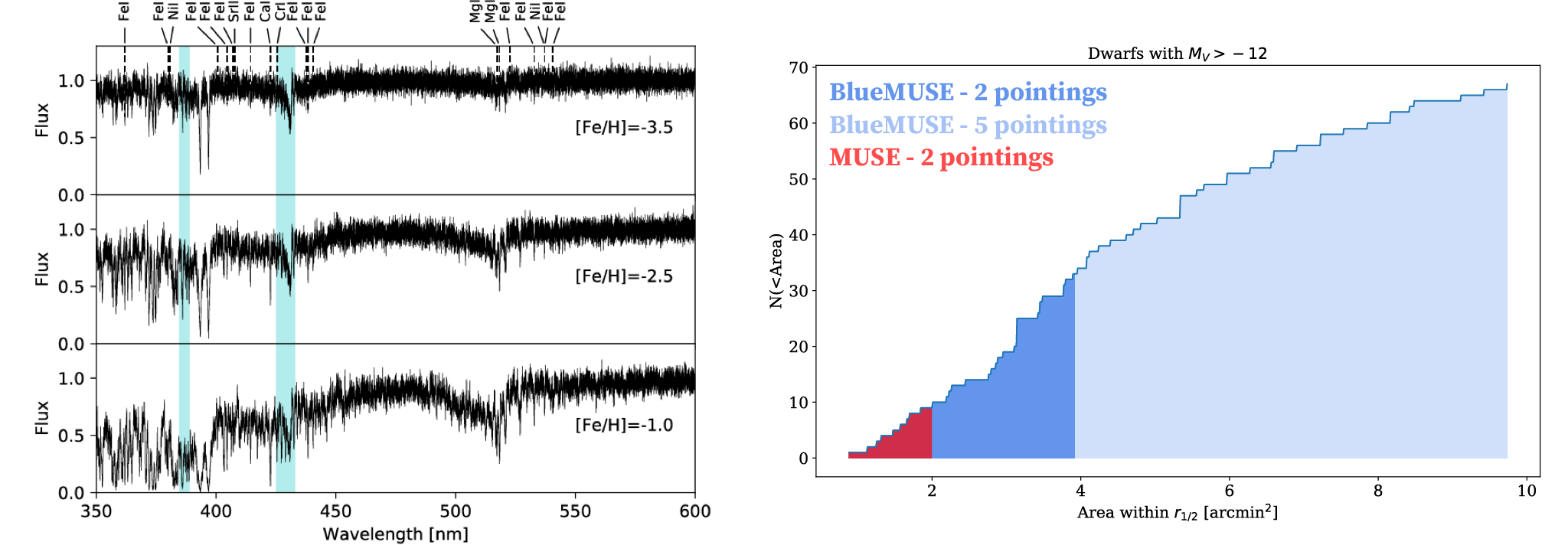}
    \caption{
    Left: Example spectra of RGB stars at three different metallicities and for
    S/N=$20$ at BlueMUSE's spectral resolution. The chemical elements and molecular bands (in blue) which are accessible at this resolution and signal-to-noise ratio are indicated. Right: The cumulative number of dwarf galaxies all sky with M$_{\rm V}>-12$ as a function of the area within the half-light radius. The dwarfs that MUSE can map with two pointings
($\sim$8~hrs) is shown in red and those by BlueMUSE in
blue. For comparison the number of galaxies mapped by five pointings of
BlueMUSE is shown in light blue. BlueMUSE will allow flexible mapping of
$\sim$3 times more faint dwarf galaxies than MUSE, allowing
statistical studies of dwarfs.}
    \label{fig:ufd}
\end{figure}

This is especially true for stellar abundance determinations. At low metallicities, [Fe/H]$<-2$, the red part of the stellar spectrum has relatively little information, while the blue part is still rich in spectral features including some molecular bands such as CN and CH (see Fig.\ref{fig:ufd}).
%For the estimation of stellar abundances and in particular the detection
%of unique enrichment episodes, it is essential to have wavelength
%coverage to 3500 \AA\  because transitions from neutron capture elements are
%%2-3 times as abundant at wavelengths shortwards of 5000 \AA\ with a peak
%between 4000 and 3500 \AA\  (\citealt{2004PASA...21..110B}, Fig.\,\ref{fig:ufdspectra}).

In order to unveil the operation mode(s) of star formation in UFDs and the extent of their chemical evolution, it is mandatory to accurately determine the  atmospheric parameters (effective temperature, gravity, [Fe/H]) of a large sample of stars, as well as their chemical patterns (e.g., abundances in $\alpha$, iron-peak, neutron capture elements etc...), their mean values and their dispersion.  
In that respect, BlueMUSE will open-up an entire new region of parameter space, allowing much better constraints, although analysis methods must also be improved as the spectral resolution is lower than is usually used for these studies. In addition BlueMUSE will provide large samples of 
secure member stars which allows for efficient follow-up with higher resolution
spectrographs, for example to measure the abundances of neutron capture elements, which are both constraining the galaxy chemical evolution path and the nucleosynthesis origin of the r-process (e.g., \citealt{2016Natur.531..610J}). 

While MUSE itself can be used to determine velocities, the velocity
uncertainties with BlueMUSE will be at least a factor of two smaller. This is
essential when the velocity dispersions of the systems are just a few
km\,s$^{-1}$.  In the most metal poor systems the rich set of absorption lines in the blue wavelength range will improve this further. Furthermore, most dwarf galaxies are larger than 1' on the side and BlueMUSE is twice as efficient in mapping as MUSE. This leads to a significant gain in the number of UFDs that can be efficiently mapped by an IFU. In order to reach the main sequence in most nearby UFDs, we need to go to r=24.5 which would require an integration time of
$\sim$5~hrs per field. If we consider 10~hrs per galaxy,
ie. two FOVs, to be a reasonable commitment, we find that BlueMUSE will
be able to map 3 times as many dwarfs as MUSE (Fig.\,\ref{fig:ufd}) given
the distribution of dwarf galaxy sizes. Since the study of dwarf
galaxies is limited by the number of these galaxies known, a jump from
$<10$ to $\sim$30 galaxies is very significant.  The capability of BlueMUSE to efficiently survey the external regions of the UFDs will serve the investigation of possible tidal features. When necessary it will allow to revisit the modeling of the galaxy dynamical mass, hence of their dark matter content.

\medskip\par
\textbf{Synergy with other facilities}:

$\bullet$ ``Classical'' MUSE: allows the coverage of the \caii-triplet as a
complementary velocity and metallicity tracer.

$\bullet$ Target selection for ELT-MOS: follow-up spectroscopy at higher
spectral resolution

$\bullet$ ERIS, MICADO, HARMONI: follow-up on red supergiants and nuclear clusters

  \hypertarget{ionized-nebulae}{%
  \subsection{Ionized Nebulae and the collisionally excited lines / optical recombination lines abundance discrepancy problem }\label{ionized-nebulae}}

\begin{tcolorbox}[colback=blue!5!white,colframe=blue!75!black,title=Science Goals]
\begin{itemize}
\item
Extend MUSE line diagnostics for the primary coolant (oxygen) of ionized gas
with inclusion of the important temperature dependent [\oiii]~$\lambda$4363\AA\ line and 
[\oii]~$\lambda$3727\AA\ doublet providing, together with [\oiii]~$\lambda$4959,5007\AA, 
inclusive diagnosis of O abundance.
\item
  Examine the ratio of forbidden to recombination line abundances
  co-spatially over extended nebular areas, towards understanding the
  abundance discrepancy factor problem
\end{itemize}
\end{tcolorbox}

The study of ionized nebulae forms one of the pillars in understanding
circumstellar, interstellar and intergalactic media. The line and
continuum from energy loss of ionized and neutral gas is a fundamental
aspect of the formation and evolution of stars and galaxies.
Spectroscopy in the UV-optical- near-infrared region is uniquely
suited to the exploration of physical conditions, light element
abundances, dust properties and the gas dynamics. An extension of the
MUSE concept to bluer wavelengths, and an increase in spectral
resolution, both have strong advantages for the study of nearby gaseous
nebulae of all types, e.g., protostellar outflows, \hii\ regions,
planetary nebulae (PNe), Wolf-Rayet nebulae, nova shells, supernova
remnants (SNR). This gain relies on the presence of many diagnostic
nebular lines in the blue wavelength range, which are astrophysically
important and have been well studied spectroscopically.

\begin{figure}
    \centering
    \includegraphics[width=16cm]{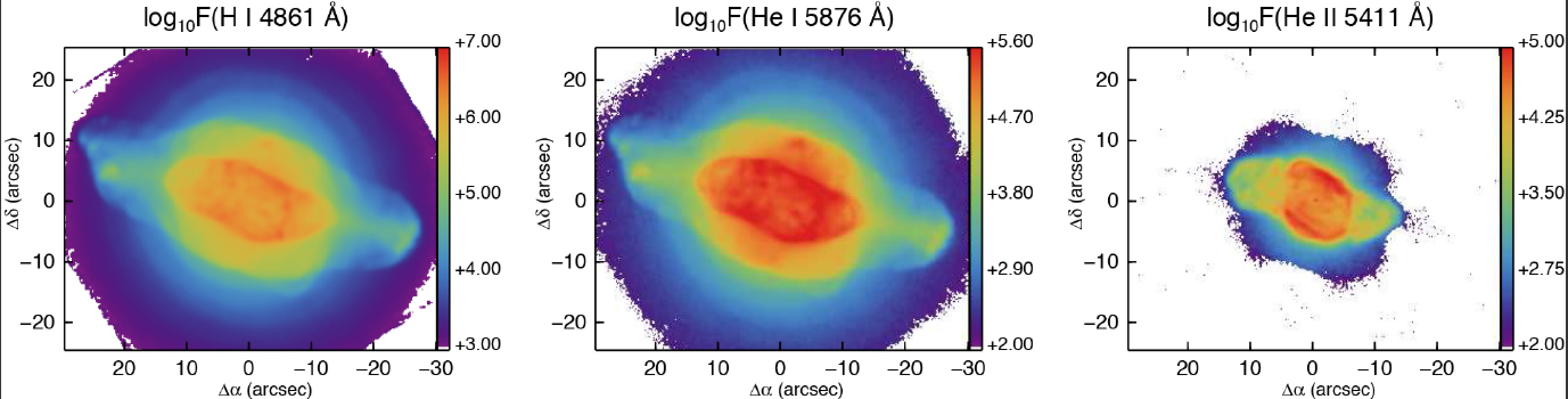}
    \includegraphics[width=12cm]{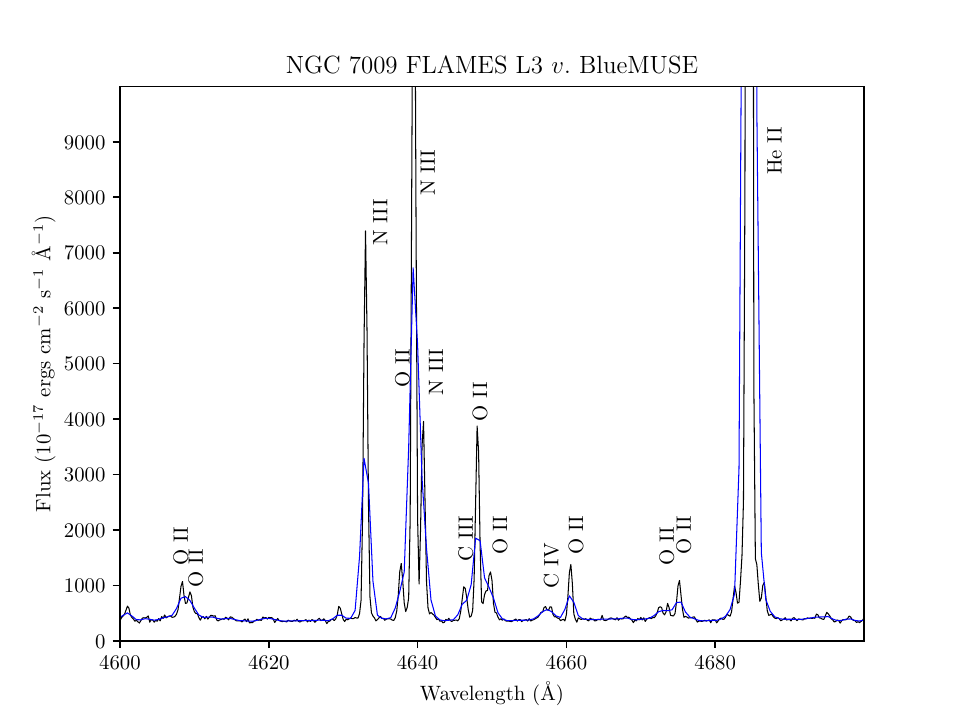} 
    \caption{(Top) Example of emission line maps across the planetary nebula NGC~7009, illustrating the spectroimaging capabilities of a large IFU for such objects \citep{2018A&A...620A.169W}. (Bottom) BlueMUSE simulated spectrum (in blue) of the rich 4600--4700\AA\ region of a 2$\arcsec^{2}$ region in the same object.  The spectrum was 
formed by smoothing and rebinning the FLAMES Medusa L3 spectrum (shown in black,
see \citealt{2008MNRAS.386...22T}) 
to the BlueMUSE resolution and pixel size. This region is rich in recombination
lines of \niii\ and \cii\ and in particular the multiplet M1 of \oii, 
which can be used to measure the O$^{++}$ recombination line abundance. It is
clear from this example that BlueMUSE will be very powerful for measuring these 
recombination lines over large areas of resolved Galactic and Local Group ionized 
nebulae.}
    \label{fig:ngc7009}
\end{figure}

A long standing controversy concerns the light element abundances of
nebulae (PNe and \hii\ regions in particular, but not restricted to these)
determined from the traditional, and strong, collisionally excited lines
(CELs) compared to the fainter recombination lines of the same elements
(mostly C, N and O). Many studies (e.g., \citealt{2006IAUS..234..219L,2003MNRAS.338..687T})
have shown differences between abundances from CEL and optical
recombination line (ORL) determinations, called the abundance
discrepancy factor (ADF), from values of a few for \hii\ regions up to $>50$ for some PNe \citep{2014RMxAA..50..329P}. Many suggestions have been made to
explain this discrepancy but no one explanation seems to be convincing,
from temperature and density fluctuations, to mixed media consisting of
inclusions of cooler high abundance, or equivalently H-poor, gas. The
key to the discrepancy seems to be that ORLs emit more strongly in
cooler (and possibly denser) media on account of the flatter emissivity
variation with temperature compared to CELs that are also affected by
collisional de-excitation at high densities. Despite intensive searches,
no spatial variation of CEL \emph{v}. ORL emission sites has been found
other than a sub-class of PNe, the born again PN, which have experienced
a late He shell flash ejecting He rich material into the pre-existing
shell. \textbf{The implications of the ADF problem cannot be over-emphasized: if
CEL abundances are wrong by large factors then the use of these lines
for studies of the ISM in general, and of distant galaxies in
particular, becomes problematic; if, on the other hand, ORL abundances
are not representative, then the physics of circumstellar media, and by
association the ISM, is poorly understood, or the atomic physics
required for recombination line abundance determination needs revising.}

The current MUSE range, while including some \nii\ and \cii\ ORLs (c.f.,
\citealt{2018A&A...620A.169W}), does not include the strongest ORLs of \oii\ and \oiii\ (4300--4700 \AA) which are crucial for exploring the ADF problem, since
ORL and CEL O\textsuperscript{+} and O\textsuperscript{++} abundances
(the latter from the very strong [\oii]~$\lambda3727$\AA, and [\oiii]~$\lambda4959,5007$ \AA, lines) can be spatially compared. Other important ORLs, of \cii\ 
and \ciii, together with \neii\ ORL's (c.f.\ the spectral compilations of
\citealt{2004ApJ...615..323S} and \citealt{2011MNRAS.415..181F}) also require a bluer spectral range (see Fig.\,\ref{fig:ngc7009}).
The other diagnostics, which BlueMUSE would open, are:

\begin{itemize}
\item the well-observed and strong [\oiii]~$\lambda4363$ \AA\ line for electron temperature (T\textsubscript{e}) determination of higher ionization gas;
\item the [\oii]~$\lambda$3726,3729 \AA\ doublet for electron density (N\textsubscript{e}) of lower ionization gas and as important tracer of O$^+$, together with O$^{++}$ providing the important oxygen abundance;
\item the [\ariv]~$\lambda4711,4740$ \AA\ ratio for N\textsubscript{e} in higher ionization gas will be better determined than with the MUSE extended mode, since the [\ariv]~$\lambda4711$ \AA\ will be resolved from \hei~$\lambda4713$~\AA;
\item the [\neiii]~$\lambda3869$ \AA\ line, entirely missing from MUSE, will be available for Ne\textsuperscript{++} abundance determination;
\item the Balmer jump at 3646 \AA\ a powerful estimator for ORL T\textsubscript{e};
\item the higher order lines of the hydrogen Balmer series can be used as ORL N\textsubscript{e} estimator.
\end{itemize}
Higher spectral resolution is a strong advantage
since the emission line spectrum is crowded, with many H and He
recombination lines in addition to the CELs, thus line blending can be a
problem particularly for the extraction of faint ORLs (see Fig.\,\ref{fig:ngc7009}). The use of an IFU is essential to sample the entire face of ionized nebulae that can be subject to different mechanisms of excitation (photoionization, shocks) and hydrodynamical effects, as well as spatially varying distributions of chemically enriched material.

  \hypertarget{comets_and_asteroids}{%
  \subsection{Comets and asteroids}\label{comets}}

\begin{tcolorbox}[colback=blue!5!white,colframe=blue!75!black,title=Science Goals]
\begin{itemize}
\item
  Study the morphology and underlying processes that shape the coma
\item
  Set constraints on the properties of the nucleus and the origin of
  chemical elements
\item
  Identify cometary activity in asteroids
\end{itemize}
\end{tcolorbox}

\medskip\par

{\bf Comets:} 

Comets are pristine relics of the protoplanetary disk, where the planets
formed and evolved, and preserve in their nucleus important clues about
the early solar nebula. One of the main questions to answer about comets
is the physical origin of the radicals (CN, C\textsubscript{2},
C\textsubscript{3}) in particular their locus of production within
comets. Ground-based observations of comets can only detect the coma,
not the nucleus. However, the nucleus strongly influence the morphology of the coma 
 via processes such as nucleus rotation, obliquity, and active regions on its surface.
 Mapping the coma morphology using an integral-field spectrograph allows
us to study the underlying processes that shape the coma and set
constraints on the properties of the nucleus, the fundamental truth we
are seeking.

To uncover the origin of radicals (CN, C\textsubscript{2},
C\textsubscript{3}) observed at optical wavelengths in the coma of
comets, one needs to study species parentage. It is a complex problem,
as some radicals can have several possible parents or be released by
different mechanisms, which are not easy to identify. However, a better
understanding of those mechanisms is crucial to link optical
observations of comets to nucleus ice abundances. Mapping the coma
morphology, using an IFU spectrograph, and comparing the
spatial distribution of the gas and the dust, allows us to study the
underlying processes that shape the coma and produce radicals such as
CN, C\textsubscript{2} or C\textsubscript{3} and also to set constraints
on the properties of the nucleus.

\begin{figure}
    \centering
    \includegraphics[width=12cm]{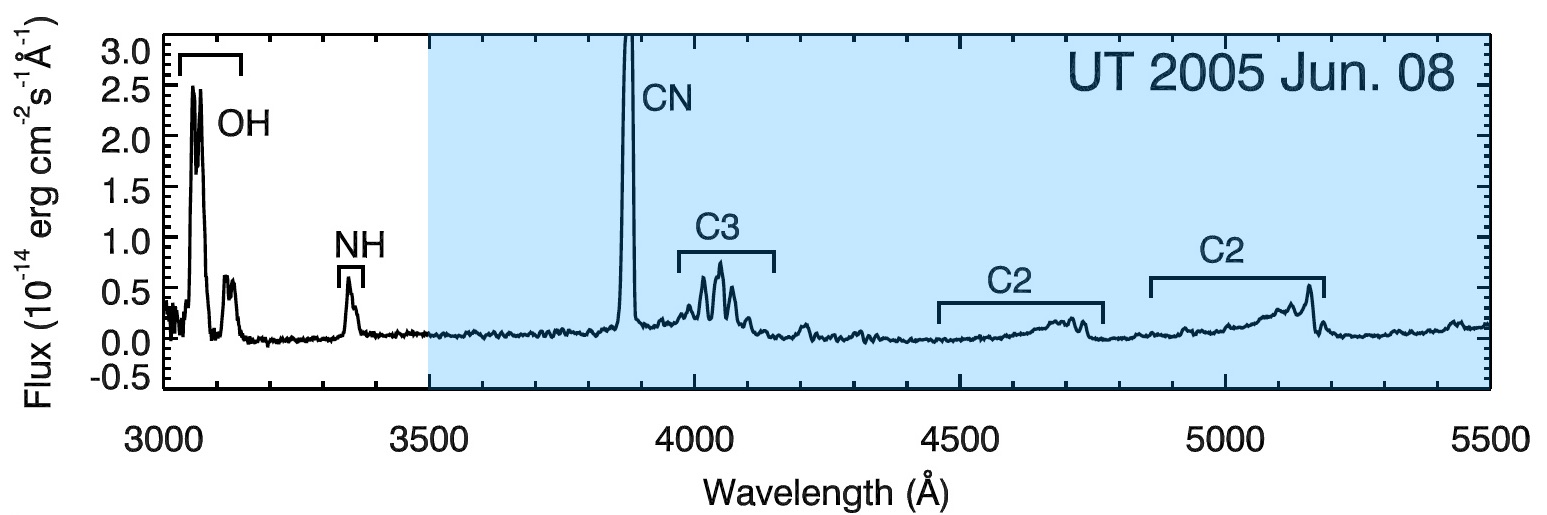}
    \includegraphics[width=4cm]{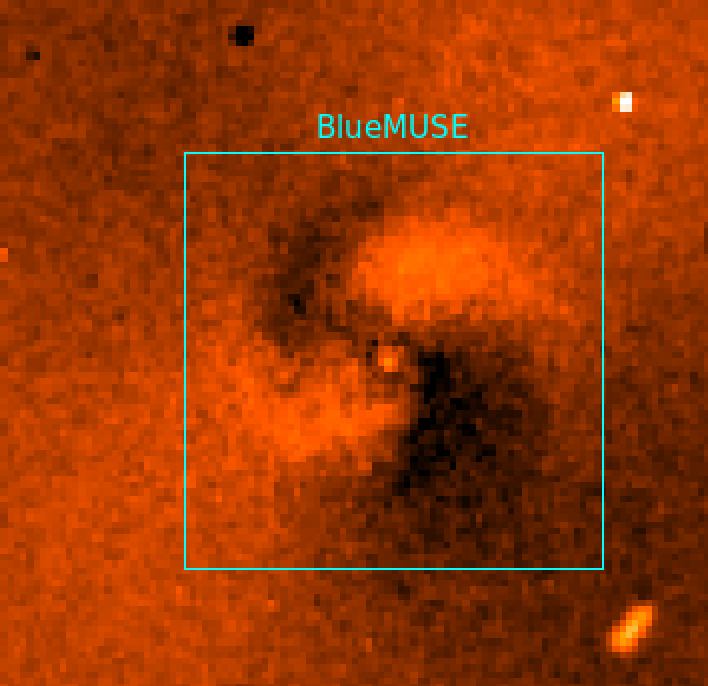}
    \caption{(Left) typical coma spectrum of comet 9P/Tempel 1, as observed
at a distance of 1 AU \citep{2011Icar..213..323M}. The BlueMUSE coverage
(blue-shaded region) covers multiple radicals like CN,
C\textsubscript{2} and C\textsubscript{3} including several groups of
transitions. (Right) Example of morphology expected in the central
region of the coma within the BlueMUSE FoV (as seen in comet Lemmon,
\citealt{2015A&A...574A..38O}). Spiral structures like this one can be observed
depending on the activity, the orientation and the geometry of the
observation.}
    \label{fig:comets}
\end{figure}

The wavelength range of BlueMUSE allows for a simultaneous coverage of
multiple radicals at $350<\lambda<500$ nm (Fig.\,\ref{fig:comets}, left). In
particular in the CN, C\textsubscript{2}, C\textsubscript{3} group, many
transitions that can be studied individually in a single IFU observation
over a large spatial scale, which is impossible to do efficiently with
narrow-band filters or long-slit spectroscopy (e.g., \citealt{2013ApJ...778..140D}).
In addition, observations with an IFU (e.g., \citealt{2017AJ....154..219V}) allow
to simultaneously study several gas species and the dust without any
concern about the effects of rotation (typically a few hours). Cometary
lines are very narrow, but moderate spectral
resolution of $\sim$4000 like BlueMUSE is sufficient to study the spatial distribution of gaseous
species.

BlueMUSE is the only instrument that enables us to study the morphology
of several gaseous species simultaneously while benefiting from such a large field of view and being
attached to an 8-m class telescope, and allows for a direct comparison
with the dust morphology (measured from the spectral continuum). The MUSE instrument, for example, does not cover the very strong CN band at 388 nm. Large
spatial structures are expected over the BlueMUSE field-of-view based on
current narrow-band observations, such as spiral-like structures (Fig.\,\ref{fig:comets}, right). One of the hypotheses that can be directly tested with
BlueMUSE is whether dust grains could be a source of production for the
CN.

At 1 AU, the scale length of a gas coma is on the order of
10$^{5}$ km, which is about 2 arcmin, and a good fit to the
1.4$\times$1.4 arcmin$^2$ FoV of BlueMUSE, as the surface
brightness flux of emission lines drops away further into the outer
coma. For a moderately active comet, the surface brightness flux for a
gas emission band is on the order of: 10$^{-14}$
erg\,s$^{-1}$\,cm$^{-2}$\,arcsec$^{-2}$. A signal-to-noise
of 10 per BlueMUSE pixel (0.3\arcsec) can be achieved in the continuum and
all spectral lines in typically 10 mins, allowing for a fine monitoring
of the coma over the full rotation of the comet (a few hours).

\medskip\par
{\bf Asteroids:} 

Most of the asteroids are inactive objects but a few of them (slightly over 30 objects known so far) present cometary activity. They are called active asteroids or Main Belt Comets. Even if they represent a small fraction of asteroids or comets their study is important for a better understanding of physical properties of both comets and asteroids, the frontier between these two categories of planetary bodies being not so obvious for scientists \citep{2017RSPTA.37560259H}. Different physical mechanisms can drive such cometary activity \citep{2015aste.book..221J}: rotational mass loss, impacts, thermal disintegration, sublimation of ice, radiation pressure sweeping, electrostatics and gardening... It is difficult to get observational clues that can permit to differentiate these possible mechanisms. In this context BlueMUSE can uniquely help in mapping gaseous emission bands of abundant species in comets like CN (388 nm) or C2 (516 nm) as well as colours scattered by the dust, allowing detailed studies of the spatially resolved physical properties of these objects.

\newpage

\hypertarget{nearby-galaxies}{%
\section{Nearby galaxies}\label{nearby-galaxies}}

  \hypertarget{key-science-case-ism-and-hii-regions-extreme-starbursts}{%
  \subsection{\texorpdfstring{\textbf{Key science case:} ISM and HII regions, extreme starbursts
  }{Key science case: ISM and HII regions, extreme starbursts }}\label{key-science-case-ism-and-hii-regions-extreme-starbursts}}

\begin{tcolorbox}[colback=blue!5!white,colframe=blue!75!black,title=Science Goals]
\begin{itemize}
\item
  Determine the physical conditions in the interstellar medium and
  diffuse haloes/circumgalactic medium of starburst galaxies by mapping
  multiple emission lines
\item
  Quantify the interplay between the populations of massive stars
  (supernovae, stellar winds and ionizing radiation) and their
  surroundings, in galaxies that drive the strongest outflows as
  analogues of high-redshift galaxies that enriched/polluted the
  intergalactic medium with metals.
\item
  Determine the opacity to Lyman continuum and Lyman-$\alpha$ radiation,
  through the ionization state and density of the gas, in the analogue
  galaxies of those that reionised the Universe.
\item
  Study the assembly history of low-mass starburst galaxies in a spatially resolved manner
\end{itemize}
\end{tcolorbox}

Low-metallicity starburst galaxies (blue compact dwarf galaxies - BCDs) provide a unique window in our
understanding of galaxy formation, and fulfill several specific and
irreplaceable roles in extragalactic astronomy and cosmology. They offer a unique opportunity to study galaxy formation under conditions approaching those of the first galaxies, prior to and during the epoch of reionization (EoR). They are: (1) ideal laboratories in which to study at low intrinsic extinction the most rare, massive and extreme stars  \citep{2009A&A...499..455C,2015ApJ...801L..28K}; (2) enable detailed studies of collective star formation and the associated feedback processes in the least chemically evolved local environments known
\citep{2000A&ARv..10....1K,2015MNRAS.448.2687J}; (3) are likely the sites of super-luminous supernova/hypernova
explosions \citep{2015MNRAS.449..917L} and long-duration GRBs \citep{2015ApJ...806..250H}; (4) their shallow gravitational potential wells provide less
resistance to galactic outflows, enabling material that is heated by the
star formation process to escape into the galaxic halo and beyond and
enrich the intergalactic medium with metals; additionally, chemical abundance patterns in BCDs 
\citep{1995A&A...294..432R} can place valuable constraints 
on the timescales for dispersal and mixing of heavy elements in protogalaxies; (5) they may have more
porous/disrupted ISM, which enhances the escape of Lyman-$\alpha$ radiation and
ionizing continua. They thereby allow us to understand how similar
galaxies at high-redshift leak ionizing photons, reionise the Universe,
and maintain the meta-galactic ionizing background; (6) they may
be the closest analogues of some of the faint galaxies identified in high-$z$ surveys, that dominate the star-formation budget at early
times; and finally (7) several lines of evidence suggest that some of the most metal-poor (12+log(O/H)$<$7.6) local BCDs 
have experienced the dominant phase of their build-up at a late cosmic epoch. For this reason, they are convenient 
laboratories to explore the main processes driving dwarf galaxy formation, as long as their morphological and dynamical relics have not had time to be erased in the course of secular galactic evolution \citep{2008A&A...491..113P}. 
For instance, the majority
of these systems shows a cometary morphology that might result from unidirectional star formation propagation. Spatially resolved spectral synthesis studies of BlueMUSE data will offer a tremendous
potential for reconstructing the assembly history of these galaxies and shedding light into the regulatory role of feedback on the synchronization of the star formation process on scales of kpc.
In contrast to objects at $z\sim2$ and beyond, such
local starbursts can be studied in enormous detail and the physical
properties determined exhaustively.

Top of this list is how feedback -- both radiative and mechanical --
from star-formation heats and disrupts the interstellar and
circumgalactic gas, and how large-scale, enriched galaxy winds develop
and evolve. In order to measure the thermal and kinetic energy in the
gas phase we need accurate mapping of temperature and density to derive
pressures and accurate masses, and well-resolved lines to identify
individual kinematic components of gas and measure the turbulent
broadening. Determination of the excitation/ionization state is vital in
order to understand how ionizing photons may penetrate the \hii\
regions to the diffuse ISM, and escape from galaxies through ionized
channels. This needs high-resolution optical spectroscopy, mainly
focused at the blue end of the spectrum, and large fields of view in
order to capture the whole extended gaseous haloes, i.e., BlueMUSE.

\begin{figure}
\begin{minipage}{8cm}
\includegraphics[width=8cm]{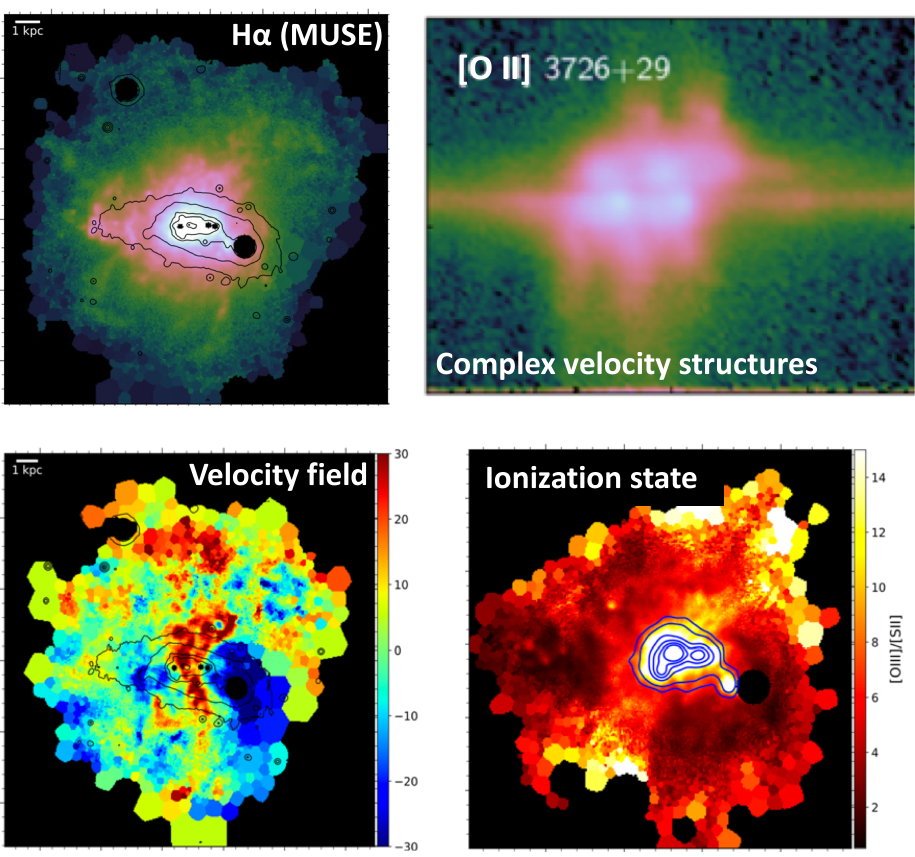}
\end{minipage}
\begin{minipage}{9cm}
   \caption{Current state-of-the-art in extragalactic ISM diagnostics, derived from VLT/MUSE and X-Shooter.   All images show the local luminous blue compact galaxy ESO 338-IG04, the nearest analogue of typical Lyman break galaxies at $z\gtrsim2$.  The \emph{upper left} panel shows H$\alpha$ emission (contours are stellar emission) which fills the field of view of MUSE, and \emph{lower left} shows the complex velocity field derived from it. \emph{Lower right} shows the [\oiii]/[\sii] ratio, which traces the ionization state of the gas (yellow is highly ionized) revealing hot channels emanating from the centre.
\emph{Upper right} shows the complex velocity structure
exhibited in the [\oii]~$\lambda$3726,29\AA\ doublet in a 2D spectral image from X-Shooter.   BlueMUSE will resolve multiple velocity components in lines like [\oii] over the entire field-of-view, and map the ionized gas density in 3 dimensions.}

    \label{fig:ism}
\end{minipage}
\end{figure}

\medskip\par
\textbf{Current status with MUSE and other instruments:}

Our current understanding of the ionized gas haloes and details of the
ISM has mainly been guided by MUSE (which easily maps diffuse gas on large scales), with
higher resolution spectrographs (e.g., X-Shooter, FLAMES) determining kinematics
and measuring vital emission lines bluewards of the MUSE bandpass -- see
Fig.\,\ref{fig:ism} \citep{2018A&A...619A.131B}. This shows ESO 338-IG04, a local luminous
blue compact galaxy and one of the nearest analogues of Lyman break galaxies at
$z\gtrsim2$. The contours show the stellar
continuum, but the ionized gas (traced by H$\alpha$) is clearly found out to
$\sim$10$\times$ the effective radius and fills the MUSE
field-of-view; its kinematics, shown in Fig.\,\ref{fig:ism} lower left, reveal a
turbulent and complex extended halo that is heated and blown out in
winds from massive stars and supernovae exploding in the central starburst. This shows
roughly biconical outflows that are highly ionized, as illustrated in
the lower right panel which shows the [\oiii]/[\sii] ratio. This
highly ionized outflow also correlates with Lyman-$\alpha$ emission \citep{2005A&A...438...71H}, which is enhanced co-spatially with these channels, which
likely also reveals the mechanism by which ionizing radiation escapes
galaxies (see `Lyman Continuum' science case \S\ref{lyman-continuum-emitters}). The upper right panel
shows the spectrum of the central region of this galaxy, demonstrating
how the complex kinematics shown to the lower left decompose into
various velocity components, as the turbulent gas is ripped up and
driven out by feedback. Only in these high resolution, blue
spectra can we see emission lines that allow us to derive the density
and temperature using the same ions, and therefore allow us to derive detailed abundances in both the ISM and galaxy winds.  This in turn provides the pressure, internal energy and mass of gas in the wind. Only
with this information can we begin to distinguish inflows and outflows, and determine the detailed energy balance and
ultimate future of galactic winds. We will therefore infer the  influence of winds on the
IGM enrichment (see `Gas Flows' science case \S\ref{key-science-case-gas-flows-around-and-between-galaxies}), and the fate of the
available gas in the ISM.

There is no doubt that high-resolution blue spectrographs and large
format IFU have lead to significant enhancements in our understanding of
the diffuse ISM and circumgalactic medium (CGM). However these existing instruments have two
main shortcomings: (1) high-resolution blue instruments (e.g., X-Shooter)
are slit-only spectrographs, and do not allow us to map CGM gas; and (2)
the only existing large format IFU (MUSE) has neither the blue coverage
nor the spectral resolution to capture the blue lines and decompose the
velocity components. This represents a major limitation because the
strongest temperature diagnostic is the [\oiii]~$\lambda$4363 \AA\ line, while
the strongest density probe is the [\oii]~$\lambda$3726,3729 \AA\ doublet.
Alternative proxies, [\nii]~$\lambda$5755 \AA\ and the [\sii]~$\lambda$6716,6731 \AA\
doublet are far too weak to be recovered on any pointings apart from the
highest surface brightness star-forming regions. This point becomes even
more true at low metallicity \citep{2017A&A...606L..11H} as we target the
analogues of the earliest dwarf galaxies to form -- the systems that
likely reionised the Universe. Without temperatures and densities in the
halo gas, we are missing key quantities needed to understand the
detailed physics of the starburst ISM as global energetic numbers are
not available.

Even blue compact galaxies, with continuum half-light radii of a few
arcsec, completely fill the MUSE field-of-view as we are detecting
diffuse circumgalactic gas. The enhanced FOV of BlueMUSE is therefore
vital. Moreover the spectral resolution, that decreases to
R$\sim$1800 at the blue end of MUSE is insufficient to
resolve multiple velocity components in nebular gas; the resolving power of BlueMUSE is ideally set to match the velocity dispersion of extragalactic nebulae.  Very importantly the [\oiii]~$\lambda4363$ \AA\ and
[\oii]~$\lambda3726,3728$ \AA\ doublet are not captured by any current 
large format IFU at redshifts where the galaxy can be spatially
well-resolved $(z<0.3)$. Like MUSE, the BlueMUSE instrument will systematically capture the  blue Wolf-Rayet bump and \heii\ 4686 \AA\ line in every observation, both of which are vital for constraining the ionizing photon
budget and contribution of these stars to feedback.  Current samples of such galaxies, selected for observation with HST \citep[e.g.][]{2014ApJ...797...11O,2017ApJ...851L...9J,2013ApJ...779...76Z,2017ApJ...847...38Y,2017MNRAS.472.2608S}, indicate that there will be hundreds of southern-hemisphere compact dwarf starbursts from which to assemble key science programs. A recent example of IFU observations of outflows in the nearest green pea galaxy analogue Mrk71 obtained with PMAS, including the [\oii] doublet in the UV, illustrates the potential that will become available with BlueMUSE at the VLT \citep{2019A&A...623A.145M}.

\medskip\par
\textbf{Synergies with other facilities }

LSST will find huge numbers of undiscovered compact starbursts, in the
same way that SDSS discovered green peas \citep{2009MNRAS.399.1191C}. Given the huge increase in depth these galaxies will extend to even
lower stellar mass and lower metallicities. These will be ideal, and
very timely systems to follow up with BlueMUSE.

Currently \hi\ observations of compact starburst galaxies are only
possible with the VLA, and are of special relevance for understanding
the Lyman continuum throughput. However beyond 50 Mpc these observations
become very challenging, and at $z\sim0.2$ are completely
hopeless at these galaxy masses. SKA will provide observations of the atomic material.

  \hypertarget{low-surface-brightness-galaxies}{%
  \subsection{Low surface brightness
  galaxies}\label{low-surface-brightness-galaxies}}

\begin{tcolorbox}[colback=blue!5!white,colframe=blue!75!black,title=Science Goals]
\begin{itemize}
\item
  Characterize star formation, dust properties, and metals through
  emission line mapping, allowing to distinguish between the different
  formation models.
\item
  Measure the distance and probe the kinematics of low surface
  brightness galaxies and ultra-diffuse galaxies, providing crucial
  information to study these populations.
\end{itemize}
\end{tcolorbox}

Low Surface Brightness galaxies (LSBs) could represent a large fraction
of local galaxies, up to 50\% according to \citet{2000ApJ...529..811O}.
Despite this large fraction, their nature and origin have remained
unknown: are they large spin disks (e.g., \citealt{2003MNRAS.343..653B,2016A&A...593A.126B}) for
instance, or the results of head-on collisions \citep{2008MNRAS.383.1223M}.
Their exceedingly low surface brightness has been hindering in-depth
studies of this important population, from which we could obtain a
census, on crucial scientific questions:

• A good constraint on the shape of the luminosity function of local
galaxies can only be reached by taking into account the LSBs \citep{2005ApJ...631..208B}.

• LSBs bring crucial elements to understand the DM nature since they may
be DM-dominated \citep{1997AJ....114.1858P}.

• LSBs allow the study of star formation in the low density regime, for
which many issues are still debated such as lower efficiency, threshold,
IMF variations, etc. This is directly comparable to the issues found in
the recently discovered phenomenon of XUV galaxies \citep{2005ApJ...627L..29G,2007ApJS..173..538T}, with extended diffuse disks found around
otherwise ``normal'' galaxies (see also \citealt{2016ApJ...826..210H}).

\textbf{The study of LSBs is thus of paramount importance for our
understanding of galaxy evolution.}

Current instruments have considerably gained in sensitivity, allowing
new studies of very diffuse objects, e.g., \citet{2015ApJ...807L...2K,2015ApJ...809L..21M}. The Virgo cluster was observed in the optical \citet{2012ApJS..200....4F}, reaching 29 mag\,arcsec$^{-2}$ in \emph{g}' band, and
in the UV \citep{2011A&A...528A.107B}. We still miss, however, comprehensive
spectroscopic surveys. In an on-going work, Madathodika et al. (in preparation)
uses Magellan-IMACS long-slit spectra in a few different places of one
of the largest LSBs (Malin 1) to put new constraints on the inner rise
of the rotation curve, crucial for the dark matter content determination
of this galaxy, and to estimate the star formation surface density in a
few knots. However, a full census of star formation regions, and a
better determination of the dynamics clearly call for large-IFU
observations. With BlueMUSE, it would be possible to include most parts
of the galaxy in a single pointing when four pointings are needed with
MUSE. Moreover, BlueMUSE will allow key observations of many emission
lines that are of utmost importance to understand the nature and origin
of these galaxies. Oxygen lines will be used to determine the
metallicity of the extended disk which will allow us to distinguish
between the different formation scenarios of LSBs (e.g., a gradient of
low metallicity gas is predicted in the models presented in \citet{2016A&A...593A.126B}, but head-on collisions should have large metallicity),
combined with stellar dynamics. Only few constraints on extinction in
LSBs have been obtained so far from far-infrared data, or from H$\beta$/H$\alpha$ in
few \hii\ regions (e.g., \citealt{2007ApJ...663..908R}), but they indicate a low
amount of attenuation. The H$\beta$ flux will thus provide a direct indicator
of the SFR on a shorter time-scale than all the other bands used so far
over the full galaxy ($\sim$10, 100, 500 Myr for H$\beta$, FUV, \emph{u},
respectively), with a spatial resolution ($\sim$1\arcsec=1.7 kpc), i.e. 5 times better
than GALEX. The UV to Balmer line ratio is a direct constraint for the
time delay between star forming events (modeled in \citealt{2009ApJ...706.1527B}),
and may also be used to constrain the massive-end slope of the IMF \citep{2012ApJ...749...20K}.

Among LSBs, the galaxy Malin 1, discovered in 1986 \citep{1987ApJ...313..629B},
is one of the prime examples of giant LSB galaxies, characterized by its
extended low surface brightness disk and high gas content. Malin 1 has
the largest radial extent of any known spiral galaxy, with a low surface
brightness disk extending out to $\sim$120 kpc \citep{2006PASA...23..165M} and an
extrapolated central surface brightness of $\mu$\textsubscript{0,V} $\simeq$ 25.5
mag\,arcsec$^{-2}$ \citep{1997ARA&A..35..267I}. Its study will
provide a reference point for the family of LSBs with similar \hi\ masses
and UV colours. Deep UV and optical imaging showed that the giant disk
behaves on long time-scales like a normal galaxy but with an angular
momentum as large as 20$\times$ the Milky Way one \citep{2015ApJ...815L..29G,2016A&A...593A.126B}. This new photometry suggests a variety of ages for the UV
emitting regions; a stochastic star formation history for LSBs was also
proposed (e.g., \citealt{2008ApJ...681..244B}), but this analysis is limited to
the GALEX resolution of 5\arcsec\ and to the timescales probed by UV filters
($>100$ Myr).

Recently, the possible existence of a new class of galaxies,
characterized by a faint central surface brightness (below 24 mag
arcsec$^{-2}$) and a large effective radius
(R\textsubscript{e} $>1.5$ kpc), has been the subject of an
active debate in the community. While such extended LSB galaxies, known
as ``Ultra-diffuse galaxies (UDGs)'', had been identified in many
optical surveys of the sky, their distance and thus intrinsic properties
have only been recently determined. This measure could be achieved with
expensive long-slit spectroscopy (33~hrs with Keck) of their stellar
populations \citep{2016ApJ...828L...6V}, done as part of a follow-up of the
Dragonfly imaging project. UDGs have been found in clusters of galaxies,
like Coma, in groups and in the field (for instance in the neighborhood
of M101, \citealt{2016ApJ...830...62M}).

A sub-class of UDGs show an excess of globular clusters (GCs), an
indication that they might be rather massive objects. Their DM content
is at the centre of a hot debate within the community with highly
DM-rich galaxies (Dragonfly 44, \citealt{2016ApJ...828L...6V}) and, on the
other hand, galaxies lacking DM \citep{2018Natur.555..629V}.

In such conditions, the true nature of UDGs is highly uncertain: they
could be inflated regular dwarfs, tidally stripped satellites, tidal
dwarf galaxies, or failed massive galaxies. The lack of information on
the age and metallicity of their stellar populations prevents us from
having a definitive answer. While on-going surveys with CFHT (NGVS,
MATLAS, CFIS), Subaru (HSC SPP) or Dragonfly (DNGS) have provided
catalogues of hundreds of UDGs candidates, less than 20 have yet
spectroscopic follow-up.

\begin{subsec}[ht]
\begin{tcolorbox}[colback=green!5!white,colframe=green!75!black,title=Why is a large field-of-view needed for BlueMUSE?]

With a FoV of 1.4$\times$1.4 arcmin$^2$, BlueMUSE will be the largest monolithic IFU ever built on an 8m-class telescope.

\begin{itemize}
\item
\begin{quote}
A large FOV will increase the survey speed and multiplexing capabilities. 

\end{quote}
\item 
\begin{quote}
  For a MUSE-like monolithic Integral Field Spectrograph such as BlueMUSE, a large area will \textbf{always be beneficial} by expanding the discovery space and the number of serendipitous discoveries, even when the main target of interest does not cover the entire FoV.
  \end{quote}
\item
  \begin{quote}
Many astrophysical sources (e.g., extended nearby galaxies \S\,\ref{low-surface-brightness-galaxies}, lensing cluster cores \S\,\ref{gravitational-lensing-in-clusters}) have  physical sizes well-suited to the BlueMUSE FoV and do not require any mosaicking.
  \end{quote}
\end{itemize}
\end{tcolorbox}

\end{subsec}

\newpage

\begin{figure}
\includegraphics[width=3.02778in,height=3.02778in]{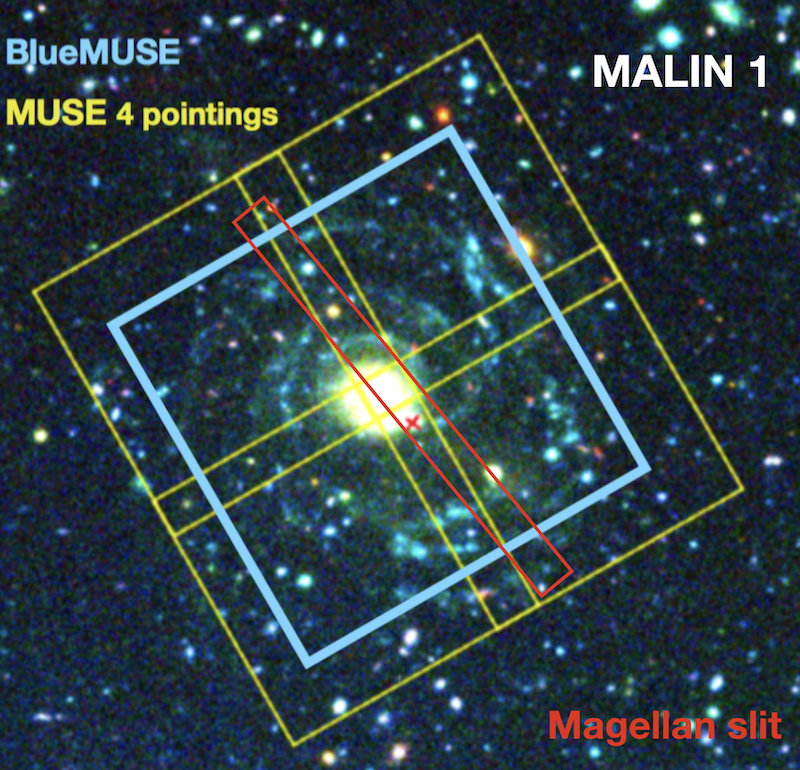}
\includegraphics[width=3.42639in,height=3.02778in]{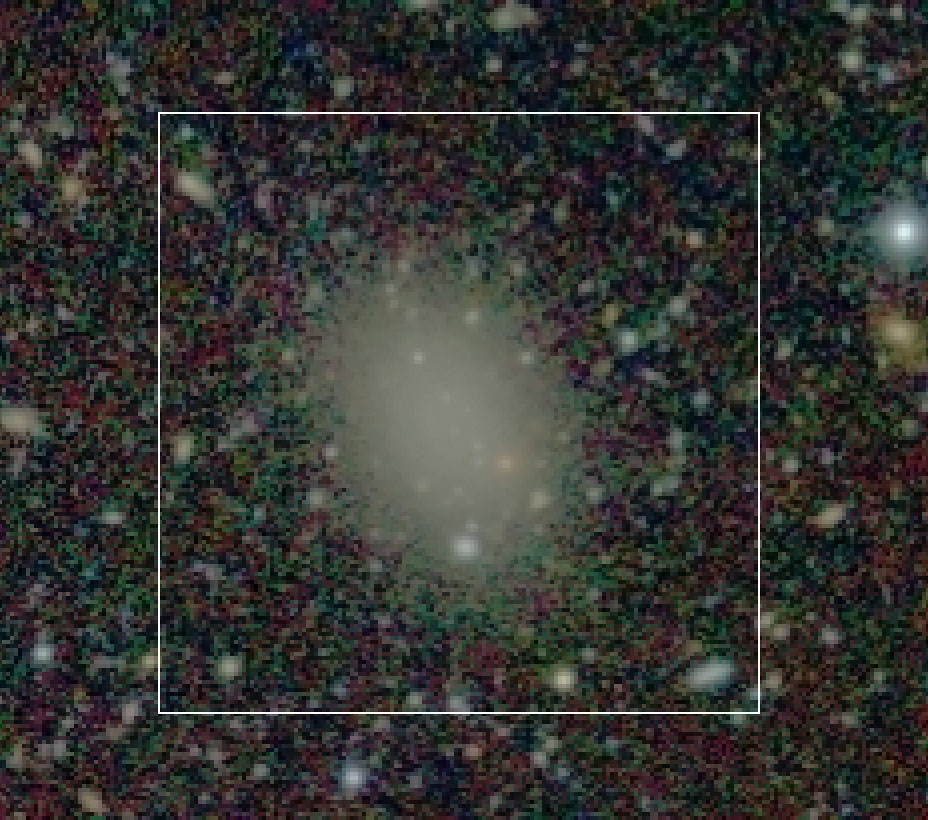}
\caption{\label{fig:UDG}Left panel: Combined UV-optical (GALEX and CFHT) images of Malin
1. The yellow squares show the four pointing mosaic that would be needed
to study this object with MUSE while the blue square marks the field of
view of BlueMUSE. Right panel: a UDG candidate in the MATLAS survey
(with CFHT, from \citealt{2015MNRAS.446..120D}), showing an excess of globular
clusters (GC). The BlueMUSE field of view (superimposed in white) will
allow spectroscopic information on the whole diffuse object and the GC
population in a single pointing.}
\end{figure}

BlueMUSE would be ideal to investigate the origin and structural
properties of UDGs, giving simultaneously information on the stellar
populations of the GC and diffuse component. BlueMUSE will revolutionize
the study of LSB and UDGs:

\begin{itemize}
\item
  Its large field of view which is perfectly suited to the size of the
  LSB/UDGs and the distribution of GCs, as seen on Fig.\,\ref{fig:UDG}.
\item
  Its wavelength coverage will provide the detections of age and
  metallicity indicator lines such as oxygen lines, only reachable in
  the blue, as well as Balmer absorption lines and H$\beta$ emission allowing
  to probe the star formation history and activity of these galaxies.
\item
  Its relatively high spectral resolution (with respect to MUSE) will
  allow us to determine the rotation curve of the star-forming gas-rich
  LSB galaxies (and more generally the full dynamics, including effects
  of eventual bars or other asymmetries), and
  for the gas-poor UDGs, the dynamics of the GC populations, together
  with the velocity dispersion of the diffuse stellar populations, thus
  testing previous hints on the DM content.
\end{itemize}

\textbf{Synergies:}

The synergy between BlueMUSE and other instruments is clear and key to
the understanding of such populations of galaxies. MUSE will bring additional emission line diagnostics, such as H$\alpha$, [\nii] and [\sii]. LSST will enable us
to detect more low surface brightness galaxies allowing us to increase
the study sample. An ALMA follow-up of the star-forming regions detected
with BlueMUSE will enable a study of dust and molecular gas in these
systems and, finally, SKA will provide a census on their \hi\ reservoir.

\clearpage

  \hypertarget{environmental-effects-in-local-clusters}{%
  \subsection{Environmental effects in local
  clusters}\label{environmental-effects-in-local-clusters}}
  
\begin{tcolorbox}[colback=blue!5!white,colframe=blue!75!black,title=Science Goals]
\begin{itemize}
\item
  Identify the dominant perturbing mechanism and quantify the
  typical timescale for the quenching process in perturbed galaxies.
\item
  Study the fate of the stripped gas in cluster and group
  galaxies
\item
  Study the star formation process in extreme environments
\end{itemize}
\end{tcolorbox}

BlueMUSE will be an ideal instrument to study the role of the
environment on galaxy evolution.

\begin{figure}
    \centering
    \includegraphics[width=6.5in]{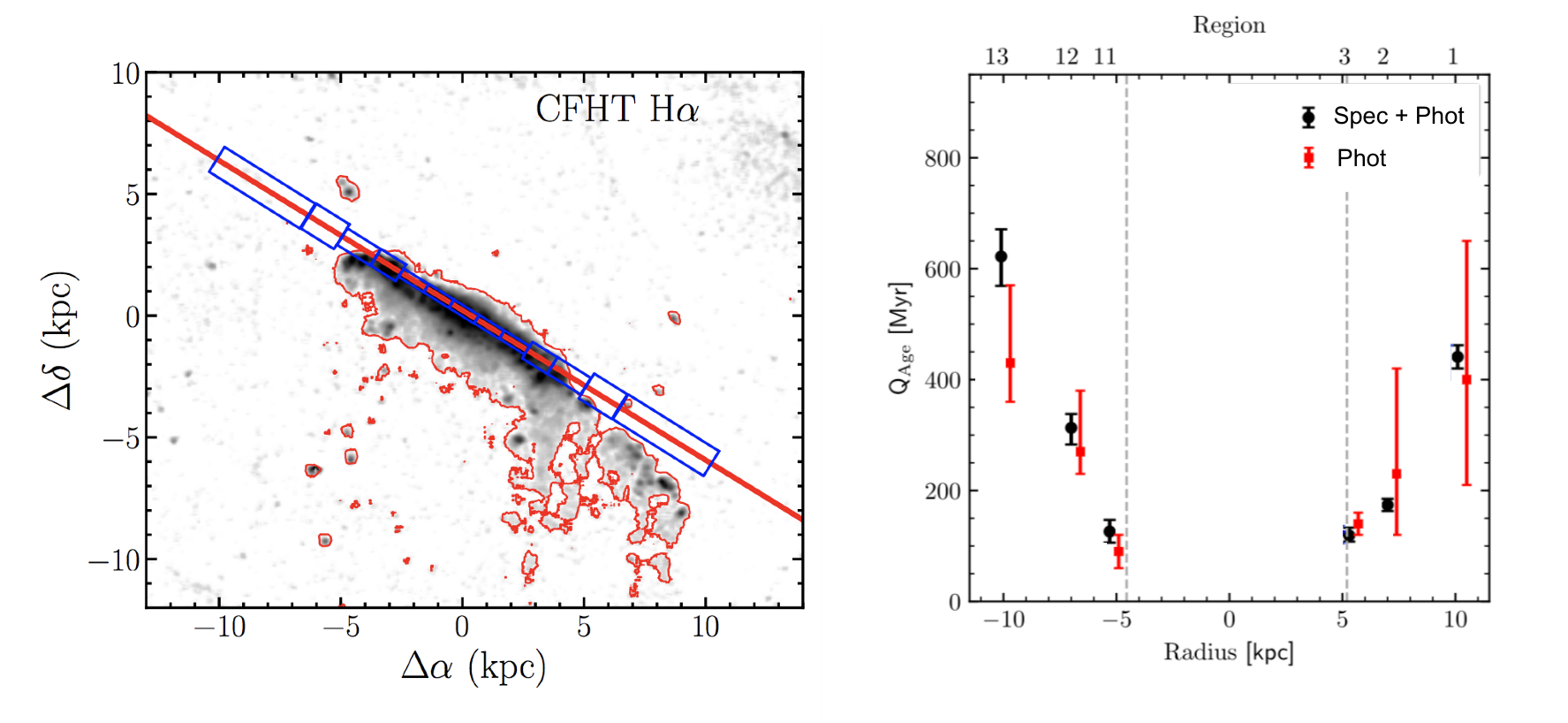}
    \caption{Left. Example of a star forming galaxy (NGC 4330) moving inside
the Virgo cluster, where ram-pressure stripping displaces the ISM from
the disc as visible from the a gas tail emitting in H$\alpha$. Right. Age of the quenching events inferred from joint fitting of multiband photometry and long-slit optical spectroscopy (black points) at blue wavelengths in different regions along the galaxy disc (blue boxes in left panel). The red points show the constraints on the quenching age from photometry alone. The inclusion of spectroscopy is critical to achieve uncertainties on the quenching age of the order of 5\% to 10\%.
By combining spectroscopy of stars and gas, BlueMUSE
will provide the ultimate view of galaxy evolution in dense
environments, including a determination of the timescale of the
quenching events and of the mass of displaced gas (adapted from \citealt{2018A&A...614A..57F}).}
\label{fig_quenching}
\end{figure}

Thanks to its large field of view, high sensitivity, spectral coverage
and resolution, it is the ideal instrument to study the quenching
mechanism in dense environments and the origin of the faint end of the
red sequence dominating the galaxy population in local clusters. Indeed,
the blue part of the spectrum of local galaxies sampled by this
instrument (3500-6000 \AA) includes several age-sensitive absorption
lines, such as the Balmer sequence produced by young stellar
populations, generally used to identify objects where the star formation
activity has been stopped abruptly (post starburst galaxies; e.g.,
\citealt{1997A&A...325.1025P}). BlueMUSE will allow us to reconstruct the 2D
spectrum of dwarf elliptical galaxies in local clusters, where the low
surface brightness of the extended stellar disc makes medium resolution
spectroscopic observations infeasible with other IFU facilities. The
observed spectrum will be fitted with state-of-the-art SED fitting codes
which allow to reconstruct the star formation history of perturbed
systems (e.g., \citealt{2016A&A...596A..11B,2018A&A...614A..57F}). This will
enable an accurate reconstruction of the quenching timescale as a
function of position in a galaxy, enabling, for example, detailed
modeling of outside-in stripping as the galaxy moves into the cluster (see Fig.\,\ref{fig_quenching}). The presence of several age-sensitive Balmer absorption features (as well as the Balmer break) in the BlueMUSE wavelength range makes this instrument far superior than MUSE in the timescale determination for recent quenching events. 

The $R\sim4000$ spectral resolution will also allow to study the kinematical
properties of the stellar component and understand whether dE galaxies
in clusters are mainly fast rotators, as expected whenever their star
formation activity has been quenched after a mild interaction with the
hot and dense intra-cluster medium emitting in X-rays, or rather by more
violent gravitational interactions with other cluster members (e.g.,
\citealt{2011A&A...526A.114T}). Combined with MUSE data at longer wavelengths,
BlueMUSE data will be crucial for detecting and analysing the properties
of the gas stripped during the interaction with the surrounding
environment (as done, for example, in ESO 137-001, \citealt{2014MNRAS.445.4335F}, \citealt{2016MNRAS.455.2028F}) now often observed during deep narrow-band
wide-field imaging surveys of nearby clusters (e.g., \citealt{2010AJ....140.1814Y,2018A&A...615A.114B}). Critically, the blue wavelengths will 
open up the detection of [\oii]$\lambda\lambda3727,3729$ \AA\ and
[\oiii]$\lambda\lambda4363,4959,5007$\AA, the most accurate indicators for density, temperature, and metallicity of the ionized gas.

\textbf{Synergies:} Once removed from the galaxy, the gas stripped from the disc
gets in contact with the hot gas trapped within the potential well of
the cluster. The stripped material, which is mainly in the cold atomic
phase, can thus be heated by different mechanisms (ionisation by the hot
gas of the intra-cluster medium, heat conduction, MHD waves, turbulent mixing, etc) and
change phase, becoming ionised or hot gas, while in other cases it can
collapse into molecular clouds to form new stars. The study of the
stripped material thus requires multifrequency observations: the cold HI
gas is observable at 21 cm (SKA), while the molecular phase through
different CO lines (with ALMA). The hot gas phase is visible in X-rays
(\textit{Athena}), while the ionised gas component is observable with MUSE and
BlueMUSE.

\newpage

\hypertarget{the-distant-universe}{%
\section{The Distant Universe}\label{the-distant-universe}}

  \hypertarget{deep-fields}{%
  \subsection{Deep fields}\label{deep-fields}}

%JB- 
%The current deep field observations performed with MUSE (HDFS \citealt{2015A&A...575A..75B}; UDF \citealt{2017A&A...608A...1B}, CDFS \citealt{2019A&A...624A.141U}, MUDF \citealt{2019MNRAS.485L..62L}) have been transformational, providing an order of magnitude increase in the number of spectroscopic redshifts \citep{2017A&A...606A..12H,2017A&A...608A...2I,2017A&A...608A...3B}, including the discovery of high equivalent width LAE without HST counterparts and extremely faint UV magnitudes (average AB $\sim$ 32) \citep{2017A&A...608A...1B,2018ApJ...865L...1M}. Thanks to this very rich data set, key information has been obtained on the high z Universe: for example the detection and characterisation of extended Lyman-$\alpha$ haloes around individual galaxies \citep{2016A&A...587A..98W,2017A&A...608A...8L}, the evolution of the LAE luminosity function \citep{2017A&A...608A...6D}, the characterisation of the LAE equivalent widths \citep{2017A&A...608A..10H}, the evolution of galaxy merger fraction \citep{2017A&A...608A...9V}, the properties of \ciiI{]} emitters \citep{2017A&A...608A...4M}) , Fe II emitters \citep{2017A&A...608A...7F} and Mg II emitters \citep{2018A&A...617A..62F}, stellar kinematics of $z\sim0.8$ galaxies \citep{2017A&A...608A...5G}, star formation at $z\sim1$ \citep{2018A&A...619A..27B} and the properties of extreme O32 emitters at $z<1$ \citep{2018A&A...618A..40P}.

Observations of deep extragalactic fields performed with MUSE (HDFS, \citealt{2015A&A...575A..75B}; UDF, \citealt{2017A&A...608A...1B}; CDFS, \citealt{2019A&A...624A.141U}; MUDF \citealt{2019MNRAS.485L..62L}) have demonstrated how a wide-field optical IFU is a game-changer for the study of distant galaxies. Two features of this instrument stand out to explain its impact in this domain. First, MUSE provides a {\it comprehensive spectroscopic view of the sky}, i.e. high quality spectra for all sources in the field of view with no prior selection. This approach has produced an order of magnitude increase in the number of spectroscopic redshifts measured in these deep fields \citep{2017A&A...606A..12H,2017A&A...608A...2I,2017A&A...608A...3B}, thereby revealing systematically groups and associations of galaxies that would never have been targeted for spectroscopic follow-up \citep[e.g.][]{2017A&A...608A...9V}.  It has been instrumental building extremely well controled and complete samples of galaxies which have allowed to set constraints e.g., on the properties of \ciii] emitters \citep{2017A&A...608A...4M}), \feii\ emitters \citep{2017A&A...608A...7F}, and \mgii\ emitters \citep{2018A&A...617A..62F}, on stellar kinematics of $z\sim0.8$ galaxies \citep{2017A&A...608A...5G}, star formation at $z\sim1$ \citep{2018A&A...619A..27B}, or on the properties of extreme [\oiii]/[\oii] emitters at $z<1$ \citep{2018A&A...618A..40P}. The second feature, which really makes MUSE a transformative instrument, is its {\it unprecedented and unique sensitivity to emission lines}. This has led to an impressive revision of the census of distant star-forming galaxies, including the discovery of very high equivalent width LAEs without HST counterparts and extremely faint UV magnitudes (average AB $\sim$ 32) \citep{2017A&A...608A...1B,2018ApJ...865L...1M}, the robust detection of numerous LAEs, the assessment of the evolution of their luminosity function \citep{2017A&A...608A...6D} and the characterisation of their equivalent widths \citep{2017A&A...608A..10H}. Perhaps uniquely, this ability to detect line emission has led MUSE to discover and characterize extended Lyman-$\alpha$ haloes around most small galaxies \citep{2016A&A...587A..98W,2017A&A...608A...8L}, or \oii\ gaseous structures in galaxy groups \citep{2018A&A...609A..40E}, and tentative evidence of ouflows seen in \feii\ fluorescent lines \citep{2017A&A...608A...7F}. The uniqueness of MUSE in this domain and the leap forward that it allowed is illustrated also by \citet{2016ApJ...831...39B} who detected and measured bright Lyman-$\alpha$ nebulae around $100\%$ of their targeted quasars when the consensus from deep narrow-band searches expected $\sim 10\%$. \textbf{MUSE has opened a new window on the physics of the CGM, and there exists no rival technique.}

%-JB

\begin{figure}
    \centering
    \includegraphics[width=8cm]{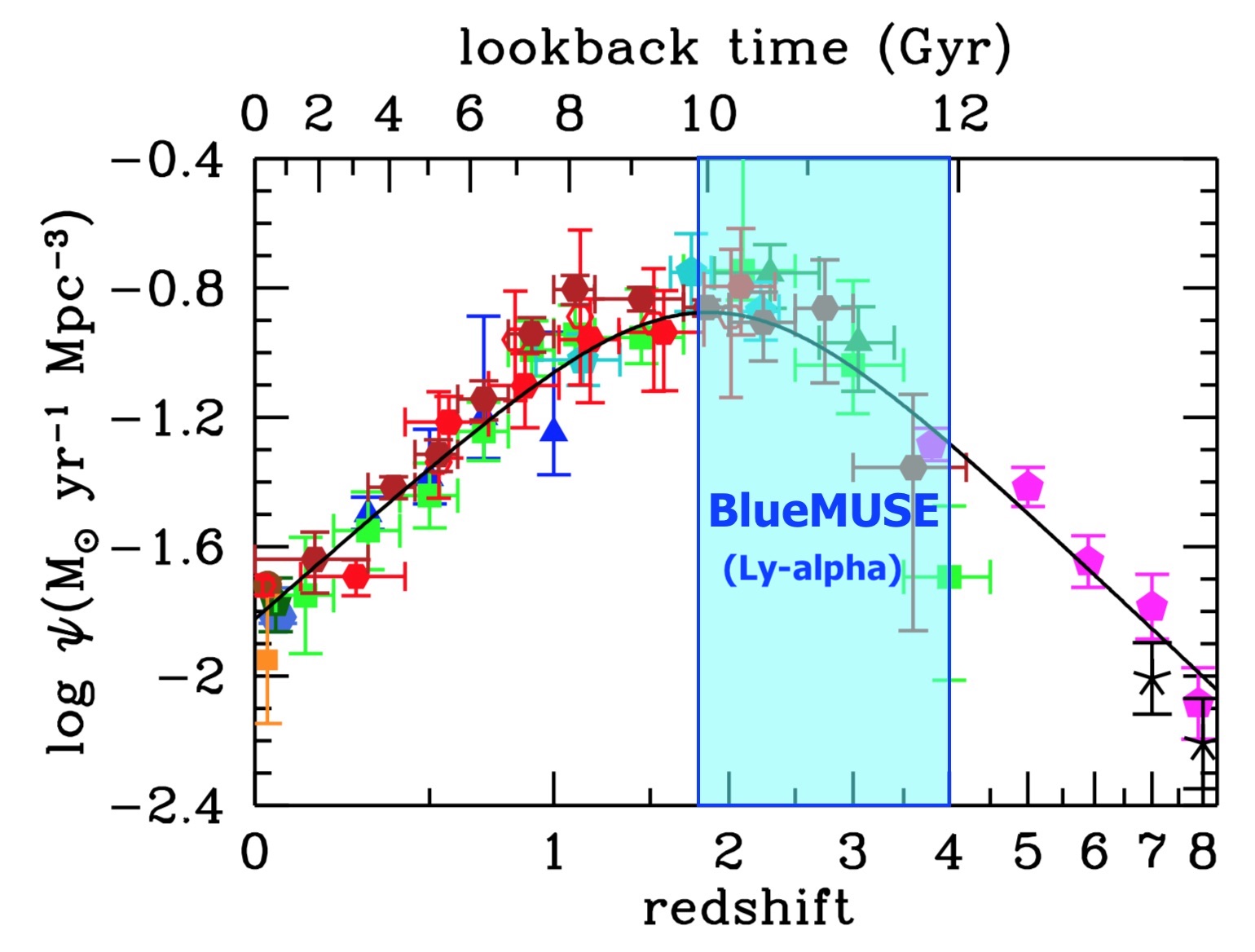}
    \includegraphics[width=3in]{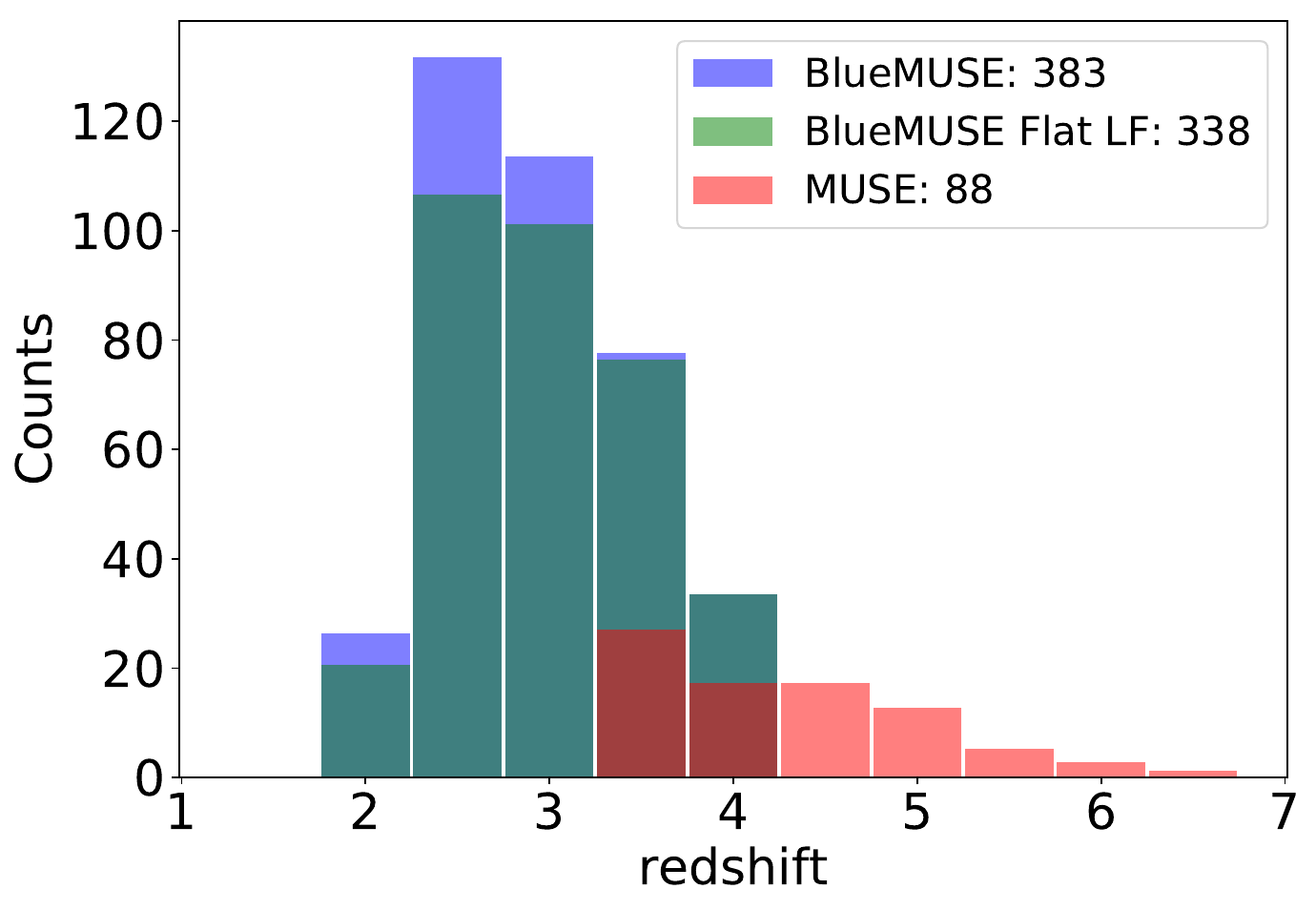}
    \caption{(Left) Evolution of the cosmic SFR (from \citealt{2014ARA&A..52..415M}) and redshift range probed in Lyman-$\alpha$ with BlueMUSE. (Right) Estimated LAE counts as function of redshift for a single BlueMUSE 10~hrs depth exposure (in blue) and a 5$\sigma$ detection limit. In green the LAE number counts is shown when the  LAE luminosity function is assumed to flatten below $10^{41.2}$ erg\,s$^{-1}$.
    MUSE counts (in red) for the same depth and SNR limit, is given for comparison.}
    \label{fig:lae_counts}
\end{figure}

%JB-
%With its blue spectral coverage and increased field of view, BlueMUSE deep fields will offer a complementary and even more powerful capability: the current MUSE redshift desert (z={[}1.5-3{]}, corresponding to the redshift of {[}O II{]} emitters at the red end and of Lyman-$\alpha$ emitters at the blue end) will be partially filled up to $z\sim2$ and the large population of {[}O II{]} emitters will be probed up to z=0. With BlueMUSE, LAEs are available in the {[}$2-4${]} redshift range, probing the peak of the star formation history at $z=2$ (Fig.\,\ref{fig:cgm_simus}) as well as the peak of AGN number counts \citep{2005MNRAS.360..839R}. The expected LAE counts for a 10 hours depth BlueMUSE single field observations (Fig.\,\ref{fig:lae_counts}) is 400, that is 4.5 times larger than MUSE observations at the same depth in the UDF (88 LAEs, \citealt{2017A&A...608A...1B}). This large gain results from the combination of field of view (factor two), the increased free spectral range  of BlueMUSE with respect to the OH impacted spectral range of MUSE in the red and another factor due to the larger volume of Universe probed at these redshifts plus a slight evolution of the faint-end LAE luminosity function at $z\sim2-3$ (based on models of \citealt{2016MNRAS.455.3436G}). The larger field of view will also be very beneficial in the IGM topology studies (see next Section) as well as clustering studies (e.g. \citealt{2017MNRAS.471.3186D}). The capability of BlueMUSE to observe a very large sample of LAEs at high spectral resolution and high SNR is unique.

With its blue spectral coverage and increased field of view, BlueMUSE will be a complementary and even more powerful facility. The current MUSE redshift desert ($z=[1.5-3]$, corresponding to the redshift of [\oii] emitters at the red end and of LAEs at the blue end) will be largely filled down to $z=1.88$, and the large population of [\oii] emitters will be probed down to $z=0$.
BlueMUSE will observe LAEs and their CGM in the
{[}1.9-4{]} redshift range. Their expected counts for a 10-hours
depth single pointing is 380, that is 4.3 times larger than
MUSE observations at the same depth in the UDF (88 LAEs,
\citealt{2017A&A...608A...1B}, see right panel of
Fig.\,\ref{fig:lae_counts}). This large gain results from the
combination of the increased field of view, the increased free
spectral range of BlueMUSE with respect to OH lines, the smaller luminosity distance which allows 
to probe much fainter galaxies and a small
increase of the volume of Universe probed at these redshifts. 
The semi-analytical model used for this prediction \citep{2016MNRAS.455.3436G} gives an accurate description of the LAE luminosity function over a large range of luminosity. However the faint end evolution is currently not constrained by the observations. We nevertheless show that, even when assuming a luminosity function flat below $10^{41.2}$ erg\,s$^{-1}$, the number of predicted LAE counts stays high (e.g., 338 galaxies with respect to 88 for the current 10~hrs depth MUSE observation).

Thanks
to a surface brightness dimming reduced by a factor $\sim 4$ at
$z\sim 2-3$ in comparison to $z\sim 3-5$ with MUSE, we expect a similar gain
in the ability of BlueMUSE to detect Lyman-$\alpha$ emission from diffuse
gas around galaxies, which may allow us to detect emission from gas
all the way out into the intergalactic medium (IGM,
\S\ref{imaging-the-intergalactic-medium-at-z23}). More than this significantly improved efficiency, the redshift range where BlueMUSE will see Lyman-$\alpha$ covers a key period of cosmic history, when the cosmic star formation rate turns around and starts to decrease with time (left panel of Fig.\,\ref{fig:lae_counts}). By allowing us to build a complete and homogeneous census of star-forming galaxies throughout this epoch, and to survey the evolution of their circum-galactic medium, BlueMUSE will show us this major transition as it happens and help us understand the emergence of strong cosmological accretion shocks, the conditions for survival of cold streams, and the effect of galactic winds (\S\ref{key-science-case-gas-flows-around-and-between-galaxies}, \ref{the-emergence-of-the-first-galaxy-clusters}). 

%-JB
%\includegraphics[width=3.57778in,height=2.25069in]{}

\begin{subsec}[ht]
\begin{tcolorbox}[colback=green!5!white,colframe=green!75!black,title=Why is a high throughput in the blue critical for BlueMUSE?]

With an average end-to-end throughput $\sim30\%$ (including telescope and atmosphere), we aim for BlueMUSE to become the most efficient instrument in the blue on an 8m-class telescope.
\begin{itemize}
\item
  \begin{quote}
 High throughput is very important for the overall survey speed, as already seen with the MUSE instrument. 
  \end{quote}
\item
  \begin{quote}
  At blue wavelengths, the instrument throughput has to maintain a high value to compensate for the decrease in atmospheric transmission, in particular at $\lambda<400$ nm (Fig.\,\ref{throughput}).
  \end{quote}
\item
  \begin{quote}
  Given the cost and pressure of 8m-class telescope time, high throughput in practice makes the difference between feasible and non-feasible observations.
  \end{quote}
\end{itemize}
\end{tcolorbox}

\end{subsec}

  \hypertarget{key-science-case-gas-flows-around-and-between-galaxies}{%
  \subsection{\texorpdfstring{\textbf{Key science case:} Gas flows
  around and between galaxies
  }{Key science case: Gas flows around and between galaxies }}\label{key-science-case-gas-flows-around-and-between-galaxies}}

\begin{tcolorbox}[colback=blue!5!white,colframe=blue!75!black,title=Science Goals]
\begin{itemize}
\item
  How do galaxies exchange baryons with the Intergalactic Medium?

  %\begin{itemize}
  %\item
  %  How is the gas around galaxies distributed, spatially and
  %   kinematically?
  %\item
  %  What is its temperature and density distribution?
  %\item
  %  How much of the gas accreted into a halo makes it to the galaxy?
  %\item
  %  What is the spatial extent and volume filling factor of metals
  %  around galaxies?
  %\end{itemize}
\item
  How do the CGM properties evolve when reaching the peak of galaxy
  star-formation?

\item  What is the spatial extent and how is the CGM gas around galaxies distributed, spatially and    kinematically? 
\end{itemize}
\end{tcolorbox}

\begin{figure}
      \centering
      \includegraphics[width=10cm]{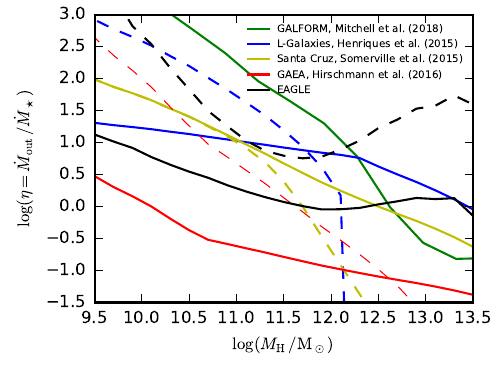}
      \caption{
      The efficiency of galactic outflows, expressed as the dimensionless ratio of outflow rate per unit star formation rate, and plotted as a function dark matter halo mass. Different colours show different contemporary theoretical models, with solid lines indicating how much gas is removed from the interstellar medium of galaxies, and dashed lines indicating how much gas is then ejected through the halo virial radius. Discrepancies spanning multiple orders of magnitude exist between different models, despite the fact that each model is able to reproduce the observed stellar properties of galaxies. Observations of gas flows around galaxies are vital to constrain the otherwise degenerate picture represented by this figure.
      }
      \label{fig:cgm_simus}
  \end{figure}

Even in the era of precision cosmology, and despite the impressive success of recent large-scale cosmological simulations at reproducing the bulk of galaxy properties (stellar mass function, clustering, ...) across cosmic time, galaxy formation is still far from understood. In particular, the fundamental questions about how galaxies acquire gas from the intergalactic medium, and how they regulate their growth through galactic winds or other preemptive processes, are mostly unconstrained from observations or theory. Figure \ref{fig:cgm_simus} (from Mitchell et al., in prep.) illustrates this by showing predictions of outflow rates from different state-of-the-art simulations, all extremely well calibrated to reproduce the stellar properties of galaxies. These predictions differ by orders of magnitudes! Clearly, constraints on the stellar properties of galaxies alone are degenerate and cannot help us decide which physical scenario among these simulations, if any, corresponds to reality. From Fig.\,\ref{fig:cgm_simus}, it is also clear that observing the Circum-Galactic Medium (CGM) and constraining the flows of gas that traverse it would be a radically new constraint on galaxy formation. 

MUSE has spectacularly demonstrated its ability to observe the CGM through its emission in the Lyman-$\alpha$ line of hydrogen (e.g.,\citealt{2018Natur.562..229W}). Observations with BlueMUSE in the redshift range $2<z<3$ will be no less spectacular, but more importantly, they will be key in constraining galaxy formation because they cover the peak of cosmic star formation (Fig.\,\ref{fig:lae_counts}, left). Among the factors that drive this turn-over in the cosmic SFR, we may expect a change in the form of accretion flows and their interactions with galactic winds, which marks the beginning of the transition from the early Universe -- where gas flows cold and collimated onto galaxies--, and the late Universe -- where star formation is sustained by cooling-regulated accretion from hot coronae. Observing the evolution of the CGM through this epoch will be a key in discriminating between all existing theories.

In the subsections below, we show how BlueMUSE, may be used to set unprecedented constraints on (1) the state of the diffuse intergalactic medium, (2) the evolution of the neutral gas content of the CGM from $z\sim 4$ to $2$, (3) the metal content of the CGM as a tracer of galactic winds.

\nocite{2018MNRAS.474.4279M}
\nocite{2015MNRAS.451.2663H}
\nocite{2015ARA&A..53...51S}
\nocite{2016MNRAS.461.1760H}

\hypertarget{imaging-the-intergalactic-medium-at-z23}{
\subsubsection{Imaging the Intergalactic Medium at
\texorpdfstring{$z\sim2-3$}{z~2-3}\label{imaging-the-intergalactic-medium-at-z23}}}

\begin{figure}
\centering
\includegraphics[scale=0.7]{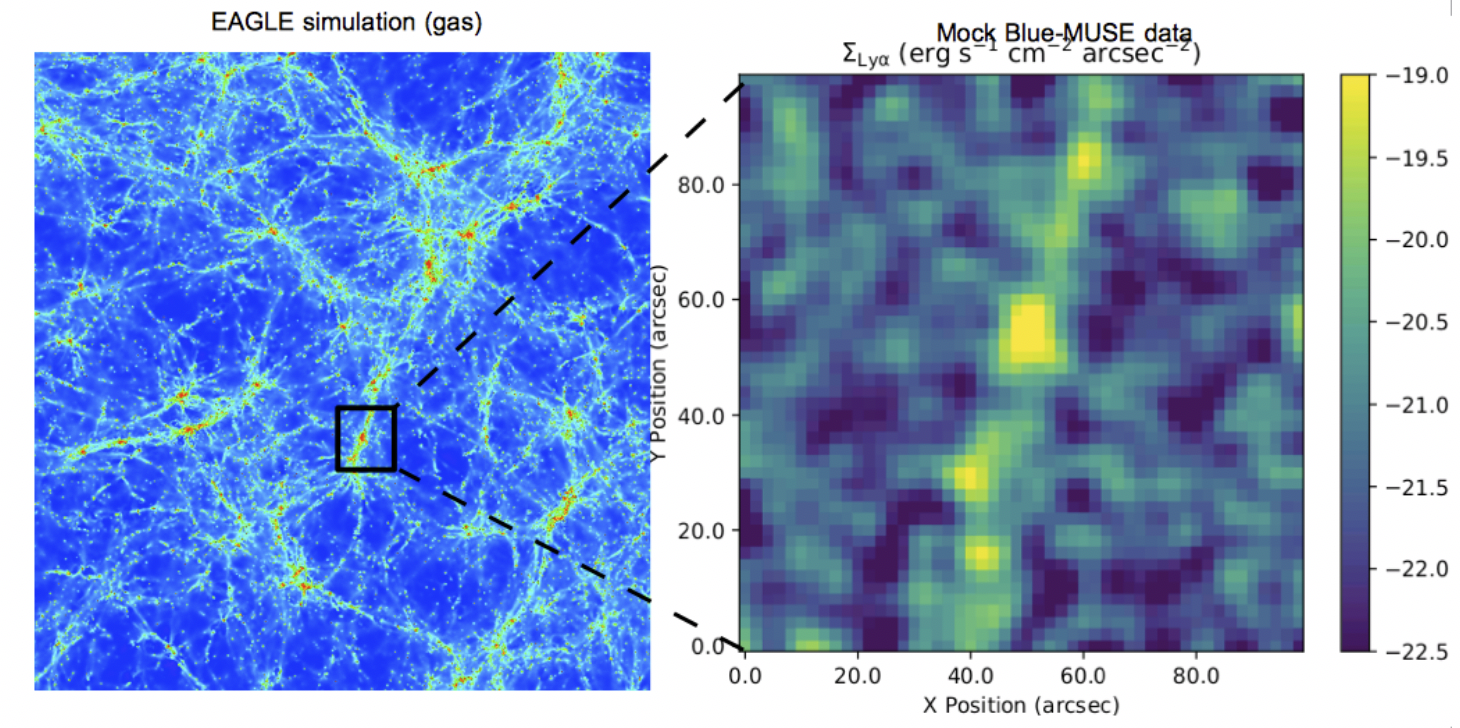}
\caption{\emph{Left}: Gas density from a snapshot of the
\textsc{eagle} simulation \citep{2015MNRAS.448.2687J} at $z\sim 2$, showing how
matter is distributed in a net of filaments connecting galaxy haloes, which is a distinctive prediction of the current $\Lambda$CDM cosmological
model. The simulation box is 25 Mpc on a side, and the region shown is 5 Mpc on a side. \emph{Right:} Mock Lyman-$\alpha$ observations of the region shown in the left-hand panel achievable with moderate integrations ($\sim 20$~hrs), demonstrating the fluorescent emission from filament connection haloes. The filament is illuminated by the UV background and local sources, and emits in fluorescent Lyman-$\alpha$. BlueMUSE will open a new discovery space,  including the prospect of imaging the diffuse intergalactic and circum-galactic medium that, so far, has eluded direct observations in emission.}\label{fig:uvb}
\end{figure}

\textbf{Aim:} A fundamental prediction of $\Lambda$CDM is that galaxies form in
overdensities that are connected by a network of filaments, which
compose the cosmic web (Fig.\,\ref{fig:uvb}). Obtaining direct images of this cosmic web at $z \sim 1.8-3$, the epoch when galaxy formation was at its peak, would represent a major breakthrough for modern cosmology, and
a goal that will be within reach of BlueMUSE with deep (tens of hours) integrations. %\citep{2019arXiv190506954W}. 
BlueMUSE will be able to mosaic the IGM over several continuous areas, a step much beyond the current attempts in single deep
pointings (30 to 100hr) with MUSE (ESO PID 1100.A-0528 PI: Fumagalli; ESO PID 1101.A-0127 PI: Bacon).

With BlueMUSE, the redshift range $z\sim 2-3$ becomes accessible where the cosmological surface brightness dimming is a factor of four lower in average compared to $z\sim3-5$ for MUSE. For a fixed signal-to-noise, this will make
BlueMUSE $\sim 16\times$ faster at mapping the IGM compared to the current MUSE (with a further gain in area covered by BlueMUSE compared to MUSE). This will mean that BlueMUSE will be able to construct large
mosaics of IGM emission in reasonable exposure times, rather than a
single 100~hrs integration at $z \sim 3-4$ with MUSE.

Furthermore, by exploiting the wavelength coverage and field of view of BlueMUSE, several independent experiments can be performed in order to image the cosmic web and measure the amplitude of the meta-galactic UV
background at $z\sim2-3$:

(1) Ionised by the UV-background, the cosmic web is predicted to radiate
in Lyman-$\alpha$, with a maximal surface brightness of $\gtrsim  3\,10^{-20} \times (2 \Gamma_{\hi} /
10^{-12}$ erg\,s$^{-1}$\,cm$^{-2}$\,arcsec$^{-2}$) 
in optically-thick gas (e.g., \citealt{1996ApJ...468..462G}) when assuming a conservative lower limit for the photoionization rate
($\Gamma_{\hi}=6\,10^{-13}~\rm s^{-1}$;
\citealt{2009ApJ...703.1416F}). By targeting known regions which contain
over-densities (e.g., galaxy pairs or quasars), BlueMUSE will have both the
area coverage and sensitivity to measure the emission line luminosity in
the large scale structure, directly imaging the cosmic web at
$z\sim 1-3$. Increased emission from hosting quasars is
expected to boost the Lyman-$\alpha$ emission of the cosmic web at $2-3$ Mpc distance
by factors $4-5$ (compared to the ``field'', \citealt{2005ApJ...628...61C}).

(2) Combining observations at the position of continuum-detected
galaxies between $z\sim2-3$ will enhance the cosmic web
where the high density peaks form. In turn, a detection of the cosmic web will yield a measurement of the intensity of the UV-background, a fundamental quantity that encodes the production and escape of ionising
photons from galaxies and AGN, but that is currently constrained only by the absorption properties of the Lyman-$\alpha$ forest and by model-dependent
radiative transfer calculations (e.g., \citealt{2012ApJ...746..125H, 2019MNRAS.484.4174K, 2019arXiv190308657F}).

\newpage

(3) Finally, BlueMUSE will map \hi\ absorption against the continuum of individual galaxies at $z\sim2-3$ with $m\lesssim 27$, which will be detected in 30~hrs with $S/N\gtrsim 1.5$ in spectral bins of $\sim 1$ \AA . This data quality is sufficient to reconstruct a tomographic map of the cosmic web at this redshift on
scales of $\sim$ 200 kpc, complementing measurements in emission.

\hypertarget{b.-the-circumgalactic-medium-of-star-forming-galaxies-in-emission}{%
\subsubsection{The Circumgalactic Medium of star-forming galaxies with Lyman-alpha emission}\label{b.-the-circumgalactic-medium-of-star-forming-galaxies-in-emission}}

Narrowband Lyman-$\alpha$ images have suggested for many years that Lyman-$\alpha$
emitters are ``fuzzy'', and image stacks have revealed significantly
extended emission \citep{Ma12,Mo14,X17}. Thanks to MUSE we know now that Lyman-$\alpha$ haloes are
ubiquitous around even low-mass galaxies at redshifts $z>3$
\citep{2016A&A...587A..98W,2017A&A...608A...8L}, with halo scale lengths
typically 10$\times$ larger than the UV sizes of the hosting galaxies. Stacking some of the deepest MUSE data reveals that these haloes extend out to the virial radius, matching the incidence rates of high column density \hi\ absorbers \citep{2018Natur.562..229W}. This extended Lyman-$\alpha$ emission thus
holds unique clues about the spatial distribution and potentially also
kinematics of circumgalactic hydrogen, but due to the resonant nature of
the Lyman-$\alpha$ transition it is a huge challenge to decode this
information. Clearly, spatially resolved spectroscopy coupled with numerical simulations and theoretical models will be crucial in
this endeavour. Yet, here MUSE is reaching its fundamental limits. Even
for the brightest known Lyman-$\alpha$ haloes at $z\gtrsim 3$ there is
barely enough signal to break up the Lyman-$\alpha$ emission into several
independent spatial elements, and the outer halo regions will remain
inaccessible to spectroscopic studies. A factor of $\gtrsim 4$ gain
in cosmological surface brightness, coming with the move from $z\gtrsim 3$ to
$z\sim2$ as well as the improved spectral resolution using BlueMUSE, will however change everything in
this game. BlueMUSE will allow us to investigate the motions of gas in
the CGM of galaxies and thus obtain crucial constraints on the balance
of inflows and outflows.

Due to the large field of view of BlueMUSE, there will be several
galaxies in each observed field bright enough to be used as background
sources for absorption line spectroscopy. Thus it will be possible to
infer \hi\ or metal line column densities for a significant number of
foreground Lyman-$\alpha$ emitters and connect this information to the detected
extended emission \citep[e.g.]{2017MNRAS.471.3686F}. While this experiment is in principle also
conceivable with MUSE, the almost total dearth of sufficiently bright
background galaxies at $z>3$ makes it practically impossible.
Again, the move to $z\sim2$ afforded by BlueMUSE will imply
an almost complete change. 

\hypertarget{c.-probing-the-cgm-with-metals}{%
\subsubsection{Tomography of the Circumgalactic Medium with metal absorption lines}\label{c.-probing-the-cgm-with-metals}}

\textbf{Aim:} Using metal absorptions lines (\mgii~$\lambda$2796,2803 \AA\ ;
\feii~$\lambda$2600 \AA) in \textbf{multiple background} sources 
%(as in Figure~\ref{fig:cgm})
to measure the full spatial extent and kinematics of the circum-galactic medium (CGM)
around \textbf{individual} star-forming galaxies ([\oii] emitters) at intermediate
redshifts ($z=0.4-1.0$).

% Figure commented - on hold -
%\begin{figure}
%\begin{minipage}{12cm}
%\centering
%\includegraphics[width=15cm]{figures/CGM_tomo.png} 
%\end{minipage}

%\begin{minipage}{4cm}
%\end{minipage}
%   \caption{{\it Left:} MUSE has been a workhorse to study the CGM with metal lines (e.g. Mg II 2796,2803) single background sources (quasars) 
% \citealt[e.g.][]{2016ApJ...833...39S,2017MNRAS.471.3686F,2019MNRAS.485.1961Z,2019MNRAS.485.1595P}
%   {\it Right:} Thanks to the blue-wavelength capabilities of Blue-MUSE, it will be possible to perform tomographic studies of the CGM around single star-forming galaxies. H-alpha emission of the wind from M82 is shown in the inset.}
%    \label{fig:cgm}
%\end{figure}

Up to now, the state-of-the-art CGM studies are limited to either single galaxy-quasar
pairs \citealt{2013Sci...341...50B,2018MNRAS.474..254R,2019MNRAS.485.1595P} on small samples or to stacking techniques with large samples of pairs (e.g., \citealt{2018ApJ...866...36L,2014ApJ...794..130B}).
Examples in individual pairs \citep{2018ApJ...868..142R} show the promise of extending this technique to the use of background galaxies as probes of foreground ones.
However, in order to understand the exact nature of gas flows around galaxies, the CGM
kinematics can \textbf{only be mapped using multiple} (5 to 10) background sources (quasars or galaxies). This is fortunately within reach with background galaxies, but this is not feasible with MUSE for the following reasons: (a) the spectral range (480-930 nm) does not allow
to study UV absorption lines below $z=0.85$, whereas $z=0.6-0.7$ is a sweet
spot for group selections as background galaxies and (b) the spectral resolution of MUSE at 5000 \AA\  is too low (R=1800 or 160 km\,s$^{-1}$; see Fig.\,\ref{fig:specres}). With BlueMUSE, we will reach many more extended background galaxies to spatially resolve absorption and address metal mixing on small scales (e.g., \citealt{2018MNRAS.479L..50P}).

With BlueMUSE, it will become possible to
study the spatial and kinematics properties extent of the CGM around {\it individual} galaxies at intermediate redshifts
$z=0.4-0.8$. This redshift range is particularly well suited as this
corresponds to the regime where the source density of background
galaxies is now becoming $>10$ arcmin$^{-2}$ to
I$=24$ mag.

\medskip\par

{\bf Feasibility:}

$\bullet$ Background source density: Down to 24(25) AB mag, there are typically about 6(14) ELGs per arcmin$^2$ at $z=0.5-1.5$ as in blank fields such as the UDF-mosaic field (\citealt{2017A&A...608A...1B}; at 10~hrs depth).
    The number of background galaxies can be suitably increased by targeting groups or clusters.
    
$\bullet$ Continuum sensitivity: 
    Down to 24(25) AB mag (V),  the SNR in the continuum (for the background galaxies) should reach SNR 5-10 in order to be sensitive to absorption lines (e.g., \feii, \mgii) down to equivalent widths of 0.5(1) \AA.  Using the preliminary ETC, the SNR in the continuum at 4500 \AA\ is $\approx 10$(5) for extended sources with $V$-band magnitudes of 24 (25)$m_{\rm AB}$, respectively. The SNR is calculated over a region of 0.8 arcsec$^2$, with a seeing of 0.8 arcsec and assuming a constant spectrum in $f_\lambda$.

\medskip\par
\textbf{Uniqueness of BlueMUSE:} 

This CGM tomography science case is currently not feasible with MUSE for several reasons. Indeed, this science case requires
\begin{itemize}
\item[(1)]  the
wavelength range ($< 5000$ \AA) in order to observe \feii\ and \mgii\ at redshifts $\lesssim0.8$ given that groups are predominantly found at $\lesssim0.7$ \citep{2012ApJ...747...15K};
\item[(2)] a spectral resolution of 60-80 km\,s$^{-1}$ in order to measure the absorption relative kinematic shifts at a precision
better than 25 km\,s$^{-1}$. MUSE only provides a low resolution (R$\sim$1800, 160 km\,s$^{-1}$) at
5000 \AA\ ;
\item[(3)] a field of view $>1$ arcmin$^2$, as the CGM is known to
extend to at least 200 kpc in radius. 
\end{itemize}

\textbf{Synergies:} While the primary science goal requires BlueMUSE,
short ($<1$~hr) MUSE snapshots for [\oii] is envisioned to
measure systemic redshifts at $z>0.6$. Large surveys like
eBOSS, DESI, 4MOST, LSST will be key for field selection, providing ever increasingly large samples
of background sources. SKA might be
able to map the cold neutral \hi\ component that we will probe in
absorption.

 \begin{figure}
      \centering
      \includegraphics[width=16cm]{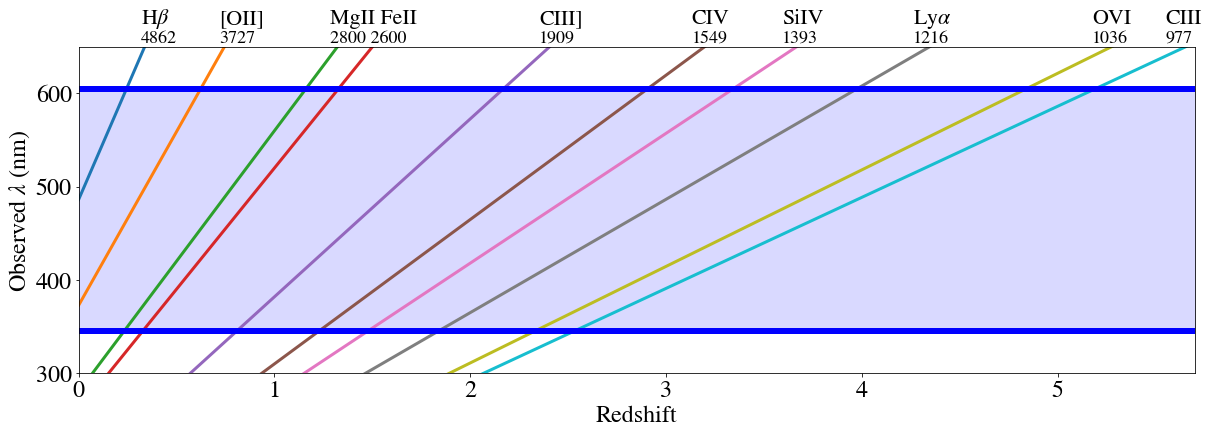}
      \caption{Main nebular lines available to BlueMUSE as a function of redshift. Rest-frame wavelengths are provided for each line in Angstroms. In particular metal lines in the rest-frame UV (\mgii, \feii, \ovi) will help probe the CGM in emission.}
      \label{fig:cgmlines}
  \end{figure}
 
 \hypertarget{c.-probing-the-cgm-in-emission-with-metals}{%
\subsubsection{Probing the Circumgalactic Medium in emission with metal  lines}\label{c.-probing-the-cgm-in-emission-with-metals}}
 
With BlueMUSE, we will have access a plethora of rest-frame UV lines in the redshift range $\sim 0-2.5$ (Fig.\,\ref{fig:cgmlines}), opening up a completely new way to map the CGM directly, also complementing spectroscopy in absorption. Metal emission lines are significantly fainter than Lyman-$\alpha$, with characteristic surface brightness predicted around $10^{-20}-10^{-21}$ erg\,s$^{-1}$\,cm$^{-2}$\,arcsec$^{-2}$ at $z\sim 2$ (\citealt{2012MNRAS.419..780B}, Augustin et al. submitted). Due to the cosmological surface brightness dimming, detection at $z>3$ with MUSE is almost prohibitive and thus limited to only extreme environments, making the case for low-redshift/shorter-wavelength observations obvious. 

The detection of such low surface brightness signal will require deep exposures, but is expected to yield a tremendous return. Indeed, while Lyman-$\alpha$ traces the bulk of the gas mass, metal transitions become key traces to map the spatial extent of the multiphase CGM. Indeed, the brightest surface brightness of each ion arises from well-defined regions of the density-temperature phase diagram, where the gas temperature approaches the peak of the ion emissivity. Thus, low and moderate ions will yield maps of the relatively cool ($T\lesssim 10^5$~K) CGM, with ions at higher ionisation states tracing the warm-hot medium at $T\gtrsim 10^5~\rm K$. Combined together, these traces will enable a complete reconstruction of the multiple phases of the CGM.

Indeed, the handful of examples of metal emission lines detected so far in the rest-frame UV (\mgii, \feii, \ovi)  \citep{2011ApJ...728...55R,2016ApJ...828...49H,2017A&A...605A.118F,2017A&A...608A...7F} already show the promise of this type of studies to infer the spatial extent and morphology of galactic outflows from distant galaxies. Moreover, the combination of ions traced in absorption and emission enables (with some assumption, e.g., on the excitation mechanism of the gas) the derivation of physical parameters (e.g., size along the line of sight, volume density and ion mass) that are weakly or unconstrained by spectroscopy in absorption alone.

While MUSE has greatly advanced our knowledge of extended Lyman-$\alpha$ emitting gas around AGN at $z>3$ (e.g., \citealt{2016ApJ...831...39B}), an important limitation has been the general rarity of detections of extended emission from other UV lines. BlueMUSE will provide a quantum leap in our ability to study Lyman-$\alpha$ halos and the CGM around AGN, by proving the wavelength coverage and sensitivity to detect the faint UV emission lines that are crucial to characterise the kinematic, ionization and chemical enrichment properties of AGN-photoionized Lyman-$\alpha$ halos and CGM, in the redshift range $1.9<z<3$. Non-resonant lines such as \heii\ 1640 and \ciii] will allow more reliable gas kinematics to be derived, free from the potential complications of Lyman-$\alpha$ line transfer effects, and the inclusion of metal lines such as \nv~$\lambda$1239,1243 and \civ~$\lambda$1548,1551 will allow the ionization properties and chemical enrichment history of the gas to be derived (e.g., \citealt{2003MNRAS.346..273V,2019A&A...621A..10H}). Mapped in two spatial dimensions thanks to the IFU technique, this information will afford detailed study of AGN feedback, the dispersion of metals via outflows, and accretion of gas in/around massive galaxies near the peak in the star formation and AGN activity histories.

%%%MF not sure how to fit this in

%\begin{itemize}
%\item
%  Competition with KCWI: \citet{2018ApJ...862L..10E} recently presented a first
%  case study at $z=2.3$ - obtained with KCWI at Keck - of how the Lyman-$\alpha$
%  spectrum of an extremely bright emitter changes over the extent of the
%  halo. However, KCWI does not have the multiplexing capability of
%  BlueMUSE and can only do one object at a time, whereas BlueMUSE could
%  provide similar or probably even better data (given the known
%  limitations of KCWI), but for several tens of objects at the same
%  time.
%  \item H$\alpha$, which is essential as an indicator of star formation and ionising
 % photon production rates, is visible from the ground until redshifted
 % out of the \emph{K}-band at $z=2.7$. For objects currently studied with
 % MUSE we basically have to wait (and compete) for JWST.

%\end{itemize}

  \hypertarget{lyman-continuum-emitters}{%
  \subsection{Lyman Continuum Emitters}\label{lyman-continuum-emitters}}

\begin{tcolorbox}[colback=blue!5!white,colframe=blue!75!black,title=Science Goals]
\begin{itemize}
\item
  Collect a statistical sample of Lyman Continuum (LyC) Emitters to investigate the analogues of the sources of cosmic reionisation.
\item
  Test observational diagnostics for LyC leakage based on the Lyman-$\alpha$ properties.
\item
  Constrain the physical properties of typical LyC Emitters.
\end{itemize}
\end{tcolorbox}

Cosmic reionisation corresponds to the period in the history of the
Universe during which the predominantly neutral intergalactic medium was
ionized by the emergence of the first luminous sources. 
%Young stars in primeval galaxies are thought to be the main sources of reionisation. 
Although star-forming galaxies in the early Universe are thought to be the main sources of reionisation, the nature of these objects remains so far totally unknown due to the increasing opacity of the intergalactic medium with redshift which renders direct LyC detections impossible at $z\gtrsim$5.
%The ionizing radiation, called Lyman continuum (LyC, $\lambda$ \textless{} 912 \AA\ ), that they produce can escape their interstellar medium: the escape fraction of LyC photons from galaxies is one of the main unknowns of reionisation studies.
%The increasing opacity of the intergalactic medium with redshift renders direct LyC detections impossible during reionisation. 
However, LyC Emitters can be observed at lower redshifts, even though these observations have proven difficult: so far, only 14 detections have been reported in the low-$z$ Universe
($<0.4$, \citealt{2006A&A...448..513B,2013A&A...553A.106L,2014Sci...346..216B,2016Natur.529..178I,2016MNRAS.461.3683I,2018MNRAS.474.4514I,2018MNRAS.478.4851I}), and a few at high redshift (\citealt{2016A&A...585A..51D,2016ApJ...825...41V,2016ApJ...826L..24S,2017ApJ...837L..12B,2018MNRAS.476L..15V,2018ApJ...869..123S}).
%Unlike previous targeted observations, BlueMUSE will carry out blind surveys and collect a statistical sample of LyC emitters from $z\sim3$ to $5$. However, in this redshift range, the IGM transmission is decreasing rapidly ($\sim0.4$ for the mean value at $z\sim3.5$, \citealt{2014MNRAS.442.1805I}, see Fig.~\ref{fig:lycont} below).

\begin{figure}
    \begin{minipage}{7cm}
    \includegraphics[width=\textwidth]{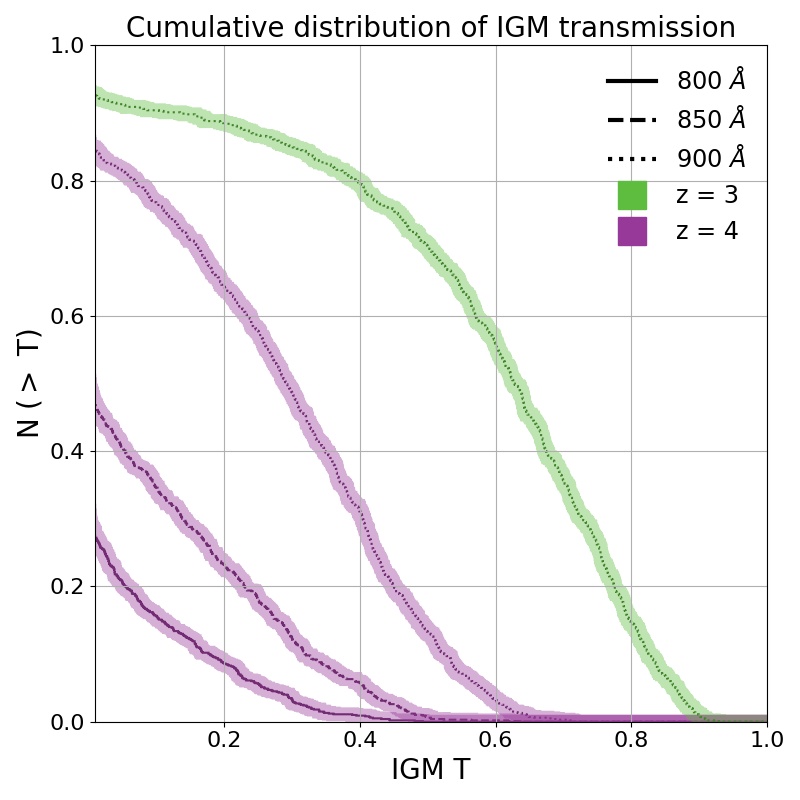}
    \end{minipage}
    \begin{minipage}{9cm}
   \includegraphics[width=\textwidth]{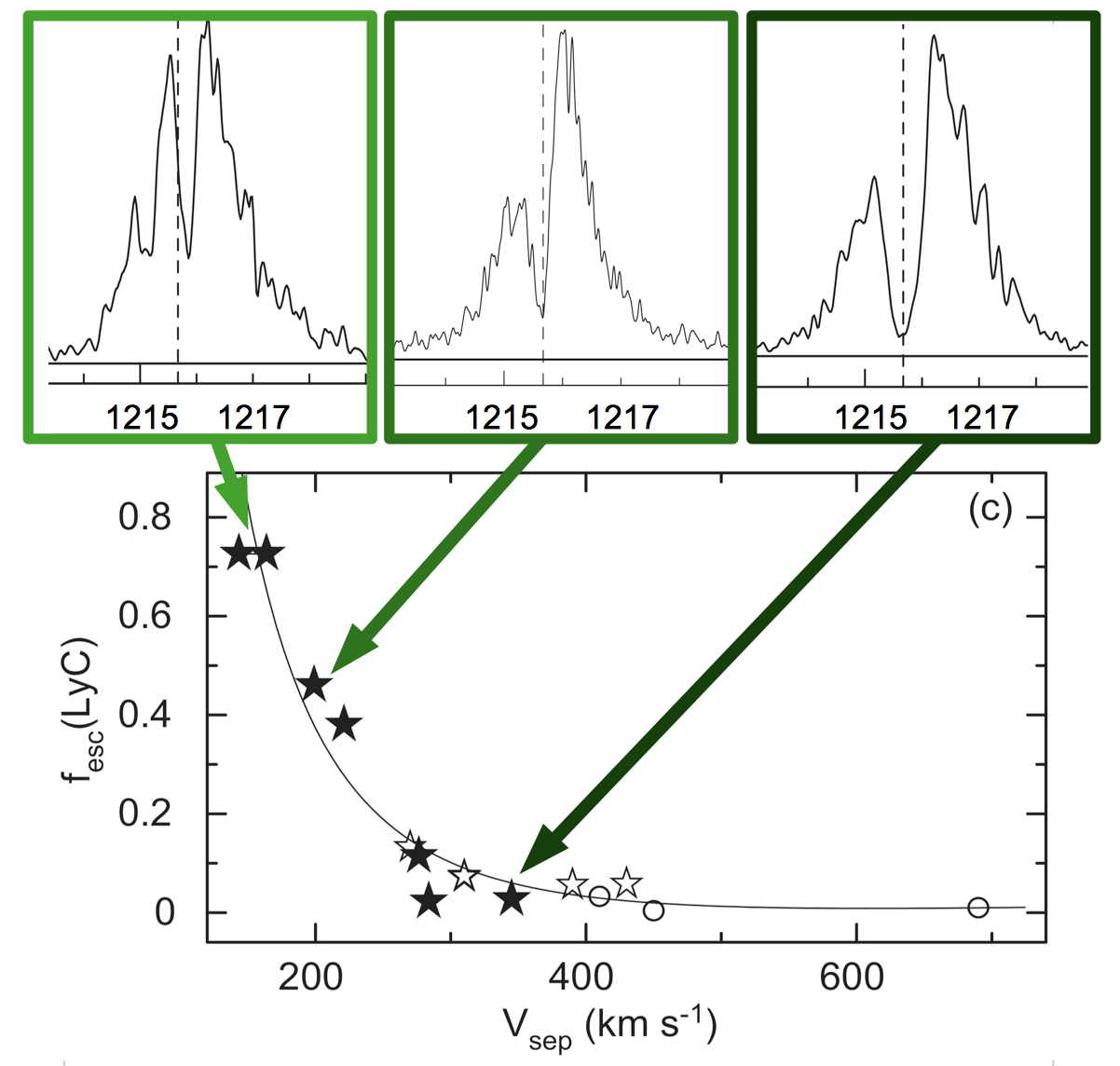}
    \end{minipage}
    \caption{Left: Distribution of IGM transmissions for different redshifts and rest wavelengths. At $z\sim3$ (green curve), $80\%$ of the lines of sight have a transmission higher than $40\%$; at $z\sim4$ (purple curves), the IGM is not completely opaque to ionizing radiation: even at $800$\AA\ rest-frame, $10\%$ of the lines of sight have a transmission higher than $20\%$. Upper right: Lyman-$\alpha$ spectra of LyC emitters of the local Universe \citep{2018MNRAS.474.4514I}. Lower right: The peak separation correlates with the escape fraction of LyC radiation in local leakers \citep{2018MNRAS.474.4514I}. }
    \label{fig:lycont}
\end{figure}

BlueMUSE will be able to directly probe the Lyman continuum from sources at $z\sim$ 3 to 5 and, unlike previous targeted observations, it will allow us to collect a statistical sample of LyC emitters as part of blind surveys. Nevertheless, due to the IGM opacity increasing rapidly towards higher redshifts, we expect most detections to fall in the redshift range $z=3-4$ where Lyman-$\alpha$ can be used to assess the spectroscopic redshift, and where the IGM transmission ($T_{\rm IGM}$) remains reasonably high: at $z\sim3$, $80\%$ of the lines of sights have a transmission higher than $40\%$, as shown on Fig.\,\ref{fig:lycont}. 
As demonstrated by MUSE studies, BlueMUSE will be particularly efficient at detecting the faintest population of $z\sim3$ galaxies \citep{2017A&A...608A...6D,2017A&A...608A..10H}. These low mass objects are plausibly strong leakers according to previous observations \citep{2018MNRAS.474.4514I} and simulations \citep{2014MNRAS.442.2560W, 2017MNRAS.470..224T,2019MNRAS.486.2215K}, and may correspond to the analogues of the sources of reionisation. 

Indirect methods are the only probes of LyC leakage in the distant Universe. Lyman-$\alpha$ escape from galaxies is expected to correlate with LyC escape \citep{2015A&A...578A...7V,2016ApJ...828...71D}, and most recent observations of the Lyman-$\alpha$ properties of LyC leaking galaxies at low 
\citep{2017A&A...597A..13V,2018MNRAS.474.4514I} and high redshift
\citep{2017A&A...601A..73M,2018A&A...614A..11M,2018ApJ...869..123S}  
have confirmed the theoretical predictions. In particular, the peak separation of the double-peaked Lyman-$\alpha$ profiles nicely anti-correlates with the LyC
escape fraction (see Fig.\,\ref{fig:lycont}). As discovered by MUSE, star-forming galaxies at high redshift show a broad diversity of Lyman-$\alpha$ halos (in terms of sizes and geometry), which trace the gas in the circumgalactic medium \citep{2016A&A...587A..98W, 2017A&A...608A...8L}. By investigating the links between LyC emission and the Lyman-$\alpha$ spectral/spatial shapes, BlueMUSE will provide unique and robust tests to probe ionizing sources during the epoch of reionization.
%correlations between the Lyman-$\alpha$ spatial extent and the escape of LyC emission, as proposed by \citet{2017A&A...601A..73M}.

Valuable information on the nature of LyC emitters could also be extracted from BlueMUSE spectra. In the red, the rest-frame far-UV emission ($\sim1400$ \AA) can be detected up to $z=3.5$, while at the blue end, BlueMUSE will probe ionising emission from the Lyman limit (912 \AA) down to 780 \AA\  to add new constraints on the typical shape of the Lyman continuum. In addition, we can expect to detect Lyman-series absorption lines to get insight into the \hi\ content of galaxies (column density, covering fraction) and its link with LyC escape. Detecting these lines at $z\sim$ 3 with BlueMUSE would be outstanding since such observations are usually limited to galaxies at much lower redshift.
BlueMUSE would also benefit greatly from ancillary data based on multiple-band imaging (e.g HDF) and MUSE existing observations (e.g.\,\ciii] emission; \citealt{2017A&A...608A...4M}) combined with SED fitting and photoionization models to assess the properties of the stellar population and the interstellar medium (mass, age, metallicity, ionization parameter) of the LyC emitters.

\begin{subsec}[ht]
\begin{tcolorbox}[colback=green!5!white,colframe=green!75!black,title=Why observing at a high spectral resolution?]

BlueMUSE will provide a spectral resolution 3000-5000 across the wavelength range, which is twice the one from MUSE in the region of overlap (Fig.\,\ref{fig:specres})

\begin{itemize}
\item
  \begin{quote}
  High resolution helps with measuring small velocity shifts and velocity dispersion, which is important in particular for the stellar science cases. (\S\ref{key-science-case-massive-stars} and \ref{globular-clusters})
  \end{quote}
\item
  \begin{quote}
  This resolution also provides more precise spectral line profiles, which allows for a more precise physical modelling of the emission or absorption, resolve line doublets, as well as study variations of line shapes across an object (\S\ref{key-science-case-ism-and-hii-regions-extreme-starbursts},\ref{lyman-continuum-emitters}).
  \end{quote}
\end{itemize}
\end{tcolorbox}

\end{subsec}

\medskip\par

\textbf{Breakthrough with BlueMUSE:} 

Thanks to a large field of view, a
blue spectral range, and a high sensitivity, BlueMUSE will
simultaneously probe LyC and Lyman-$\alpha$ for a statistical sample of low mass galaxies
with high Lyman-$\alpha$ spectral resolution at $3<z<4$,
where the IGM attenuation is still modest.
%($\sim0.4$ for the mean value, \citealt{2014MNRAS.442.1805I}, see Fig.\,\ref{fig:lycont}). 
These objects being presumably analogues of the main ionizing galaxies at higher redshift, 
%\citep{2018MNRAS.474.4514I,2018arXiv180506071S}, 
deep surveys with BlueMUSE will put unique constraints on the physical properties of the sources responsible for cosmic reionization as well as their observability with future instruments (e.g., JWST) based on indirect diagnostics. 
%As demonstrated by MUSE studies, BlueMUSE will be particularly efficient at detecting the faintest population of $z\sim3$ galaxies.

\medskip\par

\textbf{Feasibility:} 

The typical LyC flux levels of the individual detections at $z\sim3$ reported by \citet{2018MNRAS.476L..15V,2018ApJ...869..123S} are around $1-2\times10^{-19}$ erg\,s$^{-1}$\,cm$^{-2}$\,\AA$^{-1}$. This can be reached at 3$\sigma$ level in 30~hrs exposure with BlueMUSE. It is worth pointing out that the majority of sources to be found in a single BlueMUSE field will certainly be fainter than the UV-bright galaxies reported by \citet{2018MNRAS.476L..15V,2018ApJ...869..123S} but their LyC escape fraction are probably high \citep{2018MNRAS.474.4514I,2014MNRAS.442.2560W} which may lead to similar observed flux levels. In addition, detailed stacking analysis will anyway be possible thanks to the large predicted number of sources to be observed per field (see Fig.\,\ref{fig:lae_counts}).  

%This science case will benefit from any archival MUSE and HST data over the
%deep fields, allowing to extend the systematic search for Lyman
%continuum emitters over a broader redshift window.

\clearpage

  \hypertarget{gravitational-lensing-in-clusters}{%
  \subsection{Gravitational lensing in
  clusters}\label{gravitational-lensing-in-clusters}}

\begin{tcolorbox}[colback=blue!5!white,colframe=blue!75!black,title=Science Goals]
\begin{itemize}
\item
  Confirm the redshift of a large number of multiply-imaged lensed
  systems in massive cluster cores, and use them as precise constraints
  for dark matter distribution.
\item
  Construct the largest statistical sample of low mass galaxies at
  $z=2-4$ to probe the faint-end of the Luminosity Function.
\item
  Study the spectroscopic properties of typical Lyman-$\alpha$ haloes at sub-kpc scales.
\end{itemize}
\end{tcolorbox}

Massive galaxy clusters locally curve Space-Time and thus stretch and
magnify the light of background galaxies. Within the central square
arcminute, distant sources are generally multiply imaged by the lensing
effect and the magnification of these images is generally larger than a
factor of a few but can reach factors of hundreds for gravitational arcs
straddling the critical lines. Thanks to the lensing magnification we can
study lower luminosity/mass galaxies that would otherwise be
undetectable. The identification of the multiple images and the precise
determination of their redshifts offer a unique opportunity to map the
total mass distribution of the cluster cores including the unknown dark
matter.

In the best cases, such as for the Hubble Frontier Field Clusters, the
precision of the mass determination is better than the percent level
(e.g., \citealt{2014MNRAS.443.1549J}). However, this is only possible if a large
number of multiple images have measured precise redshifts \citep{2015MNRAS.446L..16R}. Measuring a large number of redshifts of multiple images in a
cluster has only been possible routinely with the current MUSE
instrument, thus revolutionising the cluster lensing modeling power
(e.g., \citealt{2017MNRAS.469.3946L}), with potential further application to put
constraints on the properties of dark matter cross-section (e.g., \citealt{2017MNRAS.472.1972H}) or cosmological parameters (e.g., \citealt{2010Sci...329..924J}). Using
these precise models, we can then correct for the lensing magnification
of distant sources, thus extracting their full properties.

\begin{figure}
    \begin{minipage}{8cm}
    \includegraphics[width=8cm]{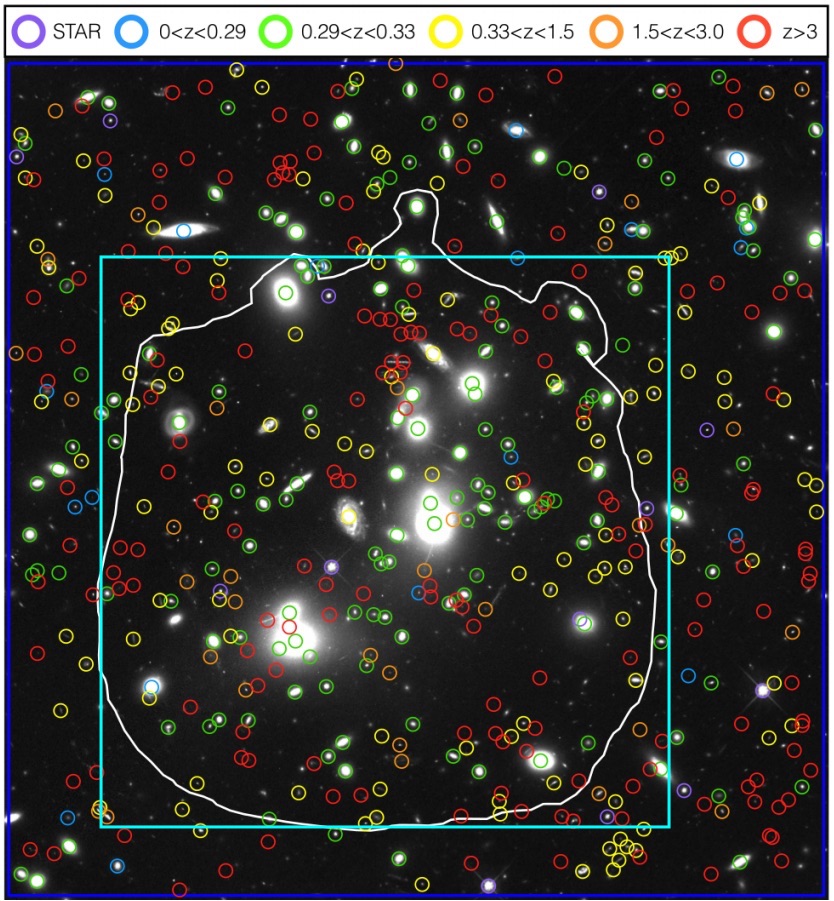}
    \end{minipage}
    \begin{minipage}{10cm}
    \centering
    \includegraphics[width=8cm]{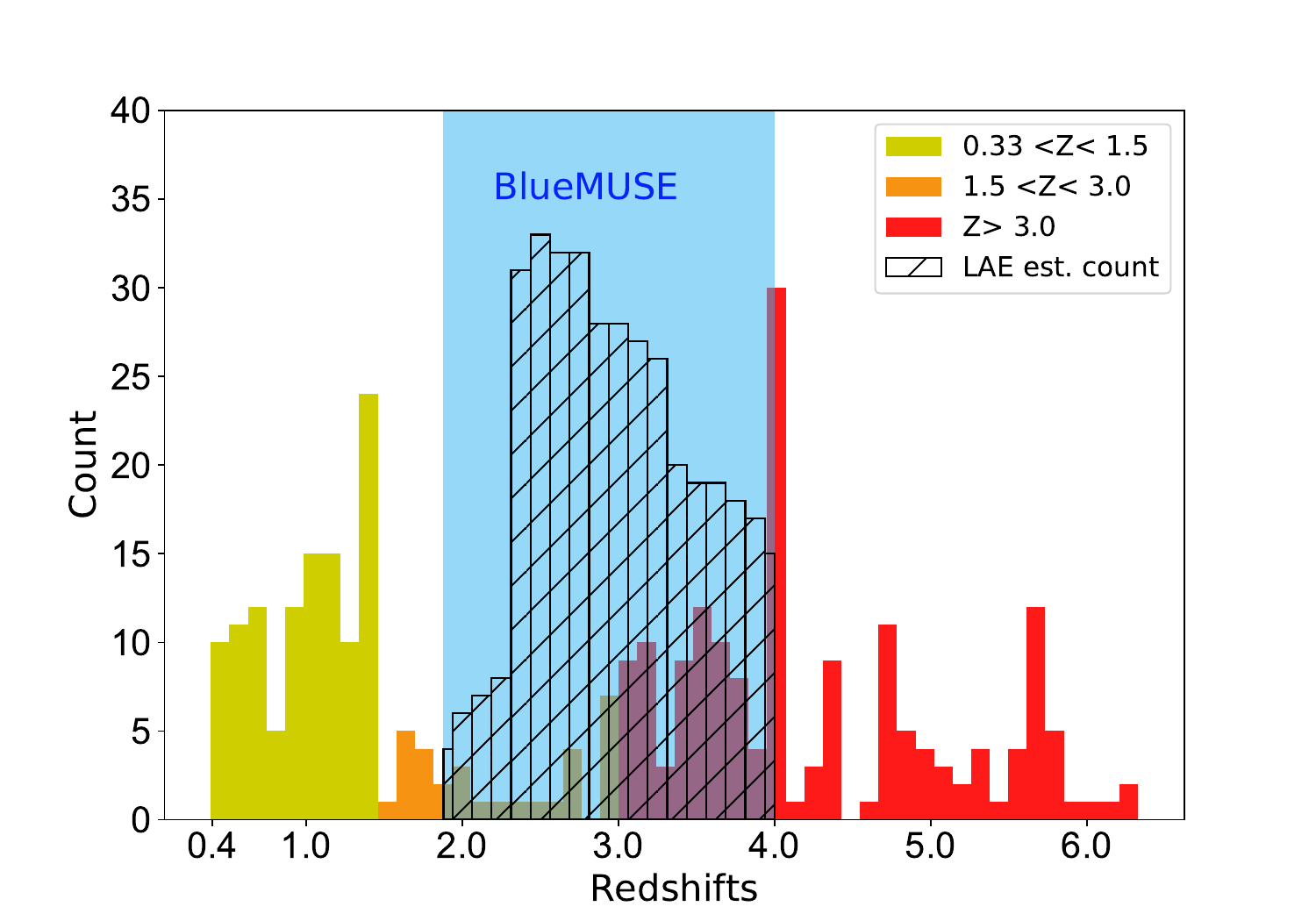}
    \caption{\label{fig:einstein}
(adapted from \citealt{2018MNRAS.473..663M}): Redshift
measurements over a 2$\times$2 MUSE mosaic of the Frontier Field cluster Abell
2744. The white line delineates the region where we expect multiple
images in the cluster core. The cyan region shows the size of BlueMUSE
FoV. Coloured circles identify spectroscopically confirmed background
galaxies lensed by the cluster. The redshift histogram from MUSE shows
that we miss most of $z=2-3$ sources and will find them with BlueMUSE (hatched histogram).}
  \end{minipage}
\end{figure}

\begin{itemize}
\item
  \textbf{Multiple images and faint LAEs:} For a typical massive cluster
  at $z\sim0.3$ the Einstein radius (characterising the
  scale of strong lensing) is $\sim$50\arcsec\ and the area of
  high magnification fits perfectly the FoV of BlueMUSE (Fig.\,\ref{fig:einstein}) and
  does not require mosaicing / multiple pointings, contrary to MUSE. In
  addition, due to the volume reduction in the source plane, the
  redshift distribution of lensed background galaxies is dominated by
  the population of LAEs, as currently seen in MUSE surveys (Fig.\,\ref{fig:einstein}).
  MUSE has uncovered and unambiguously confirmed the redshift of a large
  number of multiple systems even beyond the detection limits of deep
  HST images in the Frontier Fields survey \citep{2018MNRAS.473..663M}. By
  probing the redshift range $2<z<4$ in Lyman-$\alpha$,
  we expect a much larger number of background sources (a factor $\sim5$
  based on the LAE LF, see ``deep fields'' science case \S\ref{deep-fields}). BlueMUSE will
  therefore confirm spectroscopic redshifts for multiple images in a
  very efficient way. As demonstrated by \citet{2018arXiv181206981H}, such a high density of constraints per cluster makes it possible to test models of self-interacting dark matter. BlueMUSE will also contribute significantly in confirming the
  population of dwarf galaxies found at $z=2$ with HST down to AB
$\sim$ -14 \citep{2016ApJ...832...56A}.
\item
  \textbf{Giant arcs:} because they require an almost-perfect alignment between a large background galaxy and the center of a lens, highly magnified giant arcs are quite unique. Only a very small number ($<10$)
  of $z>4$ galaxies are known to be extremely extended (extended
  by 10 arcsec or more in a single image) in Lyman-$\alpha$. By moving to $z=2$ the number
  of such high redshift giant arcs increases significantly, allowing us
  to probe very small scales in the source plane ($<1$ kpc,
\citealt{2016MNRAS.456.4191P}, \citealt{2017ApJ...843L..21J}, Claeyssens et al. submitted). These giant arcs are
  perfect laboratories to study the mechanisms of Lyman-$\alpha$ and UV
  emission, and their high continuum level make them unique background sources for resolved CGM absorption studies (e.g., \citealt{2018Natur.554..493L}, %\citealt{2019arXiv190408186R}, 
  see also the gas flows science case \S\ref{c.-probing-the-cgm-with-metals})
\item
  \textbf{Critical line mapping:} Mapping a significant fraction of the
  surroundings of the critical lines with BlueMUSE has additional
  interests. The identification of new multiple images with
  spectroscopic redshifts along the critical lines will help further
  improve the mass distribution models of the clusters. The
  spectroscopic identification of several multiple images at different
  redshifts allows us to constrain the cosmological parameters (e.g., \citealt{2010Sci...329..924J}). Compared to blank fields, isolated emission lines
  in strong-magnification regions can be easily classified as either
  high-$z$ or low-$z$ interlopers through multiple-imaging considerations.
\end{itemize}

\textbf{Uniqueness of BlueMUSE:}

%\begin{enumerate}
%\def\labelenumi{(\arabic{enumi})}
%\item

(1)  the discovery space in redshift allowed by the blue wavelengths,
  probing a larger number of faint background galaxies and multiple
  images compared to any other instrument. The cluster members also
  contaminate less in continuum at blue wavelengths (the lens is more
  transparent). This is crucial when searching for multiple images in the very core of the clusters, which are usually demagnified and obscured by bright foreground emission but give unique constraints on the central mass distribution.
%\item

(2)  The FoV of BlueMUSE perfectly fits the typical Einstein radii of
  $z=0.2-0.5$ massive clusters (such as the Frontier Fields, e.g., \citealt{2014MNRAS.444..268R}) and 
  the region of multiple images.
%\end{enumerate}

%\includegraphics[width=3.19236in,height=3.44375in]{}

  \hypertarget{the-emergence-of-the-first-galaxy-clusters}{%
  \subsection{The emergence of the first galaxy
  clusters}\label{the-emergence-of-the-first-galaxy-clusters}}

\begin{tcolorbox}[colback=blue!5!white,colframe=blue!75!black,title=Science Goals]
\begin{itemize}
\item
  Allow the identification and characterization of Lyman-$\alpha$
  nebulae in a large number of clusters at $1.87<z<3$,
  unveiling their luminosities, kinematics, and physical properties (via
  additional \civ~$\lambda$1550 and \heii~$\lambda$1640 \AA\  lines), with crucial insights
  into cold accretion onto the most massive early structures and galaxy
  evolution models.
\item
  BlueMUSE will be a machine for cluster redshift measurements at
  $z>1.87$: via Lyman-$\alpha$ nebulae, Lyman-$\alpha$ emitters (present in
  all structures known) and UV absorption lines.
\item
  Statistical samples with BlueMUSE redshift distributions of the first
  generation of clusters from \textit{Euclid}/SKA/\textit{Athena} will crucially
  constrain cosmology, taking into account that total masses could
  be measured via X-ray (\textit{Athena}) and SZ (ALMA).
\item
  Map the evolving impact of star formation and AGN feedback into the
  clusters hot media, constraining its thermodynamic evolution.
\end{itemize}
\end{tcolorbox}

\begin{figure}
    \begin{minipage}{7cm}
    \includegraphics[width=7cm]{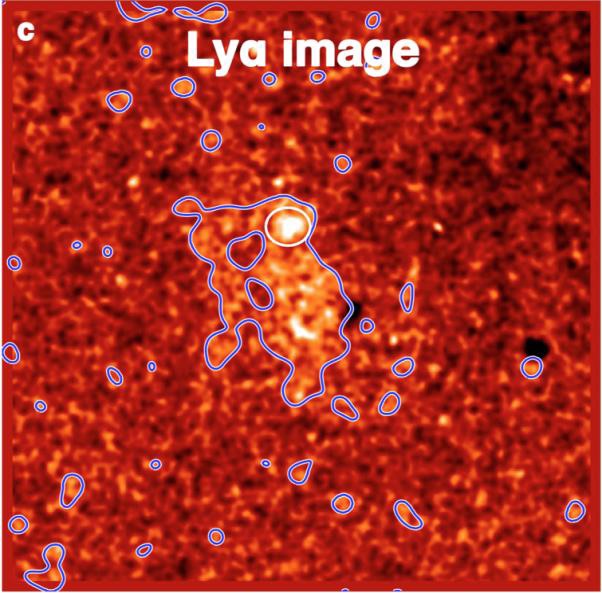}
    \end{minipage}
    \begin{minipage}{9cm}
    \includegraphics[width=9cm]{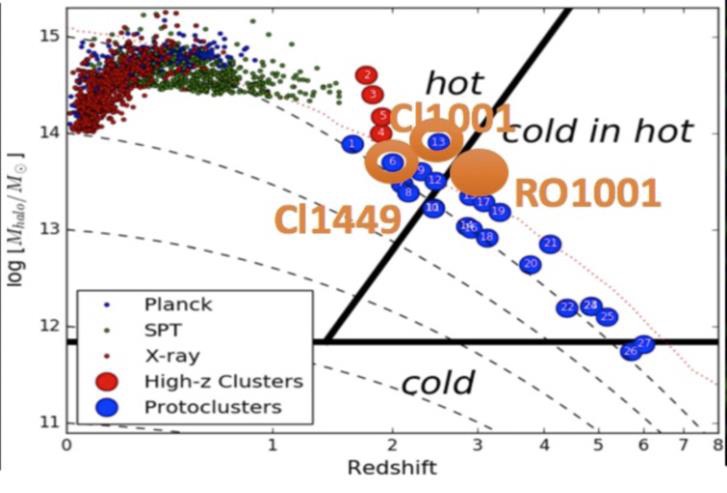}
    \end{minipage}
    \caption{Left: Lyman-$\alpha$ image of the cluster Cl 1449 at $z=1.99$ \citep{2016ApJ...829...53V}, showing the detection of a giant Lyman-$\alpha$ nebula  (extending $>100$ kpc) sitting in
its core. Right: theory prediction \citep{2009Natur.457..451D} on the region of
the redshift-dark matter mass parameter space where cold accretion
versus hot haloes are expected (adapted from \citealt{2016A&ARv..24...14O}).
Typical structures currently investigated are shown, including the
clusters in which we observed and detected giant Lyman-$\alpha$ nebulae (gold
circles with labels; Daddi et al. in preparation).}
    \label{fig:highzcluster}
\end{figure}

BlueMUSE clearly appears as the best means to achieve these goals by
discovering and characterizing \textbf{diffuse Lyman-$\alpha$ nebulae that now
we know are widespread in the early generation of galaxy clusters}.
Following the initial discovered in the dense core of a prototypical
galaxy cluster progenitor, CL J1449+0856 at $z=1.995$ (Fig.\,\ref{fig:highzcluster}-left,
 \citealt{2016ApJ...829...53V,2011A&A...526A.133G,2013ApJ...776....9G,2013ApJ...772..118S}),
we have recently used KCWI at Keck (in collaboration with Michael Rich at the University of California) to search for giant Lyman-$\alpha$ nebulae
inside several more $z>2$ structures: \textbf{we see giant Lyman-$\alpha$
nebulae in all of them}. This includes the most distant X-ray detected
cluster known (CL J1001+0220 at $z=2.506$; \citealt{2016ApJ...828...56W})
%, a radio-galaxy traced cluster at $z=1.99$ \citep{2016ApJ...830...90N}, 
and the most spectacular detection is in a
radio-selected forming cluster core  at $z=2.9$ (Daddi, Rich, et al.,
in preparation). %Interestingly, we have also targeted the much more massive z = 1.99, SZ-detected and X-ray luminous cluster XLSSC 122 \citep{2014MNRAS.440.2077M,2017MNRAS.472.2877M}, and found nothing. 
%This suggests that Lyman-$\alpha$ nebulae are
%connected with forming clusters rather than mature (and dead) ones. 
The
presence of these nebulae unambiguously demonstrates that cold gas is
co-existing with hot gas inside the deep potential well of these
structures \citep{2016ApJ...829...53V,2016ApJ...828...56W}.
The nature, origin, and fate of the cold gas is still unclear and a
matter of debate: it might be related to cold gas accretion and/or arise
from feedback between galaxy activity and the intra-cluster medium.

The identification and characterization of the most distant,
$z\gtrsim2$ galaxy clusters is currently a very
active topic of research (see \citealt{2016A&ARv..24...14O} for a recent review).
It is also a recent one, as these dense structures, already consistent
with a single massive dark matter halo (see discussion in \citealt{2015ApJ...802...31D}) as opposed to Mpc-scale loose over-densities like proto-clusters,
have only started to be discovered in the past decade. They represent
the earliest generation of massive collapsed structures, progenitors to
Coma-like $z\sim$ 0 clusters. As such, their abundance can be used to
constrain cosmological parameters. They are also unambiguous formation
sites of massive early type galaxies, and hence can shed light on the
elusive processes that lead to galaxy transformations, leading to the
build-up of this galaxy population that dominates local clusters. These
high-redshift forming clusters also offer a unique opportunity to
statistically map the feedback from SF/AGN activity through outflows and
radiation effects (see e.g., \citealt{2016ApJ...829...53V}) affecting the key,
early thermodynamic evolution of the hot intra-cluster medium (ICM)
expected by leading models to mainly happen at $z\sim1.5-3$
(e.g., \citealt{2014MNRAS.441.1270L}). At these same redshifts of
$z\sim1.5-3$, crucially the redshift range newly enabled by
BlueMUSE, simulations predict that massive $\sim$~10$^{13-14}$
M\textsubscript{$\odot$} dark matter haloes should become efficiently shielded
by this hot atmosphere, thus preventing further large-scale infall of
cold gas into their cores (\citealt{2009Natur.457..451D}; Fig.\,\ref{fig:highzcluster}-right). This
process is however still not well constrained and observations from
BlueMUSE will crucially enable us to trace the epoch and duration of
this transition. Evidence of persistent SF activity in massive
structures at $z\sim 2-2.5$ suggests that this might occur later and/or at
higher halo masses than currently expected \citep{2015ApJ...801..132V,2016ApJ...829...53V,2016ApJ...828...56W,2016A&ARv..24...14O}, but statistics from a larger
number of clusters and samples at the highest possible redshifts are
required for definitive conclusions.

BlueMUSE will uniquely allow us to discover Lyman-$\alpha$ emitting nebulae in
the first generation of forming clusters down to $z=1.87$, thus allowing
exploration of the critical redshift range 1.8-3 where we expect cold
accretion to massive haloes to peak due to the competing effects of
hierarchical mass assembly of clusters (favouring the presence of more
massive structures towards lower redshifts) and cosmic evolution of
accretion rates (rapidly increasing with redshift at fixed structure
mass). BlueMUSE will be able to statistically investigate these
competing processes for the first time.

\medskip\par
\textbf{Uniqueness of BlueMUSE:}

In the extended mode, MUSE currently allows observations down to 4650 \AA, thus allowing
detection of Lyman-$\alpha$ emission only at $z\gsim3$.
KCWI at Keck is starting to scratch the surface of this science but is
strongly limited by its small field of view: when matching the spectral
resolution and wavelength range of BlueMUSE, it covers a $40\times$ smaller
field of view, making it nearly impossible to obtain a panoramic view of
the diffuse Lyman-$\alpha$ nebulae extending to the outskirts where critically
we expect to detect connection to feeding filaments from the
intergalactic medium.

While currently the sample of high redshift clusters known is still very
limited, there are great prospects for the future: \textit{Euclid} (launch 2022)
would be able to identify all the already known $z>2$ clusters
even from its shallower, nearly full sky coverage, promising the
identification of order of thousands of candidates over
$1.87<z<3$, based on extrapolations of current
numbers. A very promising means to select the most active first
generation of clusters will be based on radio-overdensities \citep{2017ApJ...846L..31D}, and SKA (operative from 2023) will also be able to provide many
hundreds candidates over this redshift range. Finally, \textit{Athena} (2029+, \citealt{2013arXiv1306.2307N})
will have fantastic X-ray sensitivity and thus provide large number of
candidates as well as characterizing their X-ray emission in detail,
allowing to study the interplay between the hot (X-rays) and cold
(Lyman-$\alpha$) media, and unveiling cluster total masses simultaneously with
SZ from ALMA. 
ALMA will have the
power to chart star formation and cold gas distributions in these kind
of clusters, and complementary spectroscopy from JWST and ELT/HARMONI
will unveil stellar population and gas physical properties.
\medskip\par
\textbf{Feasibility:} 

Based on current KCWI observations and
luminosities recovered so far, panoramic detection and basic kinematic
characterization of the giant Lyman-$\alpha$ nebulae inside high redshift
clusters will require only of order 1~hr integration with BlueMUSE. Longer
integrations on selected targets will be able to unveil their physical
properties (from weaker lines like \civ\ and \heii) and overall
morphology down to lower surface brightness levels, with connection to
the larger scale structure.

\newpage

\hypertarget{uniqueness}{%
\section{Uniqueness}\label{uniqueness}}

By providing a large field IFU on an 8m-class telescope at blue
wavelengths, BlueMUSE is a unique instrument to tackle all the science
cases presented in this document. Among the three main characteristics which are
specific to BlueMUSE (its wavelength range, its spectral resolution and
its FoV), we have highlighted in Table\,\ref{uniquetable} the ones which make the
science case much more competitive. The only two other similar
instruments on the same telescope class are MUSE and KCWI.

\textbf{MUSE:}
by construction, BlueMUSE has a very similar architecture to MUSE and
the two instruments overlap in wavelength in the range 480-600 nm
(465-600 nm in MUSE extended mode). However, even in this overlap region
BlueMUSE provides a much higher sensitivity (see `Performance' \S\ref{expected-performance}) and twice the spectral resolution of MUSE, in
addition to the wider area (2 arcmin$^2$) making it twice
as efficient. For each of the science cases presented, Table\,\ref{uniquetable} (right
column) shows that they would be either unfeasible, or done much less
efficiently, with MUSE.

\begin{figure}
    \centerline{\includegraphics[width=9cm]{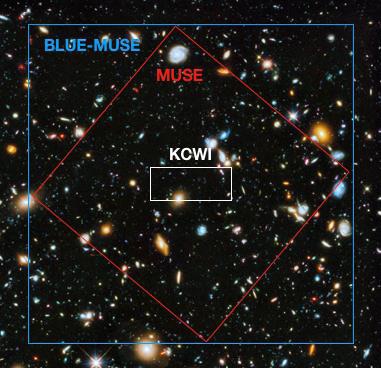}}
    \caption{Comparison between the BlueMUSE, MUSE and KCWI Fields of view, compared to an ACS image of the Hubble Ultra Deep Field. At the same spatial and spectral resolution and wavelength coverage as BlueMUSE, KCWI only provides a FoV of 8.24 $\times$ 20.4\arcsec, a factor of 40 smaller compared to the 1.4$\times$1.4 arcmin$^2$ FoV proposed for BlueMUSE.}
    \label{fig:uniqueness}
\end{figure}

\textbf{KCWI:} the Keck Cosmic Web Imager (KCWI, \citealt{2010SPIE.7735E..0MM}) is
the only other instrument which provides an IFU with similar
characteristics (wavelength range, spectral resolution) as BlueMUSE.
However there is a strong difference in FOV (Fig.\,\ref{fig:uniqueness}). Indeed, at the
same spectral range and spatial / spectral resolution as BlueMUSE, the FOV from KCWI is
8.24 $\times$ 20.4\arcsec, which is a factor of 40 smaller. This makes KCWI
unsuitable to cover large areas to a high depth and reduces its discovery potential, unlike BlueMUSE. In
addition, BlueMUSE is optimised to a single mode of operations and we
expect its transmission to be 1.5-2$\times$ higher than KCWI.

\newpage

\begin{longtable}[]{@{}lcccl@{}}
\toprule
\endhead
Science Case &
\begin{tabular}{@{}c@{}}
Wavelength \\ Coverage 
\end{tabular}
& 
\begin{tabular}{@{}c@{}}
Spectral \\ Resolution 
\end{tabular} & FoV & Why not doable
with MUSE?\\\hline \hline
Massive stars & \checkmark
& \checkmark & \checkmark & 
\begin{tabular}{@{}l@{}}
No access to Teff / log g diagnostic \\
lines in the blue
\end{tabular}\\\hline
Globular clusters & \checkmark
& \checkmark &  & 
\begin{tabular}{@{}l@{}} Cluster populations cannot be split \\ 
due to wavelength range and poorer \\
spectral resolution
\\\end{tabular}\\
\hline
\begin{tabular}{@{}l@{}}
Ultra Faint Dwarf \\ galaxies\end{tabular} & \checkmark& \checkmark& \checkmark& 
\begin{tabular}{@{}l@{}}
Precision on velocity and chemical \\ 
abundances is lower due to wavelength\\
range and poorer resolution
\end{tabular}\\\hline
Ionized nebulae & \checkmark& & \checkmark& 
\begin{tabular}{@{}l@{}}
No access to many optical\\ 
recombination lines (ORL)
\end{tabular}\\\hline
Comets & \checkmark& & \checkmark& 
\begin{tabular}{@{}l@{}}
Main radicals are outside of MUSE \\
spectral
range
\end{tabular}\\\hline
Extreme starburst galaxies & \checkmark& \checkmark& \checkmark& 
\begin{tabular}{@{}l@{}}
Not possible at this redshift\\
(T$_e$-sensitive diagnostics needed)
\end{tabular}\\\hline
\begin{tabular}{@{}l@{}}
Low surface brightness\\
galaxies\end{tabular}& \checkmark& & \checkmark& 
\begin{tabular}{@{}l@{}}
Blue lines are needed ({[}O
II{]}, Balmer)\\
and not accessible by MUSE
\end{tabular}\\\hline
Environmental effects & \checkmark & \checkmark& \checkmark& \begin{tabular}{@{}l@{}}
Not possible for lower mass galaxies\\
because of spectral resolution
\end{tabular}\\\hline
%Census of shocks and outflows & & & & Spectral resolution is required for precise measurements\tabularnewline
Deep fields & \checkmark& & \checkmark& \begin{tabular}{@{}l@{}}
4 times less efficient due to number\\
densities and field of view. \\
Different redshift range.
\end{tabular}\\\hline
Gas flows & \checkmark& \checkmark& \checkmark& SB dimming makes it 4$\times$ less efficient\\\hline
Lyman-continuum emitters & \checkmark& \checkmark& & \begin{tabular}{@{}l@{}}
IGM makes it unfeasible. \\
Too low
resolution.\end{tabular}\\\hline
Lensing clusters & \checkmark& & \checkmark& \begin{tabular}{@{}l@{}}
Less efficient ($\times$2 in FoV, \\
$\times$2 in number
density)\end{tabular}\\\hline
High redshift clusters & \checkmark& & \checkmark& \begin{tabular}{@{}l@{}}
Not feasible (no suitable targets\\and SB
dimming)\end{tabular}\\\hline
\bottomrule
\caption{\label{uniquetable}Uniqueness of the presented BlueMUSE science cases: checked cells are the instrument characteristics which
are mandatory for the feasibility of the BlueMUSE science cases. The
right columns explains why the science case would be unfeasible, or much
less efficient, with MUSE.}
\end{longtable}

\textbf{Serendipitous science: }

Astronomy is to a significant degree still driven by unexpected
discovery (e.g., dark matter and dark energy). These discoveries are
often made by pushing the limit of observations with the most powerful
telescopes and/or opening a new area of the instrumental parameter
space.

Like MUSE, a single BlueMUSE observation provides 90000 spectra over
4000 wavelengths in one go. Thanks to its 360 million voxels (volume
pixels), each BlueMUSE data cube produced by a single pointing
observation is information rich. A good example of the potential
richness of the BlueMUSE information content is given in the recent
paper by \citet{2018MNRAS.478.1595C}, which report the discovery of a new
low-redshift galaxy-scale gravitational lens, identified from a
systematic search of publicly available MUSE observations. The
probability of serendipitous discovery being proportional to the probed
volume, BlueMUSE with its 2 arcmin$^{2}$ field of view,
will have roughly twice the discovery potential of MUSE. Compared to
other MOS instruments or small field IFUs, like KCWI which can only perform pointed
observations, BlueMUSE has thus a unique potential for serendipity
discoveries.

\bigskip\par 

\textbf{High sensitivity:}

\begin{figure}
    \centering
    \includegraphics[width=16cm]{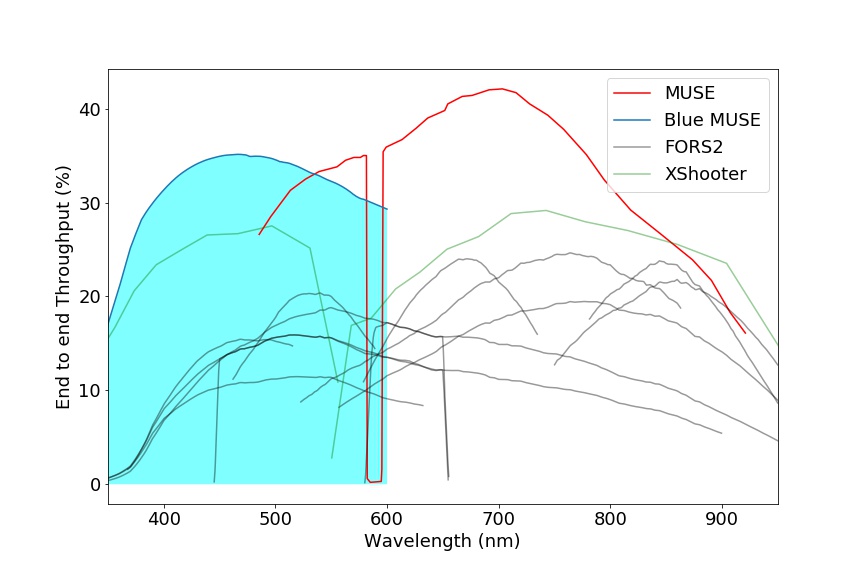}
    \caption{ Comparison between the expected BlueMUSE transmission curve
(end-to-end, including instrument, telescope and atmosphere) compared to
the same transmissions for other spectrographs currently on VLT and
working at similar wavelengths and resolution. A 15\% slit loss has been
included in all slit spectrograph transmissions when compared to the
MUSE and BlueMUSE IFUs.
}
    \label{fig:sensitivity}
\end{figure}

One of the main characteristics of the optical design of MUSE, and of the BlueMUSE design as well, is a very high sensitivity,
reaching up to $\sim$40\% end-to-end at the peak (including
atmosphere and telescope). A precursor developed by the manufacturer Winlight System (F) has already demonstrated the the expected level of efficiency in the blue is feasible \citep{2016SPIE.9912E..22M}. This makes MUSE and BlueMUSE unique
instruments, having the highest sensitivity of all optical spectrographs
on VLT at their respective wavelength (Fig.\,\ref{fig:sensitivity}). \textbf{Note also that BlueMUSE
will be more efficient in the blue-UV area than any instrument at the
ELT.}

\newpage

\hypertarget{synergies}{%
\section{Synergies}\label{synergies}}

\begin{figure}
    \centering
    \includegraphics[width=16cm]{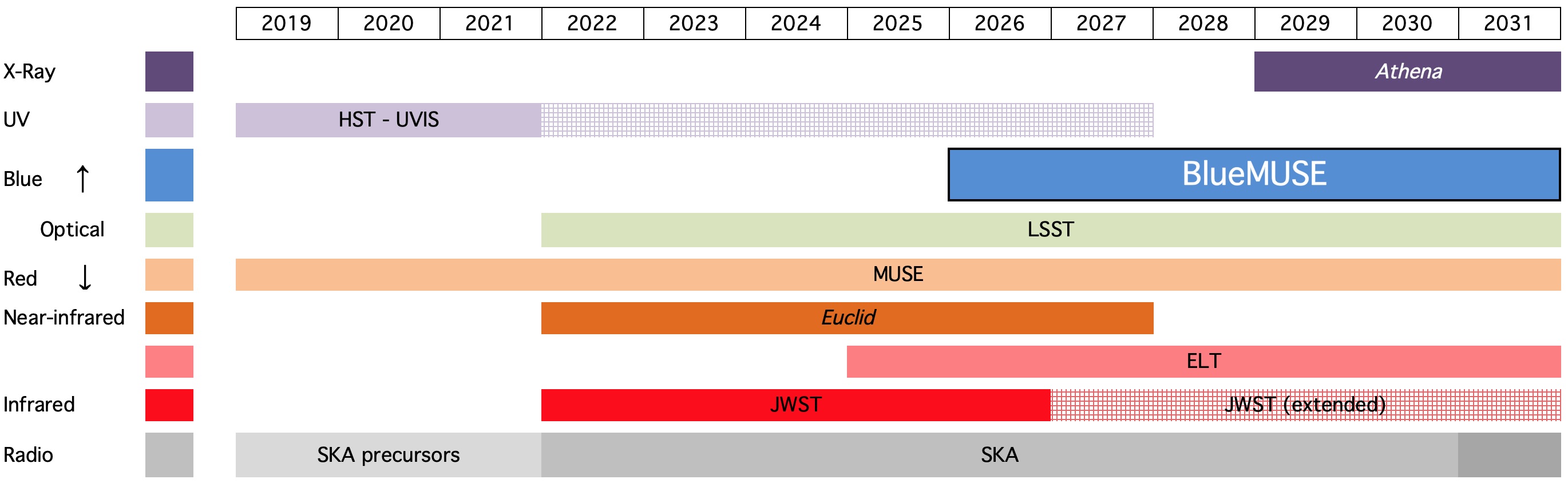}
    \caption{ Overview of the BlueMUSE timeline in the context of other major
future facilities, colour-coded as a function of their main wavelength
domain. BlueMUSE will largely complement future facilities focusing on
red and near-infrared wavelengths.
}
    \label{fig:timeline}
\end{figure}

With an expected first light around 2026, BlueMUSE will be at the
telescope at a time when all other major facilities (ELT, JWST in
particular) will focus on red and near-infrared wavelengths (see the
schematic view Fig.\,\ref{fig:timeline}). By providing a complement at shorter
wavelengths, BlueMUSE will have strong synergies with these facilities,
and will be the perfect follow-up spectroscopic instrument at short
wavelengths at a time when HST will, presumably, no longer exist. 
All
the science cases presented have synergies with future major facilities,
in particular SKA at longer wavelengths, and will benefit from all the
targets identified by \textit{Euclid} and LSST in large field imaging.

With a field of view and a sensitivity comparable to NIRSpec, BlueMUSE deep fields will be very complementary to
JWST deep fields: for example the $z=2-4$ NIRSpec population of
faint emitters in H$\alpha$ and [\oii] will also be easily detected with
BlueMUSE as LAEs providing key information such as Lyman-$\alpha$ escape
fraction. The ALMA Deep Fields are also important as they provide a
large number of targets with an overlapping redshift with BlueMUSE.

The unique survey capability and discovery potential of BlueMUSE will be
able to provide numerous targets to ELT. For example HARMONI follow-up
studies of $z=2$ extended Lyman-$\alpha$ haloes detected by BlueMUSE will enable
spatially resolved study of the Lyman-$\alpha$ escape fraction in the CGM.

\newpage

\hypertarget{conclusion}{%
\section{Conclusion}\label{conclusion}}

In this document, we have presented a selection of science cases where we believe BlueMUSE will have a very strong impact due to its unique capabilities. They however only represent a preview over the full range of science allowed by a blue-optimised wide-field monolithic IFU. Indeed, we expect BlueMUSE to be a transformative instrument for many science cases, including some that we can not predict today, like it has been the case for MUSE. By focusing on the blue/UV wavelength range, at a time where most new facilities will be operating at red/IR wavelengths, BlueMUSE will open up new discovery space, while allowing more MUSE-like science for the benefit of the community.

\section*{Acknowledgements}
\addcontentsline{toc}{section}{Acknowledgements}
We acknowledge helpful discussions with Guy Monnet during the preparation of this document. This work was supported by the Programme National Cosmologie et Galaxies (PNCG) of CNRS/INSU with INP and IN2P3, co-funded by CEA and CNES. The authors acknowledge the support of the European Research council through the H2020 - ERC-STG-2015 / 678777 ICARUS program.

%%%%%%%%%%%%%%%%%%%%%%%%%%%%%%%%%%%
%\appendix
%
%\clearpage
%
%\section{Acronyms}\label{a1:acronyms}
%\input{Misc/Acronyms}

%%%%%%%%%%%%%%%%%%%%%%%%%%%%%%%%%%%
\clearpage

\addcontentsline{toc}{section}{References}
\printbibliography
\end{document}